\documentclass[longauth]{aa}

\usepackage{graphicx}
\usepackage{natbib}
\usepackage{scalerel}
\usepackage{comment}

\usepackage{tabularx}
\usepackage{booktabs} % optional, for nicer rules
\usepackage{multicol}
\usepackage[table, rgb]{xcolor}
\usepackage{tikz}
\usepackage{lastpage}
\usetikzlibrary{
    positioning,
    arrows.meta,
    shapes.geometric,
    shapes.misc,
    shapes.symbols,
    fit,
    backgrounds,
    calc,
}

\usepackage{txfonts}
\usepackage[pdfencoding=auto,psdextra]{hyperref}
\hypersetup{
    colorlinks=true,
    linkcolor=blue,
    filecolor=magenta,      
    urlcolor=blue,
    citecolor=blue
}
\usepackage{cleveref}

\makeatletter
\renewcommand*\aa@pageof{, page \thepage{} of \pageref*{LastPage}}
\makeatother

\usepackage[utf8]{inputenc}
\usepackage{euclid}
\let\linenumbers\relax
\usepackage{xspace}

\newcommand{\ext}{OU-EXT\xspace}
\newcommand{\stack}{\texttt{DpdExtStackedFrame}\xspace}
\newcommand{\stacks}{\texttt{DpdExtStackedFrames}\xspace}
\newcommand{\sef}{\texttt{DpdExtSingleEpochFrame}\xspace}
\newcommand{\sefs}{\texttt{DpdExtSingleEpochFrames}\xspace}
\newcommand{\dear}{\texttt{DpdExtAnalysisResults}\xspace}
\newcommand{\pf}[1]{\texttt{PF#1}\xspace}
\newcommand{\equal}{\texttt{EQUAL}\xspace}
\newcommand{\swarp}{\texttt{SWarp}\xspace}
\newcommand{\sextractor}{\texttt{SEXtractor}\xspace}
\newcommand{\psfex}{\texttt{PSFEx}\xspace}
\newcommand{\sepp}{\texttt{SourceExtractor++}\xspace}

\newcommand{\gaia}{\textit{Gaia}\xspace}
\newcommand{\change}[1]{#1}

\begin{document}

\title{Euclid Data Release 1 (DR1)}
\subtitle{Colouring \Euclid with external ground-based imaging data}
%%%% Version Monday 10th of March 2025 04:53:55 PM UT							

%%%% Version Wednesday 11th of November 2026 05:41:43 PM UT
%%%% Assumes the new A&A style file from Oct 2025 or later
%%%% Please do not edit the author list -- contact ECEB Bureau for changes
\newcommand{\orcid}[1]{} %% if already defined in aa.cls: comment, or use renewcommand			   
\author{Euclid Collaboration: P.~Corcho-Caballero\orcid{0000-0001-6327-7080}\thanks{\email{p.corcho.caballero@rug.nl}}\inst{\ref{aff1}}
\and E.~Balbinot\orcid{0000-0002-1322-3153}\inst{\ref{aff1},\ref{aff2}}
\and J.~T.~A.~de~Jong\orcid{0000-0002-9511-1357}\inst{\ref{aff1}}
\and M.~Fabricius\orcid{0000-0002-7025-6058}\inst{\ref{aff3},\ref{aff4}}
\and K.~George\orcid{0000-0002-1734-8455}\inst{\ref{aff5}}
\and N.~Masoumzadeh\orcid{0000-0002-2664-8054}\inst{\ref{aff3}}
\and M.~Poulain\orcid{0000-0002-7664-4510}\inst{\ref{aff6}}
\and I.~T.~Andika\orcid{0000-0001-6102-9526}\inst{\ref{aff4}}
\and T.~Vassallo\orcid{0000-0001-6512-6358}\inst{\ref{aff7},\ref{aff5}}
\and A.~Venhola\orcid{0000-0001-6071-4564}\inst{\ref{aff6}}
\and G.~Verdoes~Kleijn\orcid{0000-0001-5803-2580}\inst{\ref{aff1}}
\and M.~Wetzstein\inst{\ref{aff3}}
\and I.~A.~Zinchenko\orcid{0000-0002-2944-2449}\inst{\ref{aff8}}
\and M.~Ruffenach\orcid{0000-0001-8772-7330}\inst{\ref{aff9}}
\and N.~R.~Stickley\orcid{0000-0003-0987-5738}\inst{\ref{aff10}}
\and I.~Das\orcid{0009-0007-7088-2044}\inst{\ref{aff11}}
\and S.~Gwyn\orcid{0000-0001-8221-8406}\inst{\ref{aff12}}
\and T.~de~Boer\orcid{0000-0001-5486-2747}\inst{\ref{aff13}}
\and E.~A.~Magnier\orcid{0000-0002-7965-2815}\inst{\ref{aff13}}
\and K.~C.~Chambers\orcid{0000-0001-6965-7789}\inst{\ref{aff13}}
\and A.W.~McConnachie\inst{\ref{aff12}}
\and T.~Saifollahi\orcid{0000-0002-9554-7660}\inst{\ref{aff14},\ref{aff9}}
\and B.~Altieri\orcid{0000-0003-3936-0284}\inst{\ref{aff15}}
\and S.~Andreon\orcid{0000-0002-2041-8784}\inst{\ref{aff16}}
\and N.~Auricchio\orcid{0000-0003-4444-8651}\inst{\ref{aff17}}
\and H.~Aussel\orcid{0000-0002-1371-5705}\inst{\ref{aff18}}
\and C.~Baccigalupi\orcid{0000-0002-8211-1630}\inst{\ref{aff19},\ref{aff7},\ref{aff20},\ref{aff21}}
\and M.~Baldi\orcid{0000-0003-4145-1943}\inst{\ref{aff22},\ref{aff17},\ref{aff23}}
\and A.~Balestra\orcid{0000-0002-6967-261X}\inst{\ref{aff24}}
\and S.~Bardelli\orcid{0000-0002-8900-0298}\inst{\ref{aff17}}
\and P.~Battaglia\orcid{0000-0002-7337-5909}\inst{\ref{aff17}}
\and A.~N.~Belikov\inst{\ref{aff1},\ref{aff25}}
\and R.~Bender\orcid{0000-0001-7179-0626}\inst{\ref{aff3},\ref{aff4}}
\and A.~Biviano\orcid{0000-0002-0857-0732}\inst{\ref{aff7},\ref{aff19}}
\and M.~Bolzonella\orcid{0000-0003-3278-4607}\inst{\ref{aff17}}
\and E.~Branchini\orcid{0000-0002-0808-6908}\inst{\ref{aff26},\ref{aff27},\ref{aff16}}
\and M.~Brescia\orcid{0000-0001-9506-5680}\inst{\ref{aff28},\ref{aff29}}
\and J.~Brinchmann\orcid{0000-0003-4359-8797}\inst{\ref{aff30},\ref{aff31},\ref{aff32}}
\and S.~Camera\orcid{0000-0003-3399-3574}\inst{\ref{aff33},\ref{aff34},\ref{aff35}}
\and G.~Ca\~nas-Herrera\orcid{0000-0003-2796-2149}\inst{\ref{aff2}}
\and V.~Capobianco\orcid{0000-0002-3309-7692}\inst{\ref{aff35}}
\and C.~Carbone\orcid{0000-0003-0125-3563}\inst{\ref{aff36}}
\and J.~Carretero\orcid{0000-0002-3130-0204}\inst{\ref{aff37},\ref{aff38}}
\and F.~J.~Castander\orcid{0000-0001-7316-4573}\inst{\ref{aff39},\ref{aff40}}
\and M.~Castellano\orcid{0000-0001-9875-8263}\inst{\ref{aff41}}
\and G.~Castignani\orcid{0000-0001-6831-0687}\inst{\ref{aff17}}
\and S.~Cavuoti\orcid{0000-0002-3787-4196}\inst{\ref{aff29},\ref{aff42}}
\and A.~Cimatti\inst{\ref{aff43}}
\and C.~Colodro-Conde\inst{\ref{aff44}}
\and G.~Congedo\orcid{0000-0003-2508-0046}\inst{\ref{aff45}}
\and C.~J.~Conselice\orcid{0000-0003-1949-7638}\inst{\ref{aff46}}
\and L.~Conversi\orcid{0000-0002-6710-8476}\inst{\ref{aff47},\ref{aff15}}
\and Y.~Copin\orcid{0000-0002-5317-7518}\inst{\ref{aff48}}
\and A.~Costille\inst{\ref{aff49}}
\and F.~Courbin\orcid{0000-0003-0758-6510}\inst{\ref{aff50},\ref{aff51},\ref{aff52}}
\and H.~M.~Courtois\orcid{0000-0003-0509-1776}\inst{\ref{aff53}}
\and M.~Cropper\orcid{0000-0003-4571-9468}\inst{\ref{aff54}}
\and A.~Da~Silva\orcid{0000-0002-6385-1609}\inst{\ref{aff55},\ref{aff56}}
\and H.~Degaudenzi\orcid{0000-0002-5887-6799}\inst{\ref{aff57}}
\and G.~De~Lucia\orcid{0000-0002-6220-9104}\inst{\ref{aff7}}
\and H.~Dole\orcid{0000-0002-9767-3839}\inst{\ref{aff58}}
\and F.~Dubath\orcid{0000-0002-6533-2810}\inst{\ref{aff57}}
\and X.~Dupac\inst{\ref{aff15}}
\and A.~Ealet\orcid{0000-0003-3070-014X}\inst{\ref{aff48}}
\and S.~Escoffier\orcid{0000-0002-2847-7498}\inst{\ref{aff59}}
\and M.~Farina\orcid{0000-0002-3089-7846}\inst{\ref{aff60}}
\and R.~Farinelli\inst{\ref{aff17}}
\and S.~Farrens\orcid{0000-0002-9594-9387}\inst{\ref{aff18}}
\and F.~Faustini\orcid{0000-0001-6274-5145}\inst{\ref{aff41}}
\and S.~Ferriol\inst{\ref{aff48}}
\and F.~Finelli\orcid{0000-0002-6694-3269}\inst{\ref{aff17},\ref{aff61}}
\and P.~Fosalba\orcid{0000-0002-1510-5214}\inst{\ref{aff40},\ref{aff39}}
\and S.~Fotopoulou\orcid{0000-0002-9686-254X}\inst{\ref{aff62}}
\and N.~Fourmanoit\orcid{0009-0005-6816-6925}\inst{\ref{aff59}}
\and M.~Frailis\orcid{0000-0002-7400-2135}\inst{\ref{aff7}}
\and E.~Franceschi\orcid{0000-0002-0585-6591}\inst{\ref{aff17}}
\and M.~Fumana\orcid{0000-0001-6787-5950}\inst{\ref{aff36}}
\and L.~Gabarra\orcid{0000-0002-8486-8856}\inst{\ref{aff63}}
\and S.~Galeotta\orcid{0000-0002-3748-5115}\inst{\ref{aff7}}
\and B.~Gillis\orcid{0000-0002-4478-1270}\inst{\ref{aff45}}
\and C.~Giocoli\orcid{0000-0002-9590-7961}\inst{\ref{aff17},\ref{aff23}}
\and J.~Gracia-Carpio\orcid{0000-0003-4689-3134}\inst{\ref{aff3}}
\and B.~R.~Granett\orcid{0000-0003-2694-9284}\inst{\ref{aff16}}
\and A.~Grazian\orcid{0000-0002-5688-0663}\inst{\ref{aff24}}
\and F.~Grupp\inst{\ref{aff3},\ref{aff4}}
\and W.~G.~Hartley\orcid{0000-0002-4563-5101}\inst{\ref{aff57}}
\and S.~V.~H.~Haugan\orcid{0000-0001-9648-7260}\inst{\ref{aff64}}
\and S.~Hemmati\orcid{0000-0003-2226-5395}\inst{\ref{aff11}}
\and H.~Hoekstra\orcid{0000-0002-0641-3231}\inst{\ref{aff2}}
\and W.~Holmes\orcid{0009-0007-8554-4646}\inst{\ref{aff65}}
\and I.~M.~Hook\orcid{0000-0002-2960-978X}\inst{\ref{aff66}}
\and F.~Hormuth\inst{\ref{aff67}}
\and A.~Hornstrup\orcid{0000-0002-3363-0936}\inst{\ref{aff68},\ref{aff69}}
\and M.~Huertas-Company\orcid{0000-0002-1416-8483}\inst{\ref{aff44},\ref{aff70},\ref{aff71}}
\and K.~Jahnke\orcid{0000-0003-3804-2137}\inst{\ref{aff72}}
\and M.~Jhabvala\inst{\ref{aff73}}
\and B.~Joachimi\orcid{0000-0001-7494-1303}\inst{\ref{aff74}}
\and S.~Kermiche\orcid{0000-0002-0302-5735}\inst{\ref{aff59}}
\and A.~Kiessling\orcid{0000-0002-2590-1273}\inst{\ref{aff65}}
\and B.~Kubik\orcid{0009-0006-5823-4880}\inst{\ref{aff48}}
\and K.~Kuijken\orcid{0000-0002-3827-0175}\inst{\ref{aff2}}
\and M.~K\"ummel\orcid{0000-0003-2791-2117}\inst{\ref{aff4}}
\and M.~Kunz\orcid{0000-0002-3052-7394}\inst{\ref{aff75}}
\and H.~Kurki-Suonio\orcid{0000-0002-4618-3063}\inst{\ref{aff76},\ref{aff77}}
\and R.~Laureijs\inst{\ref{aff1}}
\and A.~M.~C.~Le~Brun\orcid{0000-0002-0936-4594}\inst{\ref{aff78}}
\and P.~Liebing\orcid{0009-0005-8043-7150}\inst{\ref{aff54}}
\and S.~Ligori\orcid{0000-0003-4172-4606}\inst{\ref{aff35}}
\and P.~B.~Lilje\orcid{0000-0003-4324-7794}\inst{\ref{aff64}}
\and V.~Lindholm\orcid{0000-0003-2317-5471}\inst{\ref{aff76},\ref{aff77}}
\and I.~Lloro\orcid{0000-0001-5966-1434}\inst{\ref{aff79}}
\and M.~Magliocchetti\orcid{0000-0001-9158-4838}\inst{\ref{aff60}}
\and G.~Mainetti\orcid{0000-0003-2384-2377}\inst{\ref{aff80}}
\and O.~Mansutti\orcid{0000-0001-5758-4658}\inst{\ref{aff7}}
\and S.~Marcin\inst{\ref{aff81}}
\and O.~Marggraf\orcid{0000-0001-7242-3852}\inst{\ref{aff82}}
\and M.~Martinelli\orcid{0000-0002-6943-7732}\inst{\ref{aff41},\ref{aff83}}
\and N.~Martinet\orcid{0000-0003-2786-7790}\inst{\ref{aff49}}
\and F.~Marulli\orcid{0000-0002-8850-0303}\inst{\ref{aff84},\ref{aff17},\ref{aff23}}
\and R.~J.~Massey\orcid{0000-0002-6085-3780}\inst{\ref{aff85}}
\and D.~C.~Masters\orcid{0000-0001-5382-6138}\inst{\ref{aff86}}
\and N.~Mauri\orcid{0000-0001-8196-1548}\inst{\ref{aff43},\ref{aff23}}
\and H.~J.~McCracken\orcid{0000-0002-9489-7765}\inst{\ref{aff87}}
\and C.~J.~R.~McPartland\orcid{0000-0003-0639-025X}\inst{\ref{aff69},\ref{aff88}}
\and E.~Medinaceli\orcid{0000-0002-4040-7783}\inst{\ref{aff17}}
\and M.~Meneghetti\orcid{0000-0003-1225-7084}\inst{\ref{aff17},\ref{aff23}}
\and E.~Merlin\orcid{0000-0001-6870-8900}\inst{\ref{aff24}}
\and P.~Monaco\orcid{0000-0003-2083-7564}\inst{\ref{aff89},\ref{aff7},\ref{aff20},\ref{aff19}}
\and A.~Mora\orcid{0000-0002-1922-8529}\inst{\ref{aff90}}
\and M.~Moresco\orcid{0000-0002-7616-7136}\inst{\ref{aff84},\ref{aff17}}
\and C.~Moretti\orcid{0000-0003-3314-8936}\inst{\ref{aff7},\ref{aff19},\ref{aff20}}
\and L.~Moscardini\orcid{0000-0002-3473-6716}\inst{\ref{aff84},\ref{aff17},\ref{aff23}}
\and R.~Nakajima\orcid{0009-0009-1213-7040}\inst{\ref{aff82}}
\and C.~Neissner\orcid{0000-0001-8524-4968}\inst{\ref{aff91},\ref{aff38}}
\and R.~C.~Nichol\orcid{0000-0003-0939-6518}\inst{\ref{aff92}}
\and S.-M.~Niemi\orcid{0009-0005-0247-0086}\inst{\ref{aff93}}
\and C.~Padilla\orcid{0000-0001-7951-0166}\inst{\ref{aff91}}
\and S.~Paltani\orcid{0000-0002-8108-9179}\inst{\ref{aff57}}
\and F.~Pasian\orcid{0000-0002-4869-3227}\inst{\ref{aff7}}
\and W.~J.~Percival\orcid{0000-0002-0644-5727}\inst{\ref{aff94},\ref{aff95},\ref{aff96}}
\and V.~Pettorino\orcid{0000-0002-4203-9320}\inst{\ref{aff93}}
\and A.~Pezzotta\orcid{0000-0003-0726-2268}\inst{\ref{aff16}}
\and S.~Pires\orcid{0000-0002-0249-2104}\inst{\ref{aff18}}
\and G.~Polenta\orcid{0000-0003-4067-9196}\inst{\ref{aff97}}
\and M.~Poncet\inst{\ref{aff9}}
\and L.~A.~Popa\inst{\ref{aff98}}
\and L.~Pozzetti\orcid{0000-0001-7085-0412}\inst{\ref{aff17}}
\and F.~Raison\orcid{0000-0002-7819-6918}\inst{\ref{aff3}}
\and R.~Rebolo\orcid{0000-0003-3767-7085}\inst{\ref{aff44},\ref{aff99},\ref{aff100}}
\and A.~Refregier\inst{\ref{aff101}}
\and A.~Renzi\orcid{0000-0001-9856-1970}\inst{\ref{aff102},\ref{aff103},\ref{aff17}}
\and J.~Rhodes\orcid{0000-0002-4485-8549}\inst{\ref{aff65}}
\and G.~Riccio\inst{\ref{aff29}}
\and I.~Risso\orcid{0000-0003-2525-7761}\inst{\ref{aff26},\ref{aff27},\ref{aff16}}
\and F.~Rizzo\orcid{0000-0002-9407-585X}\inst{\ref{aff7}}
\and E.~Romelli\orcid{0000-0003-3069-9222}\inst{\ref{aff7}}
\and M.~Roncarelli\orcid{0000-0001-9587-7822}\inst{\ref{aff17}}
\and C.~Rosset\orcid{0000-0003-0286-2192}\inst{\ref{aff104}}
\and B.~Rusholme\orcid{0000-0001-7648-4142}\inst{\ref{aff11}}
\and R.~Saglia\orcid{0000-0003-0378-7032}\inst{\ref{aff4},\ref{aff3}}
\and Z.~Sakr\orcid{0000-0002-4823-3757}\inst{\ref{aff105},\ref{aff106},\ref{aff107}}
\and A.~G.~S\'anchez\orcid{0000-0003-1198-831X}\inst{\ref{aff3}}
\and D.~Sapone\orcid{0000-0001-7089-4503}\inst{\ref{aff108}}
\and B.~Sartoris\orcid{0000-0003-1337-5269}\inst{\ref{aff3},\ref{aff7}}
\and M.~Sauvage\orcid{0000-0002-0809-2574}\inst{\ref{aff18}}
\and M.~Schirmer\orcid{0000-0003-2568-9994}\inst{\ref{aff72}}
\and P.~Schneider\orcid{0000-0001-8561-2679}\inst{\ref{aff82}}
\and T.~Schrabback\orcid{0000-0002-6987-7834}\inst{\ref{aff109}}
\and M.~Scodeggio\inst{\ref{aff36}}
\and A.~Secroun\orcid{0000-0003-0505-3710}\inst{\ref{aff59}}
\and E.~Sihvola\orcid{0000-0003-1804-7715}\inst{\ref{aff110}}
\and C.~Sirignano\orcid{0000-0002-0995-7146}\inst{\ref{aff102},\ref{aff103}}
\and G.~Sirri\orcid{0000-0003-2626-2853}\inst{\ref{aff23}}
\and J.-L.~Starck\orcid{0000-0003-2177-7794}\inst{\ref{aff18}}
\and P.~Tallada-Cresp\'{i}\orcid{0000-0002-1336-8328}\inst{\ref{aff37},\ref{aff38}}
\and A.~N.~Taylor\inst{\ref{aff45}}
\and H.~I.~Teplitz\orcid{0000-0002-7064-5424}\inst{\ref{aff86}}
\and I.~Tereno\orcid{0000-0002-4537-6218}\inst{\ref{aff55},\ref{aff111}}
\and N.~Tessore\orcid{0000-0002-9696-7931}\inst{\ref{aff54}}
\and S.~Toft\orcid{0000-0003-3631-7176}\inst{\ref{aff112},\ref{aff88}}
\and R.~Toledo-Moreo\orcid{0000-0002-2997-4859}\inst{\ref{aff113},\ref{aff114}}
\and F.~Torradeflot\orcid{0000-0003-1160-1517}\inst{\ref{aff38},\ref{aff37}}
\and A.~Tsyganov\inst{\ref{aff115}}
\and I.~Tutusaus\orcid{0000-0002-3199-0399}\inst{\ref{aff39},\ref{aff40},\ref{aff106}}
\and E.~A.~Valentijn\inst{\ref{aff1}}
\and L.~Valenziano\orcid{0000-0002-1170-0104}\inst{\ref{aff17},\ref{aff61}}
\and J.~Valiviita\orcid{0000-0001-6225-3693}\inst{\ref{aff76},\ref{aff77}}
\and A.~Veropalumbo\orcid{0000-0003-2387-1194}\inst{\ref{aff16},\ref{aff27},\ref{aff26}}
\and Y.~Wang\orcid{0000-0002-4749-2984}\inst{\ref{aff11}}
\and J.~Weller\orcid{0000-0002-8282-2010}\inst{\ref{aff4},\ref{aff3}}
\and A.~Zacchei\orcid{0000-0003-0396-1192}\inst{\ref{aff7},\ref{aff19}}
\and G.~Zamorani\orcid{0000-0002-2318-301X}\inst{\ref{aff17}}
\and F.~M.~Zerbi\orcid{0000-0002-9996-973X}\inst{\ref{aff16}}
\and E.~Zucca\orcid{0000-0002-5845-8132}\inst{\ref{aff17}}
\and M.~Ballardini\orcid{0000-0003-4481-3559}\inst{\ref{aff116},\ref{aff117},\ref{aff17}}
\and P.~Bergamini\orcid{0000-0003-1383-9414}\inst{\ref{aff17}}
\and S.~Borgani\orcid{0000-0001-6151-6439}\inst{\ref{aff89},\ref{aff19},\ref{aff7},\ref{aff20},\ref{aff118}}
\and E.~Bozzo\orcid{0000-0002-8201-1525}\inst{\ref{aff57}}
\and C.~Burigana\orcid{0000-0002-3005-5796}\inst{\ref{aff119},\ref{aff61}}
\and R.~Cabanac\orcid{0000-0001-6679-2600}\inst{\ref{aff106}}
\and M.~Calabrese\orcid{0000-0002-2637-2422}\inst{\ref{aff120},\ref{aff36}}
\and F.~Caro\orcid{0009-0003-1053-0507}\inst{\ref{aff41}}
\and T.~Castro\orcid{0000-0002-6292-3228}\inst{\ref{aff121},\ref{aff7},\ref{aff19}}
\and J.~A.~Escartin~Vigo\inst{\ref{aff3}}
\and J.~Garc\'ia-Bellido\orcid{0000-0002-9370-8360}\inst{\ref{aff105}}
\and T.~Gasparetto\orcid{0000-0002-7913-4866}\inst{\ref{aff41}}
\and A.~Gregorio\orcid{0000-0003-4028-8785}\inst{\ref{aff89},\ref{aff7},\ref{aff20}}
\and C.~M.~Gutierrez\orcid{0000-0001-7854-783X}\inst{\ref{aff44},\ref{aff100}}
\and A.~Loureiro\orcid{0000-0002-4371-0876}\inst{\ref{aff122},\ref{aff123}}
\and J.~Macias-Perez\orcid{0000-0002-5385-2763}\inst{\ref{aff124}}
\and R.~Maoli\orcid{0000-0002-6065-3025}\inst{\ref{aff125},\ref{aff41}}
\and R.~B.~Metcalf\orcid{0000-0003-3167-2574}\inst{\ref{aff84},\ref{aff17}}
\and A.~Montoro\orcid{0000-0003-4730-8590}\inst{\ref{aff39},\ref{aff40}}
\and A.~A.~Nucita\inst{\ref{aff126},\ref{aff127},\ref{aff128}}
\and M.~P\"ontinen\orcid{0000-0001-5442-2530}\inst{\ref{aff76}}
\and E.~Sarpa\orcid{0000-0002-1256-655X}\inst{\ref{aff7}}
\and V.~Scottez\orcid{0009-0008-3864-940X}\inst{\ref{aff129},\ref{aff130}}
\and M.~Sereno\orcid{0000-0003-0302-0325}\inst{\ref{aff17},\ref{aff23}}
\and M.~Tenti\orcid{0000-0002-4254-5901}\inst{\ref{aff23}}
\and M.~Tucci\inst{\ref{aff57}}
\and M.~Wiesmann\orcid{0009-0000-8199-5860}\inst{\ref{aff64}}
\and J.~A.~Acevedo~Barroso\orcid{0000-0002-9654-1711}\inst{\ref{aff65}}
\and Y.~Akrami\orcid{0000-0002-2407-7956}\inst{\ref{aff105},\ref{aff131}}
\and G.~Angora\orcid{0000-0002-0316-6562}\inst{\ref{aff29},\ref{aff116}}
\and S.~Anselmi\orcid{0000-0002-3579-9583}\inst{\ref{aff103},\ref{aff102},\ref{aff132}}
\and M.~Archidiacono\orcid{0000-0003-4952-9012}\inst{\ref{aff133},\ref{aff134}}
\and G.~Aric\`o\orcid{0000-0002-2802-2928}\inst{\ref{aff23}}
\and F.~Atrio-Barandela\orcid{0000-0002-2130-2513}\inst{\ref{aff135}}
\and E.~Aubourg\orcid{0000-0002-5592-023X}\inst{\ref{aff104},\ref{aff136}}
\and M.~Baes\orcid{0000-0002-3930-2757}\inst{\ref{aff137}}
\and L.~Bazzanini\orcid{0000-0003-0727-0137}\inst{\ref{aff116},\ref{aff17}}
\and D.~Bertacca\orcid{0000-0002-2490-7139}\inst{\ref{aff102},\ref{aff24},\ref{aff103}}
\and M.~Bethermin\orcid{0000-0002-3915-2015}\inst{\ref{aff14}}
\and F.~Beutler\orcid{0000-0003-0467-5438}\inst{\ref{aff45}}
\and L.~Blot\orcid{0000-0002-9622-7167}\inst{\ref{aff138},\ref{aff78}}
\and M.~L.~Brown\orcid{0000-0002-0370-8077}\inst{\ref{aff46}}
\and S.~Bruton\orcid{0000-0002-6503-5218}\inst{\ref{aff10}}
\and A.~Calabro\orcid{0000-0003-2536-1614}\inst{\ref{aff41}}
\and B.~Camacho~Quevedo\orcid{0000-0002-8789-4232}\inst{\ref{aff19},\ref{aff21},\ref{aff7}}
\and P.~Carrilho\orcid{0000-0003-1339-0194}\inst{\ref{aff139}}
\and C.~S.~Carvalho\inst{\ref{aff111}}
\and R.~Claydon\orcid{0000-0003-2117-7895}\inst{\ref{aff45}}
\and F.~Cogato\orcid{0000-0003-4632-6113}\inst{\ref{aff84},\ref{aff17}}
\and T.~E.~Collett\orcid{0000-0001-5564-3140}\inst{\ref{aff140}}
\and S.~Conseil\orcid{0000-0002-3657-4191}\inst{\ref{aff48}}
\and A.~R.~Cooray\orcid{0000-0002-3892-0190}\inst{\ref{aff141}}
\and B.~Csizi\orcid{0000-0003-3227-6581}\inst{\ref{aff109}}
\and O.~Cucciati\orcid{0000-0002-9336-7551}\inst{\ref{aff17}}
\and H.~Dannerbauer\orcid{0000-0001-7147-3575}\inst{\ref{aff44},\ref{aff100}}
\and S.~Davini\orcid{0000-0003-3269-1718}\inst{\ref{aff27}}
\and F.~De~Paolis\orcid{0000-0001-6460-7563}\inst{\ref{aff126},\ref{aff127},\ref{aff128}}
\and G.~Desprez\orcid{0000-0001-8325-1742}\inst{\ref{aff1}}
\and A.~D\'iaz-S\'anchez\orcid{0000-0003-0748-4768}\inst{\ref{aff142}}
\and S.~Di~Domizio\orcid{0000-0003-2863-5895}\inst{\ref{aff26},\ref{aff27}}
\and J.~M.~Diego\orcid{0000-0001-9065-3926}\inst{\ref{aff143}}
\and P.-A.~Duc\orcid{0000-0003-3343-6284}\inst{\ref{aff14}}
\and V.~Duret\orcid{0009-0009-0383-4960}\inst{\ref{aff59}}
\and M.~Y.~Elkhashab\orcid{0000-0001-9306-2603}\inst{\ref{aff7},\ref{aff20},\ref{aff89},\ref{aff19}}
\and A.~Enia\orcid{0000-0002-0200-2857}\inst{\ref{aff144},\ref{aff17}}
\and Y.~Fang\orcid{0000-0002-0334-6950}\inst{\ref{aff4}}
\and A.~Farina\orcid{0009-0000-3420-929X}\inst{\ref{aff16},\ref{aff27}}
\and A.~M.~N.~Ferguson\orcid{0000-0001-7934-1278}\inst{\ref{aff45}}
\and A.~Finoguenov\orcid{0000-0002-4606-5403}\inst{\ref{aff76}}
\and A.~Franco\orcid{0000-0002-4761-366X}\inst{\ref{aff126},\ref{aff127},\ref{aff128}}
\and Y.~Fu\orcid{0000-0002-0759-0504}\inst{\ref{aff2},\ref{aff1}}
\and K.~Ganga\orcid{0000-0001-8159-8208}\inst{\ref{aff104}}
\and R.~Gavazzi\orcid{0000-0002-5540-6935}\inst{\ref{aff49},\ref{aff87}}
\and E.~Gaztanaga\orcid{0000-0001-9632-0815}\inst{\ref{aff39},\ref{aff40},\ref{aff140}}
\and Z.~Ghaffari\orcid{0000-0002-6467-8078}\inst{\ref{aff7},\ref{aff19}}
\and F.~Giacomini\orcid{0000-0002-3129-2814}\inst{\ref{aff23}}
\and F.~Gianotti\orcid{0000-0003-4666-119X}\inst{\ref{aff17}}
\and G.~Gozaliasl\orcid{0000-0002-0236-919X}\inst{\ref{aff145},\ref{aff76}}
\and A.~Gruppuso\orcid{0000-0001-9272-5292}\inst{\ref{aff17},\ref{aff23}}
\and J.~Haase\orcid{0000-0001-7809-3671}\inst{\ref{aff3}}
\and A.~Hall\orcid{0000-0002-3139-8651}\inst{\ref{aff45}}
\and N.~A.~Hatch\orcid{0000-0001-5600-0534}\inst{\ref{aff146}}
\and C.~Hern\'andez-Monteagudo\orcid{0000-0001-5471-9166}\inst{\ref{aff100},\ref{aff44}}
\and H.~Hildebrandt\orcid{0000-0002-9814-3338}\inst{\ref{aff147}}
\and J.~Hjorth\orcid{0000-0002-4571-2306}\inst{\ref{aff148}}
\and M.~Jauzac\orcid{0000-0003-1974-8732}\inst{\ref{aff149},\ref{aff85},\ref{aff150},\ref{aff151},\ref{aff106}}
\and J.~J.~E.~Kajava\orcid{0000-0002-3010-8333}\inst{\ref{aff152},\ref{aff153},\ref{aff154}}
\and Y.~Kang\orcid{0009-0000-8588-7250}\inst{\ref{aff109}}
\and V.~Kansal\orcid{0000-0002-4008-6078}\inst{\ref{aff155},\ref{aff156}}
\and D.~Karagiannis\orcid{0000-0002-4927-0816}\inst{\ref{aff116},\ref{aff157}}
\and J.~Kim\orcid{0000-0003-2776-2761}\inst{\ref{aff63}}
\and C.~C.~Kirkpatrick\inst{\ref{aff110}}
\and A.~Kov\'acs\orcid{0000-0002-5825-579X}\inst{\ref{aff158},\ref{aff159}}
\and I.~Kova{\v{c}}i{\'{c}}\orcid{0000-0001-6751-3263}\inst{\ref{aff137}}
\and K.~Koyama\orcid{0000-0001-6727-6915}\inst{\ref{aff140}}
\and S.~Kruk\orcid{0000-0001-8010-8879}\inst{\ref{aff15}}
\and M.~C.~Lam\orcid{0000-0002-9347-2298}\inst{\ref{aff45}}
\and M.~Lattanzi\orcid{0000-0003-1059-2532}\inst{\ref{aff117}}
\and G.~Leroy\orcid{0009-0004-2523-4425}\inst{\ref{aff149},\ref{aff85}}
\and G.~F.~Lesci\orcid{0000-0002-4607-2830}\inst{\ref{aff84},\ref{aff17}}
\and J.~Lesgourgues\orcid{0000-0001-7627-353X}\inst{\ref{aff160}}
\and T.~I.~Liaudat\orcid{0000-0002-9104-314X}\inst{\ref{aff136}}
\and L.~Linke\orcid{0000-0002-2622-8113}\inst{\ref{aff109}}
\and S.~J.~Liu\orcid{0000-0001-7680-2139}\inst{\ref{aff60}}
\and X.~Lopez~Lopez\orcid{0009-0008-5194-5908}\inst{\ref{aff17}}
\and G.~Maggio\orcid{0000-0003-4020-4836}\inst{\ref{aff7}}
\and F.~Mannucci\orcid{0000-0002-4803-2381}\inst{\ref{aff161}}
\and F.~R.~Marleau\orcid{0000-0002-1442-2947}\inst{\ref{aff109}}
\and C.~J.~A.~P.~Martins\orcid{0000-0002-4886-9261}\inst{\ref{aff162},\ref{aff31}}
\and M.~Miluzio\inst{\ref{aff15},\ref{aff163}}
\and G.~Morgante\inst{\ref{aff17}}
\and S.~Nadathur\orcid{0000-0001-9070-3102}\inst{\ref{aff140}}
\and K.~Naidoo\orcid{0000-0002-9182-1802}\inst{\ref{aff140},\ref{aff72}}
\and P.~Natoli\orcid{0000-0003-0126-9100}\inst{\ref{aff116},\ref{aff117}}
\and A.~Navarro-Alsina\orcid{0000-0002-3173-2592}\inst{\ref{aff82}}
\and A.~Nersesian\orcid{0000-0001-6843-409X}\inst{\ref{aff164},\ref{aff137}}
\and S.~Nesseris\orcid{0000-0002-0567-0324}\inst{\ref{aff105}}
\and L.~Nicastro\orcid{0000-0001-8534-6788}\inst{\ref{aff17}}
\and M.~Oguri\orcid{0000-0003-3484-399X}\inst{\ref{aff165},\ref{aff166}}
\and F.~Oppizzi\orcid{0000-0003-3904-8370}\inst{\ref{aff27},\ref{aff103},\ref{aff102}}
\and F.~Pace\orcid{0000-0001-8039-0480}\inst{\ref{aff33},\ref{aff34},\ref{aff35},\ref{aff56}}
\and D.~Paoletti\orcid{0000-0003-4761-6147}\inst{\ref{aff17},\ref{aff61}}
\and F.~Passalacqua\orcid{0000-0002-8606-4093}\inst{\ref{aff103}}
\and K.~Paterson\orcid{0000-0001-8340-3486}\inst{\ref{aff72}}
\and L.~Patrizii\inst{\ref{aff23}}
\and C.~Pattison\orcid{0000-0003-3272-2617}\inst{\ref{aff140}}
\and A.~Pisani\orcid{0000-0002-6146-4437}\inst{\ref{aff59}}
\and D.~Potter\orcid{0000-0002-0757-5195}\inst{\ref{aff167}}
\and G.~W.~Pratt\inst{\ref{aff18}}
\and S.~Quai\orcid{0000-0002-0449-8163}\inst{\ref{aff84},\ref{aff17}}
\and M.~Radovich\orcid{0000-0002-3585-866X}\inst{\ref{aff24}}
\and F.~Ricci\orcid{0000-0001-5742-5980}\inst{\ref{aff168},\ref{aff41}}
\and G.~Rodighiero\orcid{0000-0002-9415-2296}\inst{\ref{aff102},\ref{aff24}}
\and W.~Roster\orcid{0000-0002-9149-6528}\inst{\ref{aff3}}
\and S.~Sacquegna\orcid{0000-0002-8433-6630}\inst{\ref{aff169}}
\and M.~Sahl\'en\orcid{0000-0003-0973-4804}\inst{\ref{aff170}}
\and D.~B.~Sanders\orcid{0000-0002-1233-9998}\inst{\ref{aff13}}
\and J.~Schaye\orcid{0000-0002-0668-5560}\inst{\ref{aff2}}
\and A.~Schneider\orcid{0000-0001-7055-8104}\inst{\ref{aff167}}
\and D.~Sciotti\orcid{0009-0008-4519-2620}\inst{\ref{aff41},\ref{aff83}}
\and E.~Sellentin\orcid{0009-0002-2655-3458}\inst{\ref{aff171},\ref{aff2}}
\and S.~Serjeant\orcid{0000-0002-0517-7943}\inst{\ref{aff172}}
\and L.~C.~Smith\orcid{0000-0002-3259-2771}\inst{\ref{aff173}}
\and J.~G.~Sorce\orcid{0000-0002-2307-2432}\inst{\ref{aff174},\ref{aff58}}
\and M.~Talia\orcid{0000-0003-4352-2063}\inst{\ref{aff84},\ref{aff17}}
\and K.~Tanidis\orcid{0000-0001-9843-5130}\inst{\ref{aff175}}
\and C.~Tao\orcid{0000-0001-7961-8177}\inst{\ref{aff59}}
\and F.~Tarsitano\orcid{0000-0002-5919-0238}\inst{\ref{aff176},\ref{aff101},\ref{aff57}}
\and G.~Testera\orcid{0000-0003-2970-766X}\inst{\ref{aff27}}
\and R.~Teyssier\orcid{0000-0001-7689-0933}\inst{\ref{aff177}}
\and S.~Tosi\orcid{0000-0002-7275-9193}\inst{\ref{aff26},\ref{aff16},\ref{aff27}}
\and A.~Troja\orcid{0000-0003-0239-4595}\inst{\ref{aff7}}
\and C.~Uhlemann\orcid{0000-0001-7831-1579}\inst{\ref{aff178},\ref{aff179}}
\and M.~Urbano\orcid{0000-0001-5640-0650}\inst{\ref{aff146},\ref{aff14}}
\and C.~Valieri\inst{\ref{aff23}}
\and D.~Vergani\orcid{0000-0003-0898-2216}\inst{\ref{aff17}}
\and E.~Vilenius\orcid{0000-0002-6184-7681}\inst{\ref{aff54}}
\and S.~Vinciguerra\orcid{0009-0005-4018-3184}\inst{\ref{aff49}}
\and M.~von~Wietersheim-Kramsta\orcid{0000-0003-4986-5091}\inst{\ref{aff85},\ref{aff149}}
\and N.~A.~Walton\orcid{0000-0003-3983-8778}\inst{\ref{aff173}}
\and L.~Wang\orcid{0000-0002-6736-9158}\inst{\ref{aff180},\ref{aff1}}
\and A.~H.~Wright\orcid{0000-0001-7363-7932}\inst{\ref{aff147}}
\and L.~Zalesky\orcid{0000-0001-5680-2326}\inst{\ref{aff13}}
\and B.~P.~Crill\orcid{0000-0002-4650-8518}\inst{\ref{aff65}}
\and D.~Scott\orcid{0000-0002-6878-9840}\inst{\ref{aff181}}}
										   
%%%% please do not edit the affiliation list -- contact ECEB Bureau for changes
\institute{Kapteyn Astronomical Institute, University of Groningen, PO Box 800, 9700 AV Groningen, The Netherlands\label{aff1}
\and
Leiden Observatory, Leiden University, Einsteinweg 55, 2333 CC Leiden, The Netherlands\label{aff2}
\and
Max Planck Institute for Extraterrestrial Physics, Giessenbachstr. 1, 85748 Garching, Germany\label{aff3}
\and
Universit\"ats-Sternwarte M\"unchen, Fakult\"at f\"ur Physik, Ludwig-Maximilians-Universit\"at M\"unchen, Scheinerstr.~1, 81679 M\"unchen, Germany\label{aff4}
\and
University Observatory, LMU Faculty of Physics, Scheinerstr.~1, 81679 Munich, Germany\label{aff5}
\and
Space physics and astronomy research unit, University of Oulu, Pentti Kaiteran katu 1, FI-90014 Oulu, Finland\label{aff6}
\and
INAF-Osservatorio Astronomico di Trieste, Via G. B. Tiepolo 11, 34143 Trieste, Italy\label{aff7}
\and
Astronomisches Rechen-Institut, Zentrum f\"ur Astronomie der Universit\"at Heidelberg, M\"onchhofstr. 12-14, 69120 Heidelberg, Germany\label{aff8}
\and
Centre National d'Etudes Spatiales -- Centre spatial de Toulouse, 18 avenue Edouard Belin, 31401 Toulouse Cedex 9, France\label{aff9}
\and
California Institute of Technology, 1200 E California Blvd, Pasadena, CA 91125, USA\label{aff10}
\and
Caltech/IPAC, 1200 E. California Blvd., Pasadena, CA 91125, USA\label{aff11}
\and
National Research Council, Herzberg Astronomy and Astrophysics Research Centre, 5071 W. Saanich Rd. Victoria, BC, V9E 2E7, Canada\label{aff12}
\and
Institute for Astronomy, University of Hawaii, 2680 Woodlawn Drive, Honolulu, HI 96822, USA\label{aff13}
\and
Universit\'e de Strasbourg, CNRS, Observatoire astronomique de Strasbourg, UMR 7550, 67000 Strasbourg, France\label{aff14}
\and
ESAC/ESA, Camino Bajo del Castillo, s/n., Urb. Villafranca del Castillo, 28692 Villanueva de la Ca\~nada, Madrid, Spain\label{aff15}
\and
INAF-Osservatorio Astronomico di Brera, Via Brera 28, 20122 Milano, Italy\label{aff16}
\and
INAF-Osservatorio di Astrofisica e Scienza dello Spazio di Bologna, Via Piero Gobetti 93/3, 40129 Bologna, Italy\label{aff17}
\and
Universit\'e Paris-Saclay, Universit\'e Paris Cit\'e, CEA, CNRS, AIM, 91191, Gif-sur-Yvette, France\label{aff18}
\and
IFPU, Institute for Fundamental Physics of the Universe, via Beirut 2, 34151 Trieste, Italy\label{aff19}
\and
INFN, Sezione di Trieste, Via Valerio 2, 34127 Trieste TS, Italy\label{aff20}
\and
SISSA, International School for Advanced Studies, Via Bonomea 265, 34136 Trieste TS, Italy\label{aff21}
\and
Dipartimento di Fisica e Astronomia, Universit\`a di Bologna, Via Gobetti 93/2, 40129 Bologna, Italy\label{aff22}
\and
INFN-Sezione di Bologna, Viale Berti Pichat 6/2, 40127 Bologna, Italy\label{aff23}
\and
INAF-Osservatorio Astronomico di Padova, Via dell'Osservatorio 5, 35122 Padova, Italy\label{aff24}
\and
ATG Europe BV, Huygensstraat 34, 2201 DK Noordwijk, The Netherlands\label{aff25}
\and
Dipartimento di Fisica, Universit\`a di Genova, Via Dodecaneso 33, 16146, Genova, Italy\label{aff26}
\and
INFN-Sezione di Genova, Via Dodecaneso 33, 16146, Genova, Italy\label{aff27}
\and
Department of Physics "E. Pancini", University Federico II, Via Cinthia 6, 80126, Napoli, Italy\label{aff28}
\and
INAF-Osservatorio Astronomico di Capodimonte, Via Moiariello 16, 80131 Napoli, Italy\label{aff29}
\and
European Southern Observatory, Karl-Schwarzschild-Str.~2, 85748 Garching, Germany\label{aff30}
\and
Instituto de Astrof\'isica e Ci\^encias do Espa\c{c}o, Universidade do Porto, CAUP, Rua das Estrelas, PT4150-762 Porto, Portugal\label{aff31}
\and
Faculdade de Ci\^encias da Universidade do Porto, Rua do Campo de Alegre, 4150-007 Porto, Portugal\label{aff32}
\and
Dipartimento di Fisica, Universit\`a degli Studi di Torino, Via P. Giuria 1, 10125 Torino, Italy\label{aff33}
\and
INFN-Sezione di Torino, Via P. Giuria 1, 10125 Torino, Italy\label{aff34}
\and
INAF-Osservatorio Astrofisico di Torino, Via Osservatorio 20, 10025 Pino Torinese (TO), Italy\label{aff35}
\and
INAF-IASF Milano, Via Alfonso Corti 12, 20133 Milano, Italy\label{aff36}
\and
Centro de Investigaciones Energ\'eticas, Medioambientales y Tecnol\'ogicas (CIEMAT), Avenida Complutense 40, 28040 Madrid, Spain\label{aff37}
\and
Port d'Informaci\'{o} Cient\'{i}fica, Campus UAB, C. Albareda s/n, 08193 Bellaterra (Barcelona), Spain\label{aff38}
\and
Institute of Space Sciences (ICE, CSIC), Campus UAB, Carrer de Can Magrans, s/n, 08193 Barcelona, Spain\label{aff39}
\and
Institut d'Estudis Espacials de Catalunya (IEEC),  Edifici RDIT, Campus UPC, 08860 Castelldefels, Barcelona, Spain\label{aff40}
\and
INAF-Osservatorio Astronomico di Roma, Via Frascati 33, 00078 Monteporzio Catone, Italy\label{aff41}
\and
INFN -- Sezione di Napoli, Via Cinthia 6, 80126, Napoli, Italy\label{aff42}
\and
Dipartimento di Fisica e Astronomia "Augusto Righi" - Alma Mater Studiorum Universit\`a di Bologna, Viale Berti Pichat 6/2, 40127 Bologna, Italy\label{aff43}
\and
Instituto de Astrof\'{\i}sica de Canarias, E-38205 La Laguna, Tenerife, Spain\label{aff44}
\and
Institute for Astronomy, University of Edinburgh, Royal Observatory, Blackford Hill, Edinburgh EH9 3HJ, UK\label{aff45}
\and
Jodrell Bank Centre for Astrophysics, Department of Physics and Astronomy, University of Manchester, Oxford Road, Manchester M13 9PL, UK\label{aff46}
\and
European Space Agency/ESRIN, Largo Galileo Galilei 1, 00044 Frascati, Roma, Italy\label{aff47}
\and
Universit\'e Lyon 1, CNRS/IN2P3, IP2I Lyon, Villeurbanne, France\label{aff48}
\and
Aix-Marseille Universit\'e, CNRS, CNES, LAM, Marseille, France\label{aff49}
\and
Institut de Ci\`{e}ncies del Cosmos (ICCUB), Universitat de Barcelona (IEEC-UB), Mart\'{i} i Franqu\`{e}s 1, 08028 Barcelona, Spain\label{aff50}
\and
Instituci\'o Catalana de Recerca i Estudis Avan\c{c}ats (ICREA), Passeig de Llu\'{\i}s Companys 23, 08010 Barcelona, Spain\label{aff51}
\and
Institut de Ciencies de l'Espai (IEEC-CSIC), Campus UAB, Carrer de Can Magrans, s/n Cerdanyola del Vall\'es, 08193 Barcelona, Spain\label{aff52}
\and
Universit\'e Lyon 1, CNRS/IN2P3, IUF, IP2I Lyon, Villeurbanne, France \label{aff53}
\and
Mullard Space Science Laboratory, University College London, Holmbury St Mary, Dorking, Surrey RH5 6NT, UK\label{aff54}
\and
Departamento de F\'isica, Faculdade de Ci\^encias, Universidade de Lisboa, Edif\'icio C8, Campo Grande, PT1749-016 Lisboa, Portugal\label{aff55}
\and
Instituto de Astrof\'isica e Ci\^encias do Espa\c{c}o, Faculdade de Ci\^encias, Universidade de Lisboa, Campo Grande, 1749-016 Lisboa, Portugal\label{aff56}
\and
Department of Astronomy, University of Geneva, ch. d'Ecogia 16, 1290 Versoix, Switzerland\label{aff57}
\and
Universit\'e Paris-Saclay, CNRS, Institut d'astrophysique spatiale, 91405, Orsay, France\label{aff58}
\and
Aix-Marseille Universit\'e, CNRS/IN2P3, CPPM, Marseille, France\label{aff59}
\and
INAF-Istituto di Astrofisica e Planetologia Spaziali, via del Fosso del Cavaliere, 100, 00100 Roma, Italy\label{aff60}
\and
INFN-Bologna, Via Irnerio 46, 40126 Bologna, Italy\label{aff61}
\and
School of Physics, HH Wills Physics Laboratory, University of Bristol, Tyndall Avenue, Bristol, BS8 1TL, UK\label{aff62}
\and
Department of Physics, University of Oxford, Keble Road, Oxford OX1 3RH, UK\label{aff63}
\and
Institute of Theoretical Astrophysics, University of Oslo, P.O. Box 1029 Blindern, 0315 Oslo, Norway\label{aff64}
\and
Jet Propulsion Laboratory, California Institute of Technology, 4800 Oak Grove Drive, Pasadena, CA, 91109, USA\label{aff65}
\and
School of Physics and Astronomy, Lancaster University, Lancaster, LA1 4YB, UK\label{aff66}
\and
Felix Hormuth Engineering, Goethestr. 17, 69181 Leimen, Germany\label{aff67}
\and
Technical University of Denmark, Elektrovej 327, 2800 Kgs. Lyngby, Denmark\label{aff68}
\and
Cosmic Dawn Center (DAWN), Denmark\label{aff69}
\and
Universit\'e PSL, Observatoire de Paris, Sorbonne Universit\'e, CNRS, LERMA, 75014, Paris, France\label{aff70}
\and
Universit\'e Paris-Cit\'e, 5 Rue Thomas Mann, 75013, Paris, France\label{aff71}
\and
Max-Planck-Institut f\"ur Astronomie, K\"onigstuhl 17, 69117 Heidelberg, Germany\label{aff72}
\and
NASA Goddard Space Flight Center, Greenbelt, MD 20771, USA\label{aff73}
\and
Department of Physics and Astronomy, University College London, Gower Street, London WC1E 6BT, UK\label{aff74}
\and
Universit\'e de Gen\`eve, D\'epartement de Physique Th\'eorique and Centre for Astroparticle Physics, 24 quai Ernest-Ansermet, CH-1211 Gen\`eve 4, Switzerland\label{aff75}
\and
Department of Physics, P.O. Box 64, University of Helsinki, 00014 Helsinki, Finland\label{aff76}
\and
Helsinki Institute of Physics, Gustaf H{\"a}llstr{\"o}min katu 2, University of Helsinki, 00014 Helsinki, Finland\label{aff77}
\and
Laboratoire d'etude de l'Univers et des phenomenes eXtremes, Observatoire de Paris, Universit\'e PSL, Sorbonne Universit\'e, CNRS, 92190 Meudon, France\label{aff78}
\and
SKAO, Jodrell Bank, Lower Withington, Macclesfield SK11 9FT, UK\label{aff79}
\and
Centre de Calcul de l'IN2P3/CNRS, 21 avenue Pierre de Coubertin 69627 Villeurbanne Cedex, France\label{aff80}
\and
University of Applied Sciences and Arts of Northwestern Switzerland, School of Computer Science, 5210 Windisch, Switzerland\label{aff81}
\and
Universit\"at Bonn, Argelander-Institut f\"ur Astronomie, Auf dem H\"ugel 71, 53121 Bonn, Germany\label{aff82}
\and
INFN-Sezione di Roma, Piazzale Aldo Moro, 2 - c/o Dipartimento di Fisica, Edificio G. Marconi, 00185 Roma, Italy\label{aff83}
\and
Dipartimento di Fisica e Astronomia "Augusto Righi" - Alma Mater Studiorum Universit\`a di Bologna, via Piero Gobetti 93/2, 40129 Bologna, Italy\label{aff84}
\and
Department of Physics, Institute for Computational Cosmology, Durham University, South Road, Durham, DH1 3LE, UK\label{aff85}
\and
Infrared Processing and Analysis Center, California Institute of Technology, Pasadena, CA 91125, USA\label{aff86}
\and
Institut d'Astrophysique de Paris, UMR 7095, CNRS, and Sorbonne Universit\'e, 98 bis boulevard Arago, 75014 Paris, France\label{aff87}
\and
Niels Bohr Institute, University of Copenhagen, Jagtvej 128, 2200 Copenhagen, Denmark\label{aff88}
\and
Dipartimento di Fisica - Sezione di Astronomia, Universit\`a di Trieste, Via Tiepolo 11, 34131 Trieste, Italy\label{aff89}
\and
Telespazio UK S.L. for European Space Agency (ESA), Camino bajo del Castillo, s/n, Urbanizacion Villafranca del Castillo, Villanueva de la Ca\~nada, 28692 Madrid, Spain\label{aff90}
\and
Institut de F\'{i}sica d'Altes Energies (IFAE), The Barcelona Institute of Science and Technology, Campus UAB, 08193 Bellaterra (Barcelona), Spain\label{aff91}
\and
School of Mathematics and Physics, University of Surrey, Guildford, Surrey, GU2 7XH, UK\label{aff92}
\and
European Space Agency/ESTEC, Keplerlaan 1, 2201 AZ Noordwijk, The Netherlands\label{aff93}
\and
Waterloo Centre for Astrophysics, University of Waterloo, Waterloo, Ontario N2L 3G1, Canada\label{aff94}
\and
Department of Physics and Astronomy, University of Waterloo, Waterloo, Ontario N2L 3G1, Canada\label{aff95}
\and
Perimeter Institute for Theoretical Physics, Waterloo, Ontario N2L 2Y5, Canada\label{aff96}
\and
Space Science Data Center, Italian Space Agency, via del Politecnico snc, 00133 Roma, Italy\label{aff97}
\and
Institute of Space Science, Str. Atomistilor, nr. 409 M\u{a}gurele, Ilfov, 077125, Romania\label{aff98}
\and
Consejo Superior de Investigaciones Cientificas, Calle Serrano 117, 28006 Madrid, Spain\label{aff99}
\and
Universidad de La Laguna, Dpto. Astrof\'\i sica, E-38206 La Laguna, Tenerife, Spain\label{aff100}
\and
Institute for Particle Physics and Astrophysics, Dept. of Physics, ETH Zurich, Wolfgang-Pauli-Strasse 27, 8093 Zurich, Switzerland\label{aff101}
\and
Dipartimento di Fisica e Astronomia "G. Galilei", Universit\`a di Padova, Via Marzolo 8, 35131 Padova, Italy\label{aff102}
\and
INFN-Padova, Via Marzolo 8, 35131 Padova, Italy\label{aff103}
\and
Universit\'e Paris Cit\'e, CNRS, Astroparticule et Cosmologie, 75013 Paris, France\label{aff104}
\and
Instituto de F\'isica Te\'orica UAM-CSIC, Campus de Cantoblanco, 28049 Madrid, Spain\label{aff105}
\and
Institut de Recherche en Astrophysique et Plan\'etologie (IRAP), Universit\'e de Toulouse, CNRS, UPS, CNES, 14 Av. Edouard Belin, 31400 Toulouse, France\label{aff106}
\and
Universit\'e St Joseph; Faculty of Sciences, Beirut, Lebanon\label{aff107}
\and
Departamento de F\'isica, FCFM, Universidad de Chile, Blanco Encalada 2008, Santiago, Chile\label{aff108}
\and
Universit\"at Innsbruck, Institut f\"ur Astro- und Teilchenphysik, Technikerstr. 25/8, 6020 Innsbruck, Austria\label{aff109}
\and
Department of Physics and Helsinki Institute of Physics, Gustaf H\"allstr\"omin katu 2, University of Helsinki, 00014 Helsinki, Finland\label{aff110}
\and
Instituto de Astrof\'isica e Ci\^encias do Espa\c{c}o, Faculdade de Ci\^encias, Universidade de Lisboa, Tapada da Ajuda, 1349-018 Lisboa, Portugal\label{aff111}
\and
Cosmic Dawn Center (DAWN)\label{aff112}
\and
Universidad Polit\'ecnica de Cartagena, Departamento de Electr\'onica y Tecnolog\'ia de Computadoras,  Plaza del Hospital 1, 30202 Cartagena, Spain\label{aff113}
\and
European University of Technology EUt+, European Union\label{aff114}
\and
Centre for Information Technology, University of Groningen, P.O. Box 11044, 9700 CA Groningen, The Netherlands\label{aff115}
\and
Dipartimento di Fisica e Scienze della Terra, Universit\`a degli Studi di Ferrara, Via Giuseppe Saragat 1, 44122 Ferrara, Italy\label{aff116}
\and
Istituto Nazionale di Fisica Nucleare, Sezione di Ferrara, Via Giuseppe Saragat 1, 44122 Ferrara, Italy\label{aff117}
\and
ICSC - Centro Nazionale di Ricerca in High Performance Computing, Big Data e Quantum Computing, Via Magnanelli 2, Bologna, Italy\label{aff118}
\and
INAF, Istituto di Radioastronomia, Via Piero Gobetti 101, 40129 Bologna, Italy\label{aff119}
\and
Astronomical Observatory of the Autonomous Region of the Aosta Valley (OAVdA), Loc. Lignan 39, I-11020, Nus (Aosta Valley), Italy\label{aff120}
\and
Department of Mathematical Physics, Institute of Physics, University of S\~ao Paulo, R. do Mat\~ao 1371, 05508-090, S\~ao Paulo, SP, Brazil\label{aff121}
\and
Oskar Klein Centre for Cosmoparticle Physics, Department of Physics, Stockholm University, Stockholm, SE-106 91, Sweden\label{aff122}
\and
Astrophysics Group, Blackett Laboratory, Imperial College London, London SW7 2AZ, UK\label{aff123}
\and
Univ. Grenoble Alpes, CNRS, Grenoble INP, LPSC-IN2P3, 53, Avenue des Martyrs, 38000 Grenoble, France\label{aff124}
\and
Dipartimento di Fisica, Sapienza Universit\`a di Roma, Piazzale Aldo Moro 2, 00185 Roma, Italy\label{aff125}
\and
Department of Mathematics and Physics E. De Giorgi, University of Salento, Via per Arnesano, CP-I93, 73100, Lecce, Italy\label{aff126}
\and
INFN, Sezione di Lecce, Via per Arnesano, CP-193, 73100, Lecce, Italy\label{aff127}
\and
INAF-Sezione di Lecce, c/o Dipartimento Matematica e Fisica, Via per Arnesano, 73100, Lecce, Italy\label{aff128}
\and
Institut d'Astrophysique de Paris, 98bis Boulevard Arago, 75014, Paris, France\label{aff129}
\and
ICL, Junia, Universit\'e Catholique de Lille, LITL, 59000 Lille, France\label{aff130}
\and
CERCA/ISO, Department of Physics, Case Western Reserve University, 10900 Euclid Avenue, Cleveland, OH 44106, USA\label{aff131}
\and
Laboratoire Univers et Th\'eorie, Observatoire de Paris, Universit\'e PSL, Universit\'e Paris Cit\'e, CNRS, 92190 Meudon, France\label{aff132}
\and
Dipartimento di Fisica "Aldo Pontremoli", Universit\`a degli Studi di Milano, Via Celoria 16, 20133 Milano, Italy\label{aff133}
\and
INFN-Sezione di Milano, Via Celoria 16, 20133 Milano, Italy\label{aff134}
\and
Departamento de F{\'\i}sica Fundamental. Universidad de Salamanca. Plaza de la Merced s/n. 37008 Salamanca, Spain\label{aff135}
\and
IRFU, CEA, Universit\'e Paris-Saclay 91191 Gif-sur-Yvette Cedex, France\label{aff136}
\and
Universiteit Gent, Department of Physics and Astronomy, Proeftuinstraat 86 N3, 9000 Ghent, Belgium
\label{aff137}
\and
Center for Data-Driven Discovery, Kavli IPMU (WPI), UTIAS, The University of Tokyo, Kashiwa, Chiba 277-8583, Japan\label{aff138}
\and
Department of Physics, Astronomy and Mathematics, University of Hertfordshire, College Lane, Hatfield AL10 9AB, UK\label{aff139}
\and
Institute of Cosmology and Gravitation, University of Portsmouth, Portsmouth PO1 3FX, UK\label{aff140}
\and
Department of Physics \& Astronomy, University of California Irvine, Irvine CA 92697, USA\label{aff141}
\and
Departamento F\'isica Aplicada, Universidad Polit\'ecnica de Cartagena, Campus Muralla del Mar, 30202 Cartagena, Murcia, Spain\label{aff142}
\and
Instituto de F\'isica de Cantabria, Edificio Juan Jord\'a, Avenida de los Castros, 39005 Santander, Spain\label{aff143}
\and
CEA Saclay, DFR/IRFU, Service d'Astrophysique, Bat. 709, 91191 Gif-sur-Yvette, France\label{aff144}
\and
Department of Computer Science, Aalto University, PO Box 15400, Espoo, FI-00 076, Finland\label{aff145}
\and
School of Physics and Astronomy, University of Nottingham, University Park, Nottingham NG7 2RD, UK\label{aff146}
\and
Ruhr University Bochum, Faculty of Physics and Astronomy, Astronomical Institute (AIRUB), German Centre for Cosmological Lensing (GCCL), 44780 Bochum, Germany\label{aff147}
\and
DARK, Niels Bohr Institute, University of Copenhagen, Jagtvej 155, 2200 Copenhagen, Denmark\label{aff148}
\and
Department of Physics, Centre for Extragalactic Astronomy, Durham University, South Road, Durham, DH1 3LE, UK\label{aff149}
\and
Astrophysics Research Centre, University of KwaZulu-Natal, Westville Campus, Durban 4041, South Africa\label{aff150}
\and
School of Mathematics, Statistics \& Computer Science, University of KwaZulu-Natal, Westville Campus, Durban 4041, South Africa\label{aff151}
\and
Department of Physics and Astronomy, Vesilinnantie 5, University of Turku, 20014 Turku, Finland\label{aff152}
\and
Finnish Centre for Astronomy with ESO (FINCA), Quantum, Vesilinnantie 5, University of Turku, 20014 Turku, Finland\label{aff153}
\and
Serco for European Space Agency (ESA), Camino bajo del Castillo, s/n, Urbanizacion Villafranca del Castillo, Villanueva de la Ca\~nada, 28692 Madrid, Spain\label{aff154}
\and
ARC Centre of Excellence for Dark Matter Particle Physics, Melbourne, Australia\label{aff155}
\and
Centre for Astrophysics \& Supercomputing, Swinburne University of Technology,  Hawthorn, Victoria 3122, Australia\label{aff156}
\and
Department of Physics and Astronomy, University of the Western Cape, Bellville, Cape Town, 7535, South Africa\label{aff157}
\and
MTA-CSFK Lend\"ulet Large-Scale Structure Research Group, Konkoly-Thege Mikl\'os \'ut 15-17, H-1121 Budapest, Hungary\label{aff158}
\and
Konkoly Observatory, HUN-REN CSFK, MTA Centre of Excellence, Budapest, Konkoly Thege Mikl\'os {\'u}t 15-17. H-1121, Hungary\label{aff159}
\and
Institute for Theoretical Particle Physics and Cosmology (TTK), RWTH Aachen University, 52056 Aachen, Germany\label{aff160}
\and
INAF-Osservatorio Astrofisico di Arcetri, Largo E. Fermi 5, 50125, Firenze, Italy\label{aff161}
\and
Centro de Astrof\'{\i}sica da Universidade do Porto, Rua das Estrelas, 4150-762 Porto, Portugal\label{aff162}
\and
HE Space for European Space Agency (ESA), Camino bajo del Castillo, s/n, Urbanizacion Villafranca del Castillo, Villanueva de la Ca\~nada, 28692 Madrid, Spain\label{aff163}
\and
STAR Institute, University of Li{\`e}ge, Quartier Agora, All\'ee du six Ao\^ut 19c, 4000 Li\`ege, Belgium\label{aff164}
\and
Center for Frontier Science, Chiba University, 1-33 Yayoi-cho, Inage-ku, Chiba 263-8522, Japan\label{aff165}
\and
Department of Physics, Graduate School of Science, Chiba University, 1-33 Yayoi-Cho, Inage-Ku, Chiba 263-8522, Japan\label{aff166}
\and
Department of Astrophysics, University of Zurich, Winterthurerstrasse 190, 8057 Zurich, Switzerland\label{aff167}
\and
Department of Mathematics and Physics, Roma Tre University, Via della Vasca Navale 84, 00146 Rome, Italy\label{aff168}
\and
INAF - Osservatorio Astronomico d'Abruzzo, Via Maggini, 64100, Teramo, Italy\label{aff169}
\and
Theoretical astrophysics, Department of Physics and Astronomy, Uppsala University, Box 524, 751 20 Uppsala, Sweden\label{aff170}
\and
Mathematical Institute, University of Leiden, Einsteinweg 55, 2333 CA Leiden, The Netherlands\label{aff171}
\and
School of Physical Sciences, The Open University, Milton Keynes, MK7 6AA, UK\label{aff172}
\and
Institute of Astronomy, University of Cambridge, Madingley Road, Cambridge CB3 0HA, UK\label{aff173}
\and
Univ. Lille, CNRS, Centrale Lille, UMR 9189 CRIStAL, 59000 Lille, France\label{aff174}
\and
Center for Astrophysics and Cosmology, University of Nova Gorica, Nova Gorica, Slovenia\label{aff175}
\and
Kobayashi-Maskawa Institute for the Origin of Particles and the Universe, Nagoya University, Chikusa-ku, Nagoya, 464-8602, Japan\label{aff176}
\and
Department of Astrophysical Sciences, Peyton Hall, Princeton University, Princeton, NJ 08544, USA\label{aff177}
\and
Fakult\"at f\"ur Physik, Universit\"at Bielefeld, Postfach 100131, 33501 Bielefeld, Germany\label{aff178}
\and
School of Mathematics, Statistics and Physics, Newcastle University, Herschel Building, Newcastle-upon-Tyne, NE1 7RU, UK\label{aff179}
\and
SRON Netherlands Institute for Space Research, Landleven 12, 9747 AD, Groningen, The Netherlands\label{aff180}
\and
Department of Physics and Astronomy, University of British Columbia, Vancouver, BC V6T 1Z1, Canada\label{aff181}}    
\abstract{
Accurate photometric-redshift estimation and calibration of the weak-lensing point-spread function (PSF) for \Euclid require complementary ground-based optical imaging with high photometric precision and spatial homogeneity.
The Organisation Unit for External Data (OU-EXT) integrates heterogeneous ground-based surveys into the \Euclid Science Ground Segment (SGS), delivering uniformly calibrated, co-added images, and associated PSF models.
To minimise systematics in spectral energy distribution (SED) reconstruction and photometric-redshift estimation, these data must achieve relative photometric stability at the $\sim$10~mmag level on \Euclid-tile scales.
We present the ground-based imaging component of Euclid Data Release~1 (DR1), comprising $\sim2500$~deg$^2$ across the wide, deep, and auxiliary fields. 
We summarise the OU-EXT production workflow and validation framework, and quantify the achieved data quality of the co-added products in terms of depth, photometric zero-point stability, and spatial colour homogeneity.
In the Euclid Wide Survey (EWS), the point-source $10\sigma$ depths in the $griz$ bands reach 24.8/24.3/24.0/23.0 AB mag, with the deep fields extending 1--2 magnitudes deeper.
Most EWS DR1 tiles meet the 10~mmag relative photometric requirement, with 1.1\% and 1.5\% showing offsets exceeding 50~mmag in the northern and southern hemispheres, respectively.
Spatial colour variations remain typically below 5--10~mmag across the southern EWS footprint, while larger residual structures of 10--25~mmag appear in some northern regions.
The deep fields are more sensitive to calibration and stacking systematics: while most bands show low rejection rates, specific datasets exhibit systematic calibration failures.
These issues originate from OU-EXT recalibration and co-addition rather than the original survey data.
These results demonstrate that the EXT production chain provides a largely homogeneous framework for integrating ground-based datasets into \Euclid, while identifying calibration and co-addition limitations that have to be addressed in DR2 to meet the mission’s most stringent cosmological requirements.
}

%
% Provide up to five key words:
%
\keywords{}
%
% Add short versions of title and author list for page headings
%
\titlerunning{Colouring \Euclid with external ground-based imaging data}

\maketitle
%
%-------------------------------------------------------------------
%
%
%   Start the main text of your paper here
%
%\tableofcontents

\section{\label{s:Intro}Introduction}

The \Euclid mission \citep{EuclidSkyOverview} is an ongoing project of the European Space Agency (ESA), carried out in collaboration with the Euclid Consortium, with the primary goal of constraining the nature of dark energy and its role in cosmic acceleration.
Following its successful launch in 2023, the observatory has been continuously acquiring data at an unprecedented rate.

However, the onboard imaging instruments, VIS \citep{EuclidSkyVIS} and NISP \citep{EuclidSkyNISP}, provide only limited spectral coverage, resulting in coarse sampling of the spectral energy distributions (SEDs) of detected sources.
A precise characterisation of galaxy SEDs is essential for accurate photometric-redshift (photo-$z$) estimation \citep{Bolzonella+00, Abdalla+08}, which underpins galaxy clustering measurements, and weak lensing analyses -- both key methods for the \Euclid primary scientific goals.

Consequently, the nearly 14\,000 square degrees that will ultimately be covered by \Euclid must also be observed in complementary optical bands by ground-based facilities.
This multi-wavelength coverage is required to \change{effectively sample the SED of stars and galaxies, providing enough information to break the intrinsic degeneracies between redshift and stellar age, metallicity and dust extinction, and to infer robust photo-$z$ estimates.}

At the time the \Euclid mission requirements were defined \citep{Laureijs11}, several wide-field imaging surveys were either underway or being planned to map large fractions of the sky to depths comparable to those expected for \Euclid.
Notable examples include the Dark Energy Survey \citep[DES;][]{DESoverview}, the Kilo-Degree Survey \citep[KiDS;][]{deJong+13}, the Pan-STARRS surveys \citep{Chambers+16}, the Canada-France Imaging Survey \citep[CFIS;][]{Ibata+17}, the Subaru Strategic Program \citep[HSC-SSP;][]{Aihara+18}, and the Legacy Survey of Space and Time \citep[LSST;][]{Ivezic+19}.

To coordinate the external observations required by the mission, the Organisation Unit for External Data (OU-EXT) was established within the Science Ground Segment (SGS; see \citealt{DR1-TP033}).
OU-EXT coordinates the ingestion, homogenisation, calibration, and processing of the ground-based imaging used to complement the \Euclid VIS and NISP observations.
These data are required not only to constrain photometric redshifts, but also to determine the SED-weighted VIS PSF on a source-by-source basis for weak-lensing analyses \citep{EriksenHoekstra18}.

Multiple coordinated ground-based observing campaigns therefore contribute to the \Euclid survey footprint.
In the Euclid Wide Survey (EWS), the northern hemisphere is covered by the Ultraviolet Near Infrared Optical Northern Survey \citep[UNIONS;][]{Gwyn+25}, while the southern hemisphere is currently covered by DES \citep{DESoverview} and will eventually be complemented by deeper LSST observations \citep{Ivezic+19}.
The Euclid Deep Fields (EDFs) and Euclid Auxiliary Fields (EAFs) have more demanding requirements in terms of depth and wavelength coverage and are therefore supplemented by dedicated observing campaigns and archival datasets.

While \Euclid itself delivers a vast and homogeneous space-based dataset, the incorporation of ground-based ancillary data introduces an additional layer of complexity.
OU-EXT must combine observations obtained under varying atmospheric conditions, from different observing sites, instrumental passbands, survey strategies, and data-reduction pipelines into a single, consistently calibrated framework.
The resulting external dataset spans a data volume several times larger than that delivered by the spacecraft alone, making its homogenisation, photometric recalibration, and quality control a central technical challenge for the mission.
Moreover, even small spatial variations in the photometric calibration can introduce position-dependent biases in galaxy colours and colour--redshift relations, which can propagate into photometric-redshift estimates and ultimately into cosmological parameter inference.
Ensuring photometric calibration homogeneity across the \Euclid footprint is therefore not only a data-quality objective, but an essential requirement for controlling systematic uncertainties in the mission's cosmological analyses.

This paper provides a detailed description of the processing and calibration steps applied to the ground-based optical imaging datasets released in DR1 \citep{DR1cite}.
We describe the characteristics of the DR1 external data, including observations covering the Euclid Wide Survey (EWS), as well as the data used in the Euclid Auxiliary Fields (EAFs) and Euclid Deep Fields (EDFs).
We present the co-addition strategy adopted within the SGS, the generation of stacked data products, and a quantitative assessment of the resulting data quality across the DR1 footprint.

The structure of this paper is as follows.
We begin by describing the role of OU-EXT within the SGS, outlining its responsibilities in the acquisition, homogenisation, calibration, and validation of ground-based imaging data (Sect.~\ref{s:overview}).
We then present the datasets included in DR1, detailing their survey origins, sky coverage, and observational characteristics (Sect.~\ref{s:Data}).
Next, we describe the co-addition pipeline implemented within the SGS for both images and point-spread function (PSF) models (Sect.~\ref{s:pipeline}).
In Sect.~\ref{s:qa}, we assess the data quality of the released co-added images and their associated PSF models, providing a quantitative evaluation of the photometric performance achieved across the DR1 footprint.
We discuss the DR1 validation outcome, identified limitations, and improvements planned for future releases in Sect.~\ref{s:discussion}, and summarise our main conclusions in Sect.~\ref{s:conclusions}.
Appendices~\ref{appendix:sef_flagmaps}--\ref{appendix:photom_stab_sky} provide additional information on SEF flag propagation, reference catalogues, and the spatial distributions of depth and photometric stability.
\change{Magnitudes are reported using the AB magnitude system \citep{Oke+83}.}

\section{\label{s:overview}\change{Organisational Unit for External Data
overview}}

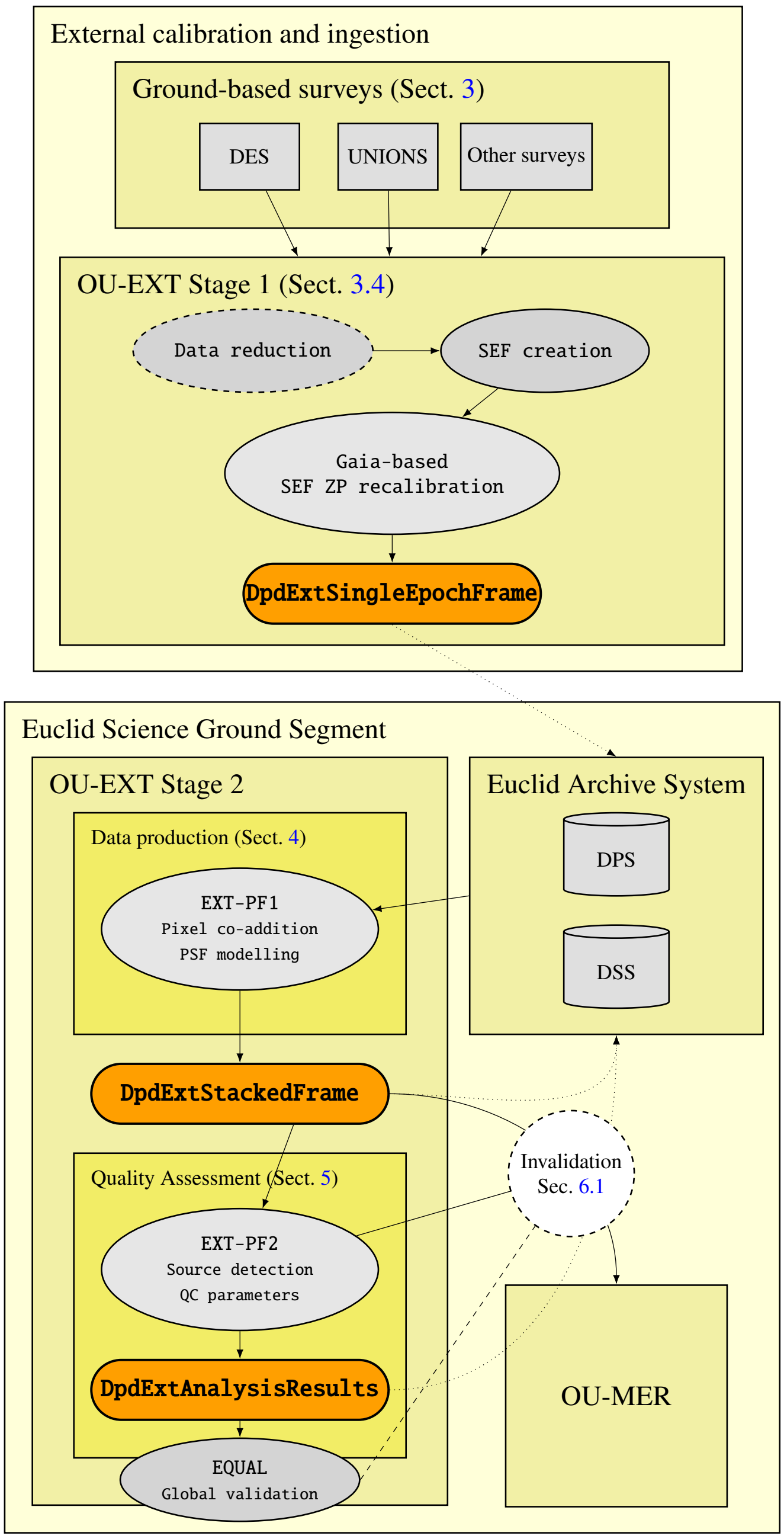
\begin{figure}
    \centering
    \resizebox{\linewidth}{!}{
    \definecolor{celement}{rgb}{0.83,0.83,0.83}
\definecolor{yellow1}{rgb}{0.95,0.94,0.65}
\definecolor{cstage2sect}{rgb}{0.95,0.93,0.4}

\begin{tikzpicture}[
  font=\large,
  >=Latex,
  node distance=10mm and 12mm,
  flow/.style={->, thick, draw=blue!55},
  flowthin/.style={->, thick, draw=blue!45},
  box/.style={draw, thick, fill=yellow!15, inner sep=8mm},
  group/.style={draw, thick, fill=yellow1, inner sep=8mm},
  subgroup/.style={draw, thick, fill=cstage2sect, inner sep=8mm},
  proc/.style={draw, thick, fill=gray!25, minimum height=12mm, minimum width=28mm, align=center},
  prod/.style={draw, rounded rectangle, very thick, fill=orange!75!yellow, minimum height=12mm, minimum width=62mm, align=center, scale=0.9, font=\Large\ttfamily\bfseries},
  oval/.style={draw, thick, fill=celement, ellipse, align=center, minimum height=15mm, font=\large\ttfamily},
  ovalbig/.style={draw, thick, fill=gray!20, ellipse, align=center, minimum height=22mm, minimum width=50mm, font=\large\ttfamily},
  cyl/.style={draw, thick, fill=gray!25, cylinder, shape border rotate=90, aspect=0.25,
              minimum height=15mm, minimum width=19mm, align=center}
]

% ============================================================
% External calibration and ingestion
% ============================================================
\node[box, minimum width=128mm, minimum height=120mm] (extcal) {};

\node[anchor=north west, font=\Large]
  at ([xshift=2mm,yshift=-2mm]extcal.north west) {External calibration and ingestion};

% % --- Ground-based surveys
\node[anchor=north] (gbs)
   at ([yshift=-8mm]extcal.north) {};

\node[group, fill=yellow1, minimum width=100mm,  minimum height=30mm, anchor=north]
   at ([xshift=2mm,yshift=-2mm]gbs.north west) (gbscontainer) {};

\node[anchor=north west, font=\Large] at ([xshift=2mm, yshift=-2mm]gbscontainer.north west) {Ground-based surveys (Sect.~\ref{s:Data})};

\node[proc, minimum width=18mm] (des)    at ([xshift=-25mm,yshift=-18mm]gbs.center) {DES};
\node[proc, minimum width=18mm] (unions) at ([xshift=0mm,yshift=-18mm]gbs.center) {UNIONS};
\node[proc, minimum width=18mm] (dots)   at ([xshift=25mm,yshift=-18mm]gbs.center) {Other surveys};

% % --- OU-EXT Stage 1
\node[group, anchor=north, minimum width=120mm, minimum height=70mm] (stage1)
       at ([yshift=-5mm]gbscontainer.south) {};

\node[anchor=north west, font=\Large]
   at ([xshift=2mm,yshift=-2mm]stage1.north west) {OU-EXT Stage 1 (Sect.~\ref{s:stage1_preprocessing})};

\node[oval, style={dashed}] (datared) at ([xshift=35mm, yshift=-17mm]stage1.north west) {Data reduction};

\node[oval, right= of datared] (sef) {SEF creation};

\node[ovalbig, below= 28mm of stage1.north] (gaiazp) {Gaia-based\\SEF ZP recalibration};

\node[prod, below= 5mm of gaiazp] (dpd_sef) {DpdExtSingleEpochFrame};

\draw [->] (des) -- (stage1);
\draw [->] (unions) -- (stage1);
\draw [->] (dots) -- (stage1);

\draw [->] (datared) -- (sef);
\draw [->] (sef) -- (gaiazp);
\draw [->] (gaiazp) -- (dpd_sef);

% % ============================================================
% % Euclid Science Ground Segment
% % ============================================================

\node[box, below=10mm of stage1, minimum width=140mm, minimum height=150mm] (sgs) {};
\node[anchor=north west, font=\Large] at ([xshift=2mm,yshift=-2mm]sgs.north west) {Euclid Science Ground Segment};

\node[group, fill=yellow1, anchor=north east, minimum width=53mm, minimum height=50mm] (eas)
       at ([xshift=-3mm, yshift=-10mm]sgs.north east) {};
\node[anchor=north, font=\Large] at ([yshift=-2mm]eas.north) {Euclid Archive System};
\node[cyl, below= of eas.north] (dps) {DPS};
\node[cyl, below= 5mm of dps.south] (dss) {DSS};

\node[group, fill=yellow1, anchor=north west, minimum width=75mm, minimum height=135mm] (stage2)
       at ([xshift=5mm, yshift=-10mm]sgs.north west) {};
\node[anchor=north west, font=\Large] at ([xshift=2mm,yshift=-2mm]stage2.north west) {OU-EXT Stage 2};

%% PRODUCTION
\node[subgroup, anchor=north, below= of stage2.north, minimum width=60mm, minimum height=40mm] (s2prod) {};
\node[anchor=north west] at ([xshift=2mm,yshift=-2mm]s2prod.north west) {Data production (Sect.~\ref{s:pipeline})};
\node[ovalbig, anchor=north, below= of s2prod.north] (pf1) {EXT-PF1\\\footnotesize{Pixel co-addition}\\\footnotesize{PSF modelling}};

\node[prod, below= 5mm of s2prod.south] (dpd_stack) {DpdExtStackedFrame};

%% VALIDATION
\node[subgroup, below= 5mm of dpd_stack, minimum width=60mm, minimum height=55mm] (s2val) {};
\node[anchor=north west] at ([xshift=2mm,yshift=-2mm]s2val.north west) {Quality Assessment (Sect.~\ref{s:qa})};
\node[ovalbig, anchor=north, below= of s2val.north] (pf2) {EXT-PF2\\\footnotesize{Source detection}\\\footnotesize{QC parameters}};

\node[prod, below= 5mm of pf2.south] (dear) {DpdExtAnalysisResults};

\node[oval, anchor=north, below= 3mm of dear.south] (equal) {EQUAL\\\footnotesize{Global validation}};

%% MER
\node[group, fill=yellow1, anchor=north, minimum width=40mm, minimum height=40mm, below= 45mm of eas.south, font=\Large] (mer) {OU-MER};

\draw[->, dotted] (dpd_sef.south) -- (eas.north);
\draw[->] (eas.west) -- (pf1);
\draw[->] (pf1.south) -- (dpd_stack.north);
\draw[->, dotted] (dpd_stack.east) to[out=0, in=270] (eas.south);

\draw[->] (dpd_stack.330) -- (pf2);
\draw[->] (pf2) -- (dear);
\draw[->] (dear) -- (equal);
\draw[->, dotted] (dear.east) to[out=0, in=270] (eas.south);
\draw[->] (dpd_stack.east) to[out=0, in=90] (mer.north);

\draw[->] (pf2) -- ([xshift=12mm,yshift=20mm]mer.north west);
\draw[->, dashed] (equal.east) -- ([xshift=12mm,yshift=20mm]mer.north west);
\node[circle, draw, dashed, thick, fill=white, minimum size=18mm, align=center] at ([xshift=12mm,yshift=20mm]mer.north west) {Invalidation\\Sec.~\ref{s:data_validation}};

\end{tikzpicture}
}
    \caption{Overview of OU-EXT data production workflow.}
    \label{fig:ext-overview}
\end{figure}

The activities carried out within OU-EXT are illustrated in Fig.~\ref{fig:ext-overview}.
The data flow is divided into two stages.
First, stage 1 is the sub-unit that i) coordinates with the external surveys the delivery of data, ii) converts the original format of the files into the \Euclid data products, iii) performs a recalibration step in terms of photometric zero-point common to all bands, and iv) delivers calibrated individual exposures into the Euclid Archive System (ESA).

Once the data are part of the \Euclid ecosystem, as shown in Fig.~\ref{fig:ext-overview}, Stage 2 is the OU-EXT sub-unit responsible for creating and validating the final EXT product (\stack), that contains co-added images and their associated PSF models, referred to as ``StackedFrames''.
These data products follow the \change{\Euclid tiling scheme, and are the input ground-based product used by the merging (MER) pipeline \citep[see][for details]{Q1-TP004, DR1-TP009}, which produces the final multi-wavelength dataset released to the public.}

\subsection{End-of-mission OU-EXT PF requirements}

After the final Euclid Data Release, the EXT processing function\footnote{\change{Within the \Euclid pipeline, a processing function refers to an individual processing element responsible for a specific task.}}(PF; see Sect.~\ref{s:pipeline}) must produce ground-based data compliant with the following requirements:

\begin{itemize}
    \item PF EXT shall verify that the raw external imaging science and calibration data meet the Euclid EXT imaging depth requirements\change{, ensuring a sufficient signal-to-noise ratio across the full survey footprint}.
    
    \item The post-calibration relative photometric bias shall be $<10$ mmag between any two points in the survey.
    
    \item PF EXT shall provide photometric zero-point calibrations enabling colour measurements with a systematic error $<5$ mmag on spatial scales relevant for modelling the \Euclid VIS PSF \citep[see][]{EriksenHoekstra18}.
    
    \item PF EXT shall calibrate the external imaging data such that the post-calibration median $1\sigma$ systematic error on individual stars used for VIS PSF modelling is better than 5 mmag.
    
    \item PF EXT shall achieve an astrometric accuracy in the post-calibration external imaging data that matches the VIS astrometric performance, ensuring that astrometry does not limit the cross-matching success rate with VIS sources.
    
    \item The astrometric solution shall be corrected for velocity aberration through calibration to the \gaia reference frame.
    
    \item PF EXT shall provide PSF characterisation with sufficient accuracy to meet the colour and photometric zero-point requirements listed above.
    
    \item PF EXT shall characterise the passbands of the external imaging data with sufficient accuracy to meet the colour and photometric requirements listed above.
\end{itemize}

Some of these requirements pose an unprecedented challenge, as several of the ground-based surveys were not originally designed to achieve \Euclid levels of accuracy.
Throughout the remainder of this paper, we present the first EXT data release, assess the currently achieved data quality, and outline the improvements that will be implemented to ensure that the EXT data meet the final requirements.

\section{\label{s:Data} DR1 ground-based optical imaging}

\begin{table}[htp!]
	\caption{EXT StackedFrames produced in DR1.}
    \centering
	\label{tab:stack_stats}
	\begin{tabular}{crrr}
		\hline\hline\\[-8pt]
		Band & \shortstack{No. wide stacks\\(EWS)}& \shortstack{No. deep stacks\\(EDFs/EAFs)} & Total\\[2pt]
		\hline\\[-8pt]
        DECam-$g$ & 7512 & 1043 & 8555\\
        DECam-$r$ & 7511 & 1046 & 8557\\
        DECam-$i$ & 7511 & 1046 & 8557\\
        DECam-$z$ & 7511 & 1047 & 8558\\
        HSC-$g$ & 2729 & 867 & 3596\\
        HSC-$r$ & 64 & 195 & 259\\
        HSC-$r2$ & 130 & 743 & 873\\
        HSC-$i$ & 0 & 234 & 234\\
        HSC-$i2$ & 147 & 708 & 855\\
        HSC-$z$ & 2527 & 780 & 3307\\
        MegaCam-$u$ & 1391 & 82 & 1473\\
        MegaCam-$r$ & 2603 & 65 & 2668\\
        Pan-STARRS-$i$ & 2718 & 82 & 2800\\
        Total & 42\,354 & 7938 & 50\,292\\
        \hline
	\end{tabular}
\end{table}

The DR1 \change{production campaign resulted in a} ground-based dataset \change{covering nearly} 2500 deg$^2$ \citep[][]{DR1-TP033} that spans both hemispheres \citep[see also][for the final released sample footprint]{DR1-TP001}.
In terms of the EWS, this translates into $10\,322$ MER tiles.
The EDFs and EAFs correspond to about 130 deg$^2$ and 1651 MER deep tiles.
For each tile typically multiple ground-based imaging bands are available covering the $ugriz$ wavelength range.
\change{Table~\ref{tab:stack_stats} summarises the number of tiles processed in each photometric band for DR1.
In the following sections, we describe the origin and characteristics of the ground-based data contributing to the different \Euclid survey components.}

\subsection{EWS-North}

The northern component comprises around 30\% of the EWS MER tile footprint in the current release.
The imaging data are provided by the UNIONS\footnote{Data available at \url{https://www.skysurvey.cc}} collaboration \citep[see][for a detailed overview]{Gwyn+25}, and include observations from multiple facilities and surveys:

\begin{itemize}
    \item $u$ and $r$ bands (CFHT/MegaCam):
    Imaging in the $u$ and $r$ bands was obtained with the MegaCam instrument on the Canada--France--Hawaii Telescope (CFHT) as part of the CFHT survey.
    The data were processed using the \textsc{MegaPipe} pipeline \citep{Gwyn+08}.

    \item $g$ band (Subaru/HSC):
    $g$-band observations were acquired with the Hyper Suprime-Cam (HSC) instrument on the Subaru Telescope and are provided by the Waterloo--Hawaii--IfA $g$-band Survey (WHIGS).
    This programme combines archival HSC-$g$ imaging with targeted follow-up observations conducted through several Canada- and Hawaii-led observing programmes.
    Data reduction was performed using the LSST pipeline \citep[v23, based on][]{Bosch+18}.

    \item $z$ band (Subaru/HSC):
    $z$-band imaging is provided by the Wide Imaging with Subaru Hyper Suprime-Cam for Euclid Skies (WISHES) programme.
    This dataset combines archival HSC-$z$ observations, including data from HSC-SSP \citep{Aihara+18}, with optimized follow-up imaging from multiple observing campaigns involving Canadian and Hawaiian institutions.
    The data were reduced using the same HSC pipeline \citep{Bosch+18} version adopted in \citet{Aihara+22}.

    \item $i$ band (Pan-STARRS/GPC1):
    $i$-band observations were obtained with the Gigapixel Camera (GPC) on the Panoramic Survey Telescope and Rapid Response System (Pan-STARRS; \citealt{Kaiser+02}).
    The data reduction and calibration procedures are described in detail by \citet{Magnier+20, Magnier+20b}.
\end{itemize}

\subsection{EWS-South}

The southern component of the EWS comprises the largest fraction of the release ($\simeq70$\%) and covers the $griz$ bands.
It is entirely covered by observations performed with the Dark Energy Camera (DECam) on the Blanco telescope.
The majority of the observations are part of DES\footnote{Data available at \url{https://www.darkenergysurvey.org}} \citep[][]{Abbott+21}, complemented by archival DECam imaging in selected regions.
At variance with the northern counterpart, EWS-South data is processed, de-trended and calibrated starting from the raw observations by the OU-EXT Stage 1 team.

\subsection{EDFs and EAFs}

Ground-based observations of EDF-N are covered by the Hawaii Twenty Square Degree Survey\footnote{Data available at \url{https://dawn.ipac.caltech.edu/page/data}} \citep[H2O;][]{EP-McPartland, EP-Zalesky}.
This survey combines dedicated HSC observational campaigns with archival HSC data such as the HEROES \citep{Taylor+23} and AKARI-NEP \citep{Oi+21} programs\change{, covering the HSC-$g,\,r,\,r2,\,i,\,i2,\,\rm{and}~z$ bands}.
Data were processed with \texttt{hscPipe} v8.4 \citep{Bosch+18}.
\change{Owing to the large number of overlapping exposures, the calibration did not adopt the standard HSC-SSP PDR3 configuration, in which \texttt{jointcal} is used for astrometric calibration and the Forward Global Calibration Method (FGCM; \citealt{Burke+18}) for photometric calibration \citep{Aihara+22}.
Instead, \texttt{jointcal} was used to derive both the astrometric and photometric calibration, providing a computationally efficient global solution across the large number of available exposures \citep[see][for details]{Bosch+18}.}

\change{EDF-F is also covered by the H2O programme and follows the same HSC processing strategy as EDF-N.}
In addition, EDF-F includes DECam imaging \change{($griz$)} from archival observations and the DES supernovae survey \citep{Abbott+21} \change{, processed by the EXT Stage~1 team following the same procedure as for EWS-South.}

EDF-S is entirely covered by DES $griz$ imaging \citep{Abbott+21}, processed by the EXT Stage~1 team following the same procedure and to the same nominal depth as EWS-South.

The COSMOS field combines both HSC ($g$, $r$, $r2$, $i$, $i2$, $z$) imaging from the SSP program \citep{Aihara+22}, and DECam ($griz$) archival imaging.

\subsection{\label{s:stage1_preprocessing}OU-EXT Stage~1 processing and calibration}

\change{After the ground-based data are delivered by the external teams, the OU-EXT Stage~1 pipelines perform the survey-specific processing and harmonisation required to produce the standardised data products ingested by the \Euclid system.}
The outcome of Stage~1 is a homogeneous data format consisting of single-CCD exposures, and their associated catalogues of sources and PSF models (\sef, SEFs).
In practice, each SEF contains the calibrated science image, an inverse-variance or RMS map, a per-pixel flag map, astrometric WCS metadata, and survey-specific PSF information and source catalogues used for the Stage~1 recalibration.

Despite this harmonisation, \change{some differences remain between the survey-specific Stage~1 processing chains contributing to DR1} listed across Sect.~\ref{s:Data}.
These differences mainly concern
\begin{itemize}
    \item the definition and usage of data-quality flags, in particular those employed for masking during co-addition
    (see Appendix~\ref{appendix:sef_flagmaps} for details);
    \item the specification of PSF models (e.g. \textsc{PSFEx} polynomial models versus instrument-native PSF representations), including differences in the stamp normalisation conventions and metadata required to reconstruct PSF photometry;
    \item the choice of photometric reference catalogue and the photometric-flux definition used for calibration (e.g. PSF flux normalisation aperture, aperture-correction strategy, and treatment of spatial variations);
    \item the use of \gaia for astrometric calibration \citep[e.g.,][]{George+20}, in particular whether stellar proper motions are propagated to the epoch of observation, which can matter at the tens of milliarcsecond level for multi-year baselines.
\end{itemize}
These survey-dependent differences are accounted for by the Stage 2 processing described in Sect.~\ref{s:pipeline}.

\change{
For DECam data covering the southern hemisphere, the Stage~1 pipeline transforms the raw exposures into fully calibrated, science-ready images, producing one image for each CCD within a DES science exposure.
The pipeline is a modified version of the \texttt{CosmoDM} data-reduction pipeline \citep{Desai+15} and comprises the following steps:
\begin{enumerate}
\item Single-epoch image detrending, including bias subtraction, flat-fielding, and cross-talk correction.
\item Astrometric calibration to the \gaia reference frame.
\item Single-epoch masking and interpolation of defective pixels affected by, for example, cosmic rays, saturation, or scattered light.
\item Background modelling.
\item PSF modelling.
\item Source-catalogue extraction.
\end{enumerate}
}

\change{
For the external surveys covering the northern DR1 footprint, the input observations have already undergone the main image-processing and calibration steps within their respective survey pipelines.
OU-EXT Stage~1 therefore primarily performs the required data-format harmonisation and the common photometric recalibration described below.
}

\subsubsection{Photometric recalibration}

\change{Photometric calibration is refined at two points in the EXT processing workflow.
The first is performed during Stage~1, where a zero-point recalibration pipeline uses the source catalogues associated with the SEFs to derive a homogeneous photometric solution across the EXT datasets.
}

The recalibration is based on a weighted linear least-squares optimisation of the photometric zero-points of individual CCD images over regions spanning tens of degrees, defined as \texttt{SuperTiles}.
This technique employs synthetic magnitudes, derived from \gaia XP spectra \citep{Montegriffo+23} as a uniform photometric reference.
Absolute constraints are obtained by comparing the photometry of individual CCD images with \gaia spectrophotometry, while relative constraints are derived from repeated stellar measurements in overlapping CCD images.
Together, these constraints define an overdetermined sparse linear system, whose weighted least-squares solution provides the optimal relative zero-point calibration of the contributing CCD images.

The second photometric adjustment is performed during Stage~2 processing within the SGS.
It consists of a single photometric adjustment factor at the mmag level, determined independently for each filter from the underlying SEFs and applied during the production of the \stack products.
This calibration step is described in Sect.~\ref{s:photomAdjustment}.

\section{\label{s:pipeline}EXT StackedFrame production PF}

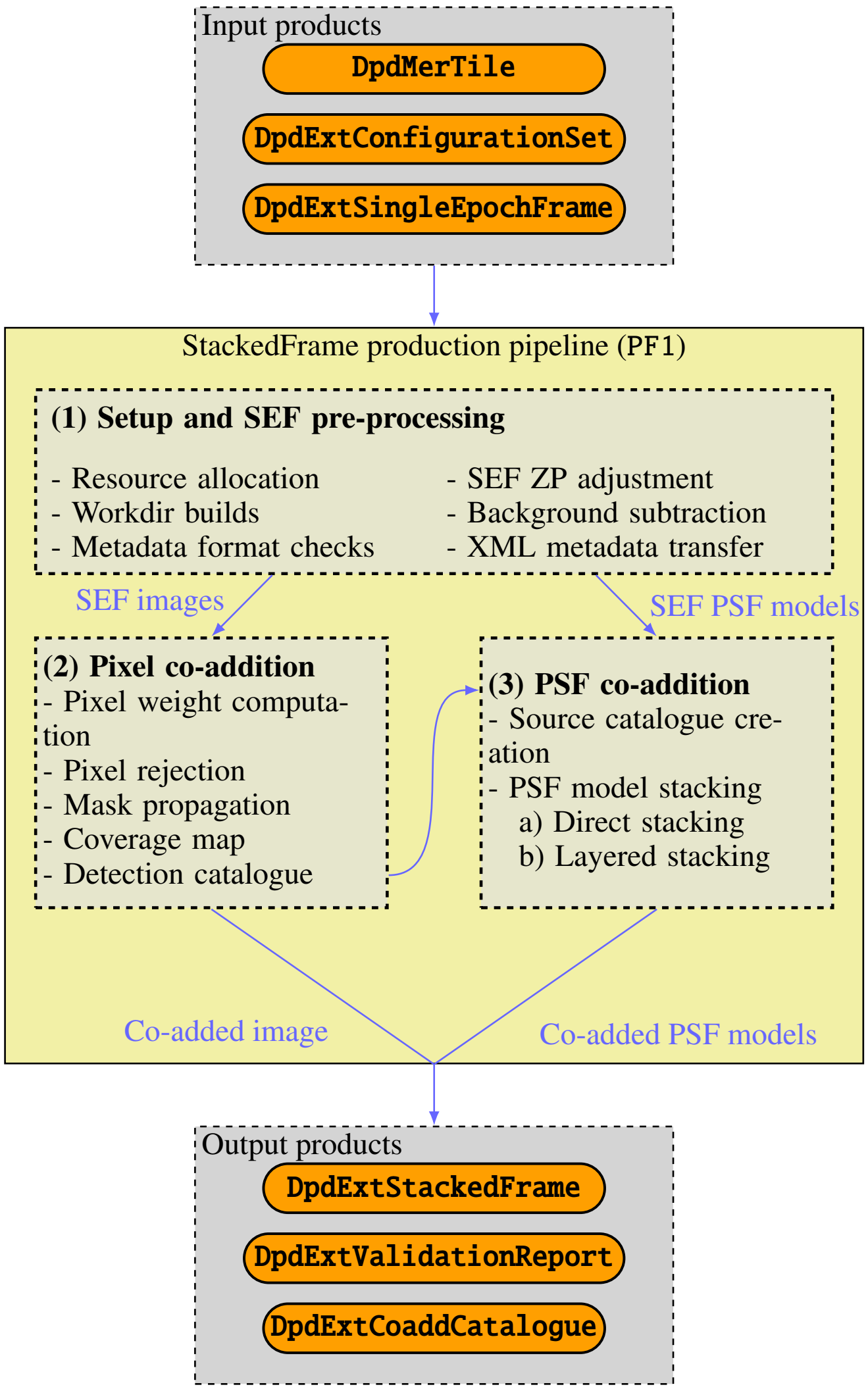
\begin{figure}
    \centering
    \resizebox{\linewidth}{!}{
    \definecolor{celement}{rgb}{0.83,0.83,0.83}
\definecolor{yellow1}{rgb}{0.95,0.94,0.65}
\definecolor{cstage2sect}{rgb}{0.95,0.93,0.4}

\begin{tikzpicture}[
    font=\Large, % <-- usually needed for single-column; change back if desired
    >=Latex,
    node distance=10mm and 10mm,
    process/.style={
        draw,
        dashed,
        line width=1.4pt,
        fill=olive!20,
        minimum height=44mm,
        text width=55mm,
        align=left
    },
    groupbox/.style={
        draw,
        dashed,
        thick,
        fill=celement,
        inner sep=6mm,
    },
    pfbox/.style={
        draw,
        thick,
        fill=yellow1,
        inner sep=6mm
    },
    flow/.style={
        ->,
        thick,
        draw=blue!60
    },
    prod/.style={
        draw,
        rounded rectangle,
        very thick,
        fill=orange!75!yellow,
        minimum height=8mm,
        minimum width=58mm,
        align=center,
        font=\Large\ttfamily\bfseries
    },
]

% -------------------------
% INPUT PRODUCTS (TOP)
% -------------------------
\node[groupbox,
      anchor=north,
      minimum width=78mm,
      minimum height=42mm,
      label={[anchor=north west]north west:Input products}
] (inputbox) {};

\node[prod, anchor=north] (tile)  at ([yshift=-6mm]inputbox.north) {\textbf{DpdMerTile}};
\node[prod, below=3mm of tile]   (config) {\textbf{DpdExtConfigurationSet}};
\node[prod, below=3mm of config] (single) {\textbf{DpdExtSingleEpochFrame}};

% -------------------------
% PF1 GROUP (MIDDLE)
% -------------------------
\node[pfbox,
      anchor=north,
      minimum width=140mm,
      minimum height=120mm,
      label={[anchor=north]north :StackedFrame production pipeline (\texttt{PF1})}
] (pfgroup) [below=10mm of inputbox.south] {};

% Step 1 (top, centered in PF1)
\node[process, anchor=north, minimum width=130mm, minimum height=30mm, text width=125mm, align=left] (step1)
    at ([yshift=-10mm]pfgroup.north)
{
\textbf{(1) Setup and SEF pre-processing}
\begin{multicols}{2}
- Resource allocation\\
- Workdir builds \\
- Metadata format checks\\
- SEF ZP adjustment\\
- Background subtraction\\
- XML metadata transfer\\
\end{multicols}
};

% Steps 2 & 3 (same vertical height, below step 1)
\node[process, anchor=north west, align=left] (step2)
    at ([xshift=-0mm, yshift=-10mm]step1.south west)
{
\textbf{(2) Pixel co-addition}\\
- Pixel weight computation\\
- Pixel rejection\\
- Mask propagation\\
- Coverage map\\
- Detection catalogue
};

\node[process, anchor=north east, align=left] (step3)
    at ([xshift=0, yshift=-10mm]step1.south east)
{
\textbf{(3) PSF co-addition}\\
- Source catalogue creation\\
- PSF model stacking\\
\hspace{5mm}a) Direct stacking\\
\hspace{5mm}b) Layered stacking
};

% -------------------------
% OUTPUT PRODUCTS (BOTTOM)
% -------------------------
\node[groupbox,
      anchor=north,
      minimum width=78mm,
      minimum height=42mm,
      label={[anchor=north west]north west:Output products}
] (outputbox) [below=10mm of pfgroup.south] {};

\node[prod, anchor=north] (out1) at ([yshift=-6mm]outputbox.north) {\textbf{DpdExtStackedFrame}};
\node[prod, below=3mm of out1]   (out2) {\textbf{DpdExtValidationReport}};
\node[prod, below=3mm of out2]   (out3) {\textbf{DpdExtCoaddCatalogue}};

% -------------------------
% FLOWS (VERTICAL)
% -------------------------
\draw[flow] (step1.210) -- (step2.north);
\node[anchor=south, text=blue!60] at ([xshift=-10mm, yshift=2mm]step2.north) {SEF images};

\draw[flow] (step1.330) -- (step3.north);
\node[anchor=south, text=blue!60] at ([xshift=16mm, yshift=2mm]step3.north) {SEF PSF models};
\draw[flow] (step2.330)  to[out=0, in=180] (step3.155);

\draw[thick, draw=blue!60] (step2.south) -- (pfgroup.south);
\node[anchor=east, text=blue!60] at ([xshift=-16mm, yshift=5mm]pfgroup.south) {Co-added image};
\draw[thick, draw=blue!60] (step3.south) -- (pfgroup.south);
\node[anchor=west, text=blue!60] at ([xshift=16mm, yshift=5mm]pfgroup.south) {Co-added PSF models};

\draw[flow] (inputbox.south) -- (pfgroup.north);
\draw[flow] (pfgroup.south)  -- (outputbox.north);

\end{tikzpicture}
}
    \caption{
    Workflow of the EXT StackedFrame production processing function (\pf{1}).
    Input products are shown on the top, the three main processing steps in the centre,
    and the resulting output products at the bottom.
    }
    \label{fig:pf1-workflow}
\end{figure}

In this section we describe the \change{Stage~2} \ext PF responsible for the creation of \stack products (\pf{1}).
\change{The pipeline} performs two principal tasks: (1) the co-addition of calibrated \sef images, and (2) the co-addition of their PSF models.
In the following sections, we provide an overview of the PF design and requirements (Sect.~\ref{s:pipeline-overview}), as well as a detailed description of the internal steps for the creation of StackedFrames: pipeline setup and pre-processing (Sect.~\ref{s:pf_step1}), pixel co-addition (Sect.~\ref{s:pf_step2}), and PSF model stacking (Sect.~\ref{s:pf_step3}).

\subsection{\label{s:pipeline-overview}Pipeline design and workflow}

\pf{1} provides a unified framework for processing data from multiple instruments while delivering homogeneous stacked products.
Instrument-specific requirements are accommodated through configurable processing parameters, including background subtraction and pixel-masking strategies.

Figure~\ref{fig:pf1-workflow} provides a schematic view of the \pf{1} workflow.
The pipeline uses the following input products:
\begin{itemize}
\item a \texttt{DpdMerTile} object defining the target stacked-frame geometry,
\item a set of configuration files specifying instrument-specific processing options,
\item a collection of calibrated \sef\ exposures whose spatial footprints intersect the tile area.
\end{itemize}

\subsection{Step 1: pipeline set-up and configuration}
\label{s:pf_step1}

During initialisation, \pf{1} validates and prepares the input SEFs, applies the photometric zero-point adjustment described in Sect.~\ref{s:photomAdjustment}, and, where applicable, performs local background subtraction with \swarp \citep{Bertin+02}.\footnote{Background subtraction is applied only to MegaCam and DECam datasets, using \texttt{BACK\_SIZE}=128 and 256 pixels, respectively, and \texttt{BACK\_FILTERSIZE}=3.}
The pipeline then constructs the SEF pixel weights from the background-only \texttt{RMS} or \texttt{WEIGHT} extensions and incorporates the relevant pixel flags (Appendix~\ref{appendix:sef_flagmaps}).

\subsubsection{SEF zero-point photometric adjustment}
\label{s:photomAdjustment}

\begin{table*}[htp!]
	\centering
	\caption{OU-EXT Stage 2 SEF zero-point adjustment.}
	\label{tab:sef_zp_adjustment}
	\begin{tabular}{crrrrr}
		\hline\hline\\[-8pt]
		Band & $\Delta\rm ZP$ (mmag)& $\sigma\left(\Delta\rm ZP\right)$ & NMAD (mmag) & Ref. catalogue & Ref. field\\[2pt]
		\hline\\[-8pt]
        DECam-$g$ & --0.2 & 0.2 & 2.4 & \gaia-XP synthetic photometry & EDF-S\\
        DECam-$r$ & --2.1 & 0.2 & 1.8 & \gaia-XP synthetic photometry & EDF-S\\
        DECam-$i$ & --3.4 & 0.2 & 2.1 & \gaia-XP synthetic photometry & EDF-S\\
        DECam-$z$ & --2.2 & 0.4 & 4.0 & \gaia-XP synthetic photometry & EDF-S\\
        HSC-$g$ & --0.1 & 0.7 & 6.8 & \gaia-XP synthetic photometry & EDF-N\\
        HSC-$r$ & 31.1 & 0.9 & 7.4 & SEF composite photometry & EDMS\\
        HSC-$r2$ & 25.5 & 0.5 & 4.6 & SEF composite photometry & EDMS\\
        HSC-$i$ & --35.3 & 0.5 & 1.4 & SEF composite photometry & COSMOS\\
        HSC-$i2$ & 17.0 & 0.2 & 1.6 & SEF composite photometry & EDMS\\
        HSC-$z$ & 5.8 & 0.7 & 5.2 & \gaia-XP synthetic photometry & EDF-N\\
        MegaCam-$u$ & 0.0 & 0.0 & 0.0 & -- & --\\
        MegaCam-$r$ & 2.3 & 0.4 & 4.3 & \gaia-XP synthetic photometry & EDF-N\\
        Pan-STARRS-$i$ & --2.0 & 0.5 & 3.9 & \gaia-XP synthetic photometry & EDF-N\\
        \hline
	\end{tabular}
\end{table*}

\change{The SEF zero-point corrections applied by \pf{1} consist of an \textit{a posteriori} per-filter rescaling designed to minimise residual photometric differences between SEF and co-added images. These differences can originate either during Stage 1 SEF recalibration, for example from variations in the reference aperture adopted for total-flux estimation in the PSF models, or from small systematic effects introduced during the Stage 2 pixel and PSF co-addition used to construct the StackedFrames.
This second-level correction therefore does not redefine the global photometric system; instead, it brings the catalogue-level SEF calibration and the pixel-level StackedFrame photometry onto a consistent internal scale.}

The \change{zero-point (ZP)} offsets, $\Delta\rm ZP$, were computed independently for each band (DECam, HSC, MegaCam, PS).
Their estimation is based on a comparison between StackedFrame photometry, measured by \sepp, from a \change{previous} test run spanning multiple fields (EDF-N/S/F, EDMS, COSMOS), against two reference catalogues: \gaia-XP synthetic photometry and SEF composite photometry (see Appendix~\ref{appendix:ref_catalogues} for more details about reference catalogues).
For each band, we used unsaturated stellar sources to determine the median photometric residual per StackedFrame.
The magnitude range depends on the reference catalogue used: $[16,\,17.6]$ for \gaia synthetic photometry, and $[18,\,21]$ for SEF composite photometry.
From the resulting distribution of $\Delta\rm ZP_{\rm stack}$ after combining all StackedFrames of a given band, we used the global median as estimator of $\Delta\rm ZP$.
We also used the normalised median absolute deviation, $\rm NMAD\equiv1.4826\,MAD$, and the median uncertainty, estimated as $\sigma\left(\Delta\rm ZP\right)\approx 1.25~\rm \frac{NMAD}{\sqrt{N}}$, with $N$ denoting the number of tiles, as proxies of the stability of $\Delta\rm ZP$.
Table~\ref{tab:sef_zp_adjustment} provides the summary of the statistics of $\Delta\rm ZP$ obtained for all bands included in DR1, together with the corresponding reference catalogues and calibration fields.
In the EWS, the ZP corrections are always below the 1\% relative flux level, whereas in the EDFs/EAFs the corrections range between 2--3\%.
\pf{1} uses these reference $\Delta\rm ZP$ values to correct the original SEF zero-point prior co-addition starts, ensuring consistency between catalogue- and pixel-level photometric definitions.

\subsection{Step 2: pixel co-addition}
\label{s:pf_step2}

For each SEF, we compute a flux scaling factor such that the resulting zero-point of the StackedFrame is 30 mag (i.e. 3.631 nJy/ADU).
\change{This reference zero-point is adopted as an internal convention and keeps the multiplicative flux scaling of the input SEFs close to unity for the majority of the datasets.}
This information is passed to \swarp, which first resamples the SEF images onto the MER tile grid using a Lanczos-3 kernel, \change{preserving the original instrument-specific pixel scale,} and combines them using a weighted mean (\texttt{COMBINE\_TYPE=WEIGHTED}) with per-pixel weights derived as described in the previous section.
A similar procedure is performed to propagate the weights and RMS extensions.
In addition, a selection of SEF bitmasks is resampled and merged into the final flag map through a logical OR operation (see Appendix~\ref{appendix:sef_flagmaps}).
The output StackedFrame science image comprises four extensions: flux, weights, RMS, and flag bitmask.
This step also generates a SEF coverage map, storing the number of contributing SEFs per pixel into the final StackedFrame.

Finally, a first catalogue of detected sources is extracted from the co-added image using \sextractor \citep{Bertin+96}.
When multiple photometric bands are stacked simultaneously, one or multiple bands are designated as the detection image, while the others are used for measurement.
For DECam data, we use $i$ and $z$ as detection bands, whereas for MegaCam and HSC we use $u/r$ and $g/z$, respectively.

\subsection{Step 3: PSF model stacking}
\label{s:pf_step3}

The detection catalogue built during the previous step is used by \psfex \citep{Bertin+11} to generate PSF model stamps from the StackedFrame image.
This information is used to build a final source catalogue, created after another \sextractor call, including PSF magnitudes based on the PSF models previously generated.
The StackedFrame PSF magnitudes will be used as baseline for re-scaling the SEF PSF models during co-addition.

The stacking of PSF models can proceed via two complementary strategies:
\begin{enumerate}
    \item Source-by-source stacking: a PSF model grid is created by stacking all SEF PSF models at the corresponding location of each detected source.
    \item Ensemble stacking: for very dense fields where a significant fraction of sources overlap, the PF organises the PSF models of target SEFs in multiple layers.
    Each layer keeps the original dimensions of the SEF image, and contains the maximum possible number of non-overlapping stamps, located at the same positions as the original sources.
    These layers are co-added independently, in the exact same fashion as the SEF science images, and the resulting stacked PSF stamps are subsequently combined into the final grid.
\end{enumerate}

Both methods only differ in the number of PSF models that are stacked per \swarp call, yielding identical stacked PSF models.
The source-by-source approach is used to produce DECam stacked PSF models, while the layered-approach was employed to build the PSF models of HSC and MegaCam data.
Further details about the PSF model co-addition algorithm are described in Euclid Collaboration: Vassallo et al. (in prep.).
After the three steps mentioned above are completed, the resulting \stack product is used by the MER pipeline in combination with the \Euclid data \citep{DR1-TP009}.

\section{\label{s:qa} EXT StackedFrame quality assessment and validation}

\begin{figure}
    \centering
    \resizebox{\linewidth}{!}{
    \definecolor{celement}{rgb}{0.83,0.83,0.83}
\definecolor{yellow1}{rgb}{0.95,0.94,0.65}
\definecolor{cstage2sect}{rgb}{0.95,0.93,0.4}

\begin{tikzpicture}[
    font=\Large,
    >=Latex,
    node distance=12mm,
    data/.style={
        draw,
        thick,
        fill=orange!55!yellow,
        minimum height=8mm,
        minimum width=35mm,
        align=center
    },
    process/.style={
        draw,
        dashed,
        line width=2pt,
        fill=olive!20,
        align=left,
        text width=110mm
    },
    groupbox/.style={
        draw,
        dashed,
        thick,
        fill=celement,
        inner sep=10mm,
    },
    pfbox/.style={
        draw,
        thick,
        fill=yellow1,
        inner sep=8mm
    },
    flow/.style={
        ->,
        thick,
        draw=blue!60
    },
    prod/.style={
        draw,
        rounded rectangle,
        very thick,
        fill=orange!75!yellow,
        minimum height=12mm,
        minimum width=80mm,
        align=center,
        font=\Large\ttfamily\bfseries
    },
]

% =====================================================
% INPUT
% =====================================================

\node[groupbox, minimum width=100mm, minimum height=30mm,
      label={[anchor=north west]north west:Input products}] (inputbox) {};

\node[prod] (tile) {DpdExtStackedFrame};

\node[above=5mm of inputbox] {OU-EXT Stage 2 data validation};

% =====================================================
% PF GROUP
% =====================================================

\node[pfbox,
      below=5mm of inputbox,
      minimum width=130mm,
      minimum height=105mm,
      label={[anchor=north]north:StackedFrame validation pipeline (\texttt{PF2})}
] (pfgroup) {};

% ---- Steps stacked vertically inside PF ----

\node[process, anchor=north] (step1)
  at ([yshift=-10mm]pfgroup.north) {
\textbf{Step 1: Setup}\\
- Pre-processing\\
- Read metadata\\
- Parse filter/band\\
- Source Extraction (\sepp)
};

\node[process, below=5mm of step1] (step2) {
\textbf{Step 2: Validation}\\
- \texttt{GaiaAstromPhotomValidation}\\
- \texttt{FwhmValidation}\\
- \texttt{PsfValidation}\\
- \texttt{PixelStatsValidation}\\
% - (\texttt{EffWavelengthValidation})\\
% - (\texttt{ColorValidation})
};

\node[process, below=5mm of step2] (step3) {
\textbf{Step 3: Report}\\
- Merge analysis\\
- Build dashboards
};

% =====================================================
% OUTPUT
% =====================================================

\node[groupbox,
      below=5mm of pfgroup,
      minimum width=100mm,
      minimum height=30mm,
      label={[anchor=north west]north west:Output products}
] (outputbox) {};

\node[prod, anchor=north]
  at ([yshift=-10mm]outputbox.north)
  (dear) {DpdExtAnalysisResults};

% =====================================================
% EQUAL
% =====================================================
\node[pfbox,
      below=5mm of outputbox,
      minimum width=120mm,
      minimum height=60mm,
      label={[anchor=north]north:EXT Quality Assurance Tools (\texttt{EQUAL})}
] (equal) {};

\node[process, anchor=north] (globaldepth)
  at ([yshift=-10mm]equal.north) {
\textbf{Global depth} (Sect.~\ref{s:stack_depth})\\
\textbf{Global single-band photometric stability} (Sect.~\ref{s:stack_photomstab})\\
- Stack PSF vs. AUTO\\
- Stack vs. GRC\\
- Stack vs. SCC\\
- Stack vs. ERC\\
\textbf{Global colour stability} (Sect.~\ref{s:stack_colourstab})
};

\node[groupbox,
      below=5mm of equal,
      minimum width=100mm,
      minimum height=15mm,
      label={[anchor=center]center:\textbf{Data flagging and invalidation} (Sect.~\ref{s:data_validation})}
] (invalidation) {};

% =====================================================
% FLOWS
% =====================================================

\draw[flow] (inputbox.south) -- (pfgroup.north);
\draw[flow] (step1.south) -- (step2.north);
\draw[flow] (step2.south) -- (step3.north);
\draw[flow] (pfgroup.south) -- (outputbox.north);
\draw[flow] (outputbox.south) -- (equal.north);

\draw[flow] (equal.south) -- (invalidation.north);

\end{tikzpicture}
}
    \caption{Overview of \ext data validation workflow.}
    \label{fig:pf2-workflow}
\end{figure}
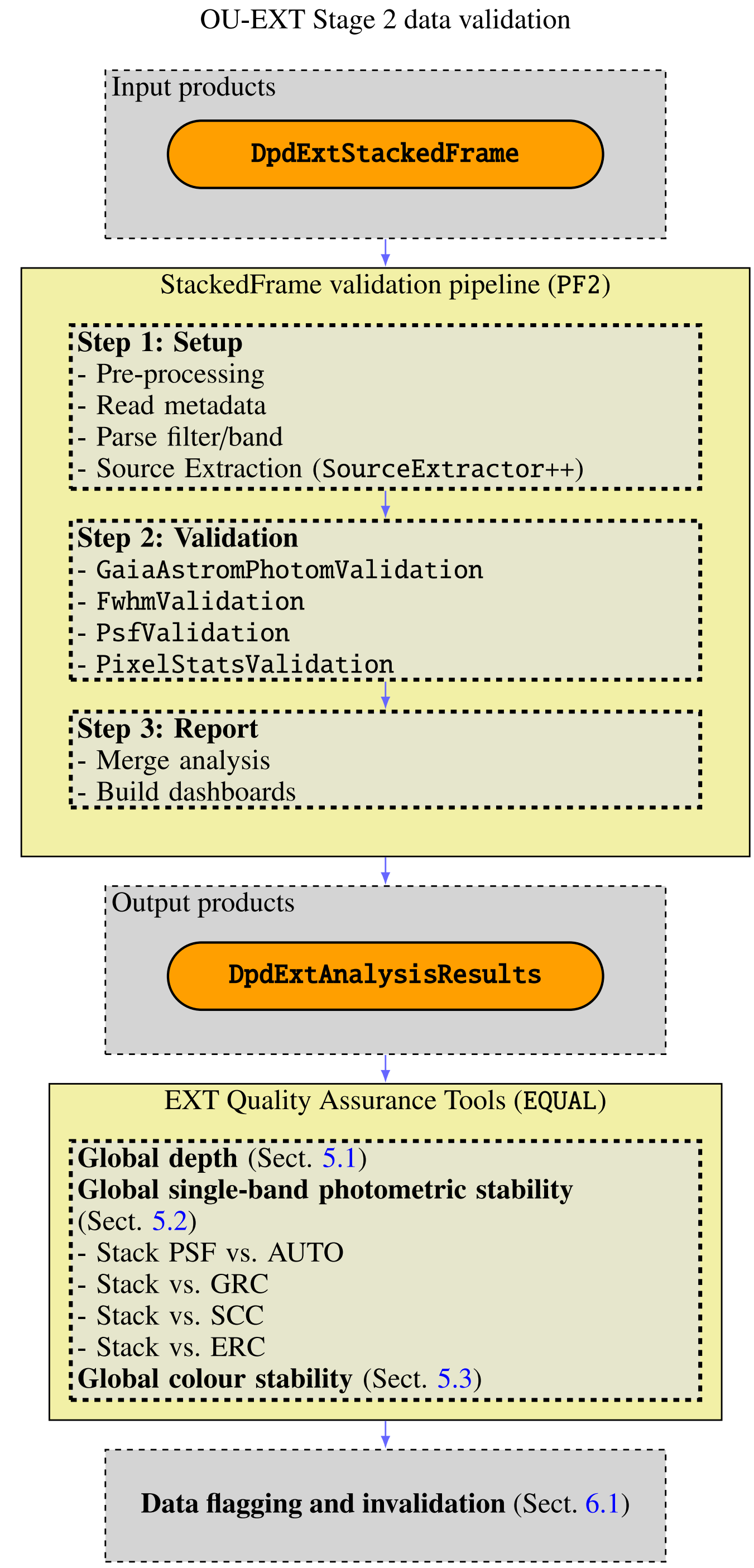

The OU-EXT \change{Stage 2} data validation process is described in Fig.~\ref{fig:pf2-workflow}.
Every \stack product produced by \pf{1} is used as input by the post-validation processing function (\pf{2}), that performs the assessment of the achieved pixel and PSF model quality.

The procedure starts by generating a source catalogue from each StackedFrame using \sepp \citep{Bertin+20}. 
The associated PSF models are used to measure PSF-based point-source photometry, together with circular-aperture photometry (diameters from $1\arcsec$ to $10\arcsec$) and \change{Kron} \citep{Kron80} photometry.
Validation is then performed through a set of independent checks, primarily using stellar sources:
\begin{enumerate}
    \item Astrometry and photometric validation with respect to \gaia.% based on BP-RP photometry and colour transformations.
    \item PSF model quality by comparing the full-width half maximum (FWHM) of 2D Gaussian fits to PSF models and source cutout images.
    \item PSF model quality by comparing the brightness profile of stellar sources and their associated PSF models.
    \item Pixel analysis in terms of the background RMS noise and depth.
\end{enumerate}
The resulting measurements are combined into a \dear\ product and stored in DPS/DSS, together with the relevant quality-control parameters, validation products, and the \sepp\ source catalogue.

At survey level, the results produced by \pf{2} for individual StackedFrames are further analysed with the EXT Quality Assurance Tools (\equal) library. \equal\ uses the Metadata Access Layer (MAL) to access the \Euclid DPS/DSS database services \citep{Williams+19,DR1-TP033} and derive the survey-level diagnostics used to define the validation quality flags.

\subsection{\label{s:stack_depth}StackedFrame depth quality}

For each StackedFrame, the photometric depth is characterised in terms of the $10\sigma$ (background only, without accounting for correlated noise) point-source limiting magnitude measured in $2\arcsec$-diameter apertures.
The EWS depth requirements for the EXT bands are 24.7, 24.4, 23.8, and 23.1 mag for DECam $griz$ bands, 23.9 mag for MegaCam $r$, 24.5 and 23.4 mag for HSC $g$ and $z$, and 23.6 mag for Pan-STARRS $i$.

\subsubsection{EWS photometric depth}

\begin{figure*}[hpt!]
    \includegraphics[width=\linewidth]{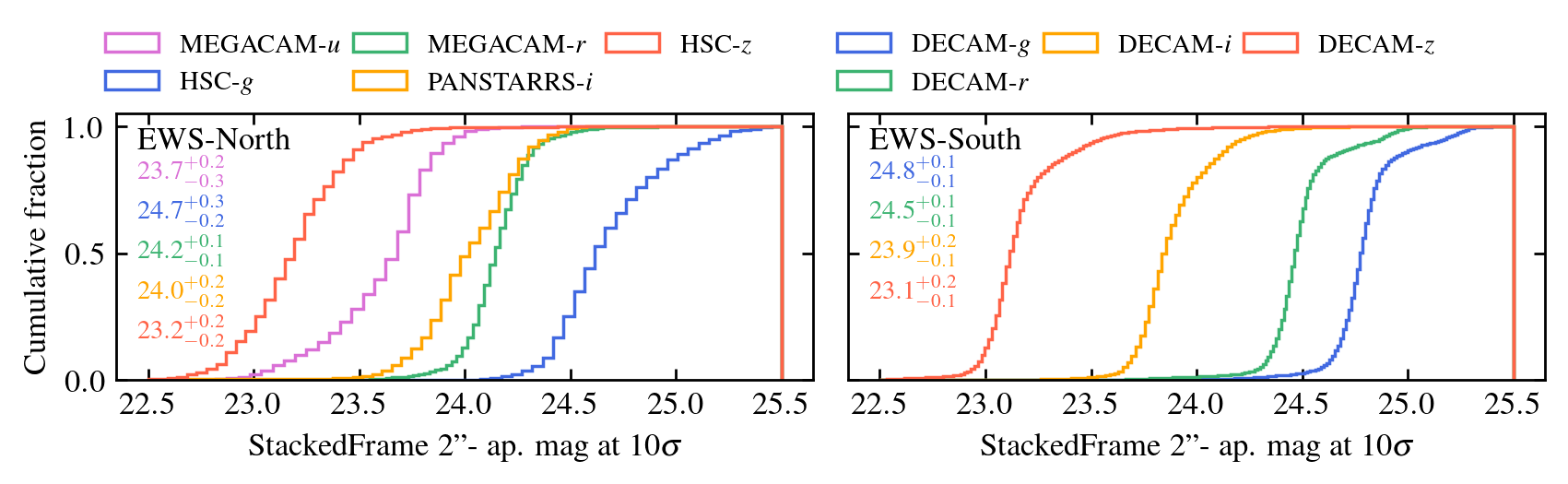}
    \caption{DR1 EXT EWS StackedFrame photometric depth distribution. \emph{Left:} Cumulative distribution of StackedFrames as function of achieved depth in the northern hemisphere of the DR1 EWS footprint. \emph{Right:} Same for the southern component of the DR1 EWS. Median depth values and 68\% intervals are included for each band following the same colour scheme.}
    \label{fig:dr1_wide_depth}
\end{figure*}

Figure~\ref{fig:dr1_wide_depth} illustrates the achieved photometric depth per StackedFrame across the EWS footprint.
The left and right panels show the cumulative distribution of the achieved depth per StackedFrame in the areas covered by UNIONS (north) and DECam (south), respectively.

In the northern hemisphere, the fraction of co-added images exceeding the end-of-mission requirements is 78\%, 96\%, 98\%, and 20\% for the $griz$ bands (see legend).
The depth achieved across both hemispheres is broadly consistent.
The southern component is slightly deeper in the $g$ and $r$ bands, whereas the northern component shows marginally greater depth in $i$ and $z$.
This behaviour reflects the different survey origins: the EWS-South footprint is based on a single-instrument dataset (DECam), leading to a relatively uniform exposure strategy, while EWS-North combines multiple instruments and programmes with heterogeneous observing conditions.
In terms of depth homogeneity, both hemispheres exhibit a tile-to-tile scatter around the median value of each band (shown in each panel of Fig.~\ref{fig:dr1_wide_depth}) in the range 0.1--0.3 mag.

Note that there are no explicit mission requirements for the achieved depth in the $u$ band.
In Appendix~\ref{appendix:depth_sky} we provide a spatial view of the photometric depth distribution across the DR1 footprint.

\subsubsection{EDFs and EAFs photometric depth}

\begin{figure*}[hpt!]
    \includegraphics[width=\linewidth]{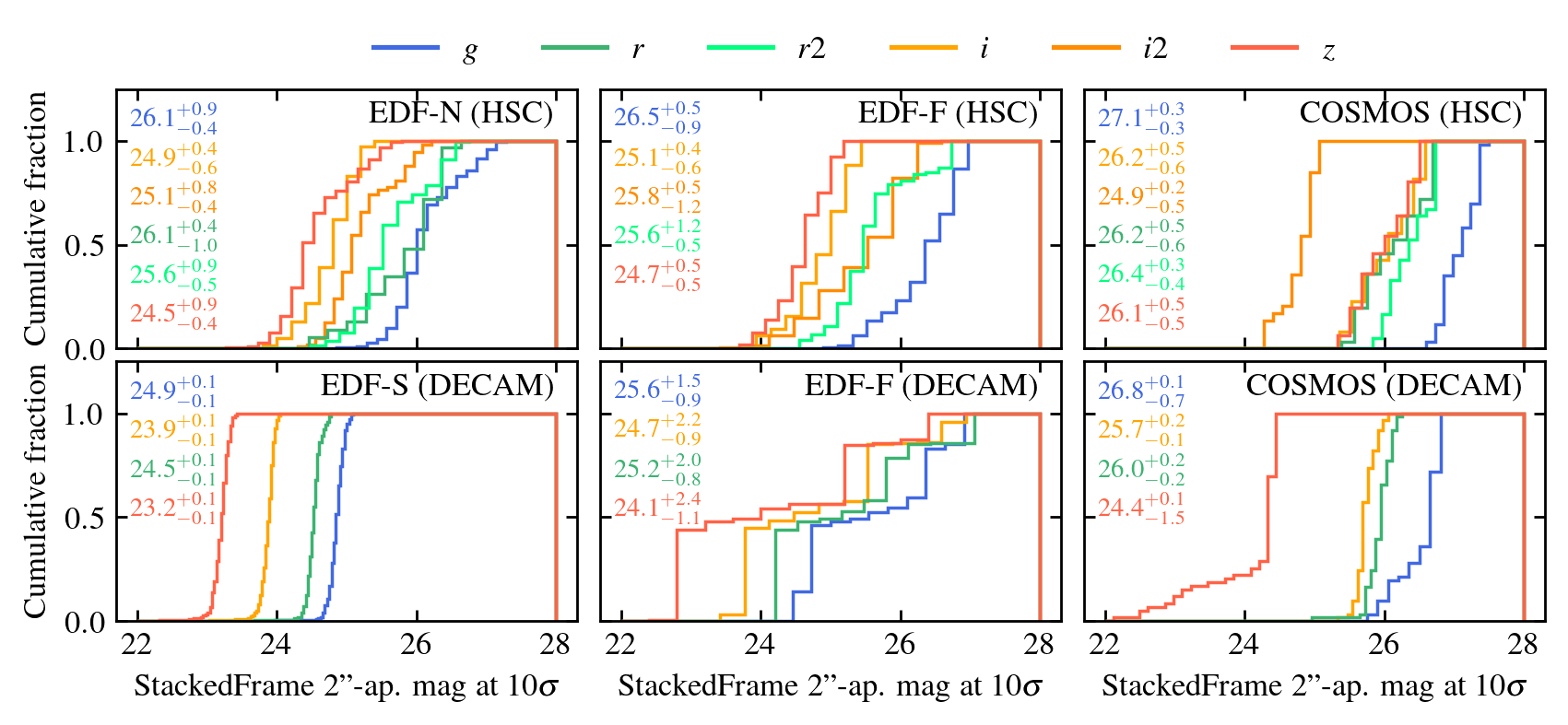}
    \caption{Same as Fig.~\ref{fig:dr1_wide_depth}, but now showing the EDFs and COSMOS.}
    \label{fig:dr1_deep_depth}
\end{figure*}

Figure~\ref{fig:dr1_deep_depth} illustrates the achieved photometric depth per StackedFrame in the EDFs and EAFs.
Each panel corresponds to a different field and instrument (HSC or DECam).

The achieved depth varies significantly from field to field due to the heterogeneous combination of observations, typically exceeding the EWS depth by 1 to 2 magnitudes (except in the case of EDF-S, which remains the same as the EWS depth).
Within individual fields, the depth distribution exhibits a broader range than in the EWS, in some cases exceeding 1 magnitude.
This behaviour reflects the complex layering of archival and dedicated observations, variable seeing conditions, and spatial variations in the number of contributing SEFs (several hundred per StackedFrame).

\subsection{\label{s:stack_photomstab}StackedFrame photometric stability}

\begin{figure*}[hpt!]
    \includegraphics[width=\linewidth]{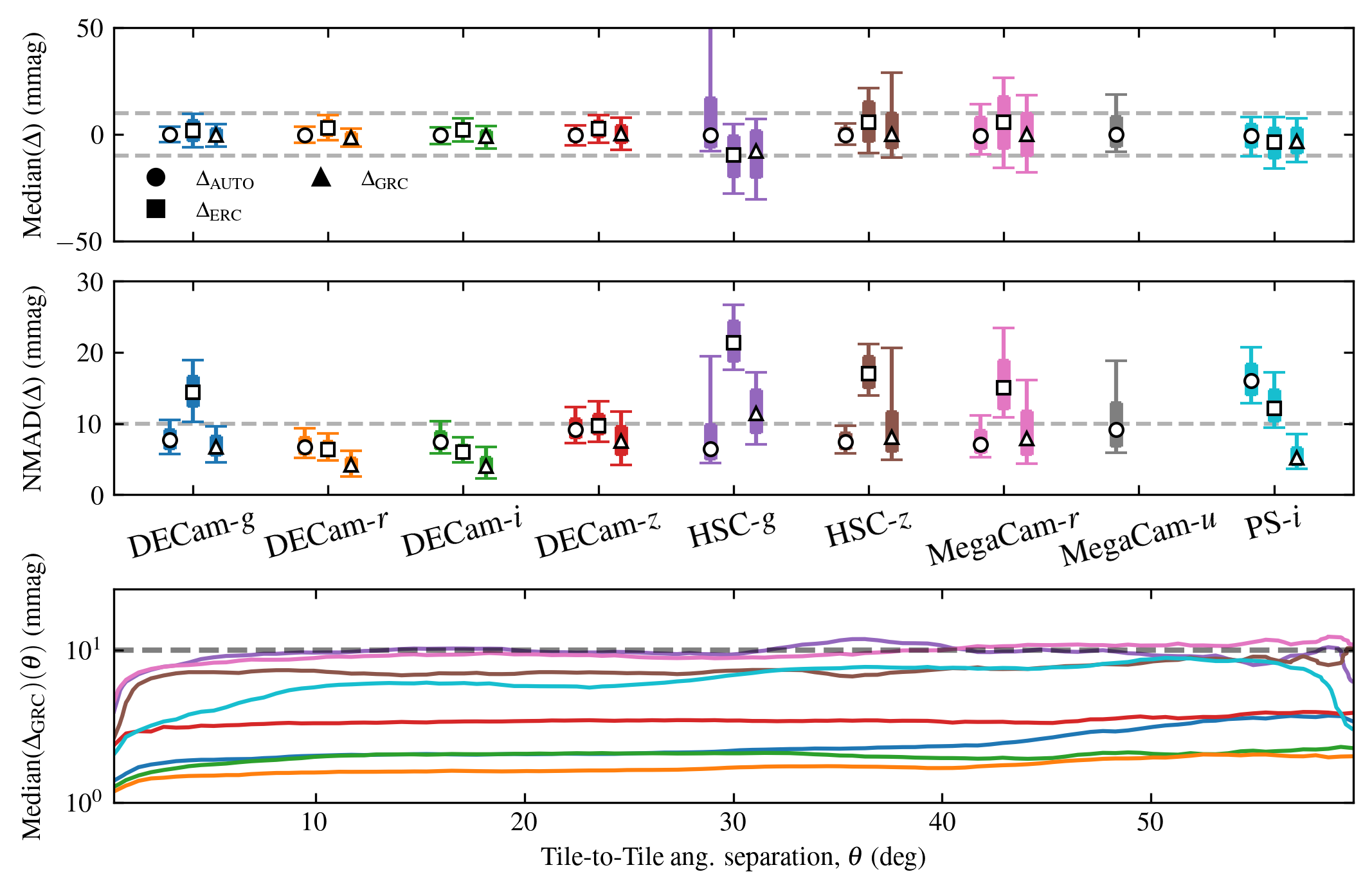}
    \caption{EXT EWS photometric stability summary.
    \emph{Top:} Box-plot showing the distribution of the StackedFrame median photometric offset, computed from the differences between PSF and Kron photometry, PSF photometry and the GRC, and PSF photometry and the ERC (see legend). Symbols denote the median offset per band, thick and thin error bars indicate the 68\% and 90\% intervals of the distribution, respectively.
    \emph{Middle:} Box-plot showing the distribution of the StackedFrame internal photometric scatter, quantified by the NMAD of the magnitude offsets.
    \emph{Bottom:} Two-tile median photometric offset differences as function of angular separation, derived from PSF magnitudes and the \change{GRC}.}
    \label{fig:dr1_wide_photom}
\end{figure*}   

The \Euclid mission requires the relative photometric accuracy for all galaxies to be better than 1.5\% after full data processing (i.e. including EXT and MER production steps) over the whole survey footprint \citep{Laureijs11}.
This stringent requirement is driven by the need for highly accurate source SEDs to enable OU-PHZ to derive reliable photo-$z$ estimates \citep{Desprez-EP10}.
At OU-EXT level, we adopt a goal of 10 mmag (i.e. $\simeq 1\%$ relative flux accuracy) based on stellar photometry across the EWS and EDFs/EAFs.

Therefore, in order to assess the photometric calibration stability of the DR1 stacks, we performed a suite of tests based on the StackedFrame photometry.
These tests are carried out on a MER-tile basis ($\simeq0.25$~deg$^2$) and independently for each filter.
Three different photometric reference catalogues are used to evaluate the calibration of the stacks.

The first reference is the catalogue produced by the Stage~1 team, based on synthetic photometry derived from \gaia XP spectra (hereafter GRC).
This catalogue constitutes the primary reference of EXT data.
It is used for the zero-point recalibration of the SEFs, and therefore allows us to verify whether additional systematics are introduced during the stacking process.
Its main limitation is the relatively bright magnitude range of the available stars, typically $m \lesssim 17.7$, which makes the comparison more sensitive to saturation effects.

The second reference catalogue, also produced during the Stage~1 recalibration phase, is constructed by combining the individual SEF catalogues into a SEF composite catalogue (hereafter SCC), where the photometry of each source is estimated as an inverse-variance weighted mean.
In this work, this catalogue is used to assess the photometric stability of the deep fields rather than the GRC in order to avoid saturation effects.

Finally, we constructed an additional reference catalogue based on external photometry (hereafter ERC), combining DES~DR2 \citep{Abbott+21} and UNIONS photometry (UNIONS team, priv.~comm.).
To build this catalogue, the original photometry from DES and UNIONS was corrected to match the GRC by means of colour-transformations derived from a Random Forest regression (see Appendix~\ref{appendix:ref_catalogues} for more details).
The main advantage of this catalogue is that it probes a fainter magnitude regime, $18 \lesssim m \lesssim 20.5$, and it is not tied to the internal EXT calibration procedure.

Two of the main sources of photometric errors come from either wrong ZP calibrations, resulting in an overall offset, or bad co-added PSF models that fail to properly describe the light profiles of the stacked images.
To examine both issues, we defined the following photometric quality metrics based on magnitude differences between the StackedFrame \sepp PSF-fitting photometry and several reference measurements:
\begin{itemize}
\item $\Delta_{\rm GRC} \equiv m_{\rm PSF} - m_{\rm GRC}$, computed for stars in the range $17 \leq m_{\rm PSF} \leq 17.7$;
\item $\Delta_{\rm SCC} \equiv m_{\rm PSF} - m_{\rm SCC}$, computed for $18 \leq m_{\rm PSF} \leq 20.5$;
\item $\Delta_{\rm ERC} \equiv m_{\rm PSF} - m_{\rm ERC}$, computed for $18 \leq m_{\rm PSF} \leq 20.5$;
\item $\Delta_{\rm AUTO} \equiv m_{\rm PSF} - m_{\rm AUTO}$, computed for $18 \leq m_{\rm PSF} \leq 20.5$, where $m_{\rm AUTO}$ denotes \sepp version of Kron photometry performed on the StackedFrames. This quantity is used as a proxy for the internal consistency between the PSF model and the observed point-source light distribution\footnote{Note that Kron flux depends on PSF wings and background/segmentation.}.
\end{itemize}

From these differences, we estimate the median offset and NMAD per StackedFrame, which we adopt as proxies for the absolute and relative ZP calibration accuracy, respectively.

\subsubsection{EWS photometric stability}

Figure~\ref{fig:dr1_wide_photom} summarizes the photometric calibration performance achieved for the DR1 EWS ground-based data.
The top panel shows, for each filter, the distribution of median offsets measured per tile.
Individual symbols indicate the global median offset obtained for each metric, while the error bars represent the 68\% and 90\% percentiles of the corresponding distributions.
For $\Delta_{\rm AUTO}$, the global median value computed for each filter has been subtracted, since only relative variations are of interest; this normalisation causes the symbols to appear centred at zero.
The middle panel displays the distribution of the StackedFrame internal NMAD scatter.

In terms of zero-point consistency, illustrated by the top panel of Fig.~\ref{fig:dr1_wide_photom}, the results derived from $\Delta_{\rm GRC}$ and $\Delta_{\rm ERC}$ are highly compatible.
For EWS-South, more than 95\% of the stacks satisfy the 10 mmag calibration requirement.
In contrast, the EWS-North results exhibit a larger degree of heterogeneity, with 55\%, 68\%, 85\%, and 78\% of the stacks in the $griz$ bands, respectively, satisfying the 10 mmag requirement.
For MegaCam-$r$ and the HSC-$g$ and $z$ bands in particular, extended tails reaching 20--30 mmag are observed in a subset of tiles.
Appendix~\ref{appendix:photom_stab_sky} provides a spatial view of the photometric offsets based on the comparison against the GRC.

The internal ZP stability, traced by the NMAD scatter of the StackedFrames photometry in the middle row of Fig.~\ref{fig:dr1_wide_photom}, also shows relatively good consistency between the various proxies.
On the one hand, the median NMAD($\Delta_{\rm GRC}$) value is below the requirements for almost all bands, except HSC-$g$ that shows $\rm NMAD(\Delta_{GRC})\simeq11.4$ mmag, and MegaCam-$u$ where there is no reference photometry available.
Note that the HSC data are more sensitive to saturation effects, that can lead to biased results when using bright samples such as the GRC.
On the other hand, the values of NMAD($\Delta_{\rm ERC}$) show systematically higher internal scatter.
This discrepancy can partially be attributed to additional systematic noise introduced in the ERC when transforming the external photometry into the GRC system.
Finally, NMAD($\Delta_{\rm AUTO}$) suggests that the internal calibration of the StackedFrame images and PSF models is typically within 1\% in almost all bands.
The only band whose median NMAD value appears beyond the 10 mmag requirement is Pan-STARRS-$i$, which is known to have extended PSF tails that challenge the Kron aperture to capture a stable fraction of the total flux.

Lastly, we assess the spatial coherence of the stack ZPs across the survey footprint.
This test is particularly relevant for PHZ, as large-scale spatial calibration variations may imprint artificial cosmic structure.
For each pair of tiles, we compute the difference in their absolute median ZP offsets and build the distribution of ZP offset as a function of angular separation.
The median of this distribution at fixed separation is shown in the bottom panel of Fig.~\ref{fig:dr1_wide_photom}.
For all bands, the measured variations remain within the 10 mmag requirement, indicating a statistically homogeneous photometric calibration across the EWS.
In the case of DECam data, the results stay below the 5 mmag level, well within the requirements.
The largest spatial variations are observed in the HSC-$g$ and MegaCam-$r$ bands, where the median differences approach the 10 mmag threshold.
As shown in Fig.~\ref{fig:dr1_ews_photom_stab_sky}, $\Delta_{\rm GRC}$ presents a noticeable structure across some of the UNIONS bands at the level of 10--30 mmag.
\change{Some of these problems arise from a combination of varying depth and seeing conditions between individual SEF exposures, which were not properly handled during the Stage~1 zero-point recalibration and \pf{1} stacking steps, and should not be interpreted as intrinsic photometric defects of the contributing survey data.}

\subsubsection{EDFs and EAFs photometric stability}

\begin{figure}[hpt!]
    \includegraphics[width=\linewidth]{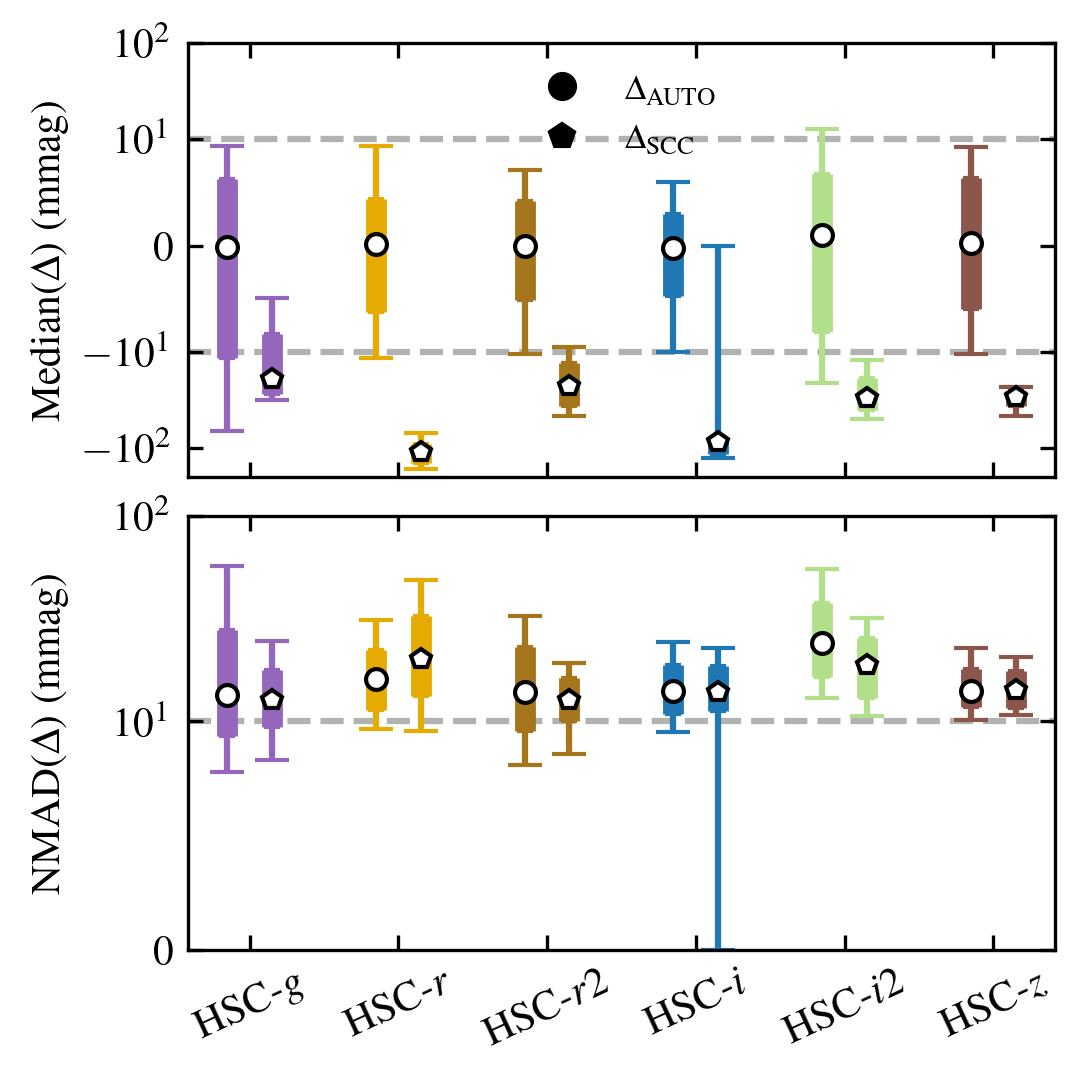}
    \caption{Per-band EXT EDF-N photometric quality summary.
    \emph{Top:} Distribution of the median photometric offset per MER tile, computed from the differences between PSF and Kron photometry, PSF photometry, and the SCC (see legend). Symbols denote the median offset per band, thick and thin error bars indicate the 68\% and 90\% intervals, respectively.
    \emph{Bottom:} Distribution of the intra-tile photometric scatter, quantified by the NMAD of the magnitude offsets per MER tile.}
    \label{fig:dr1_edfn_photom}
\end{figure}

\begin{figure}[hpt!]
    \includegraphics[width=\linewidth]{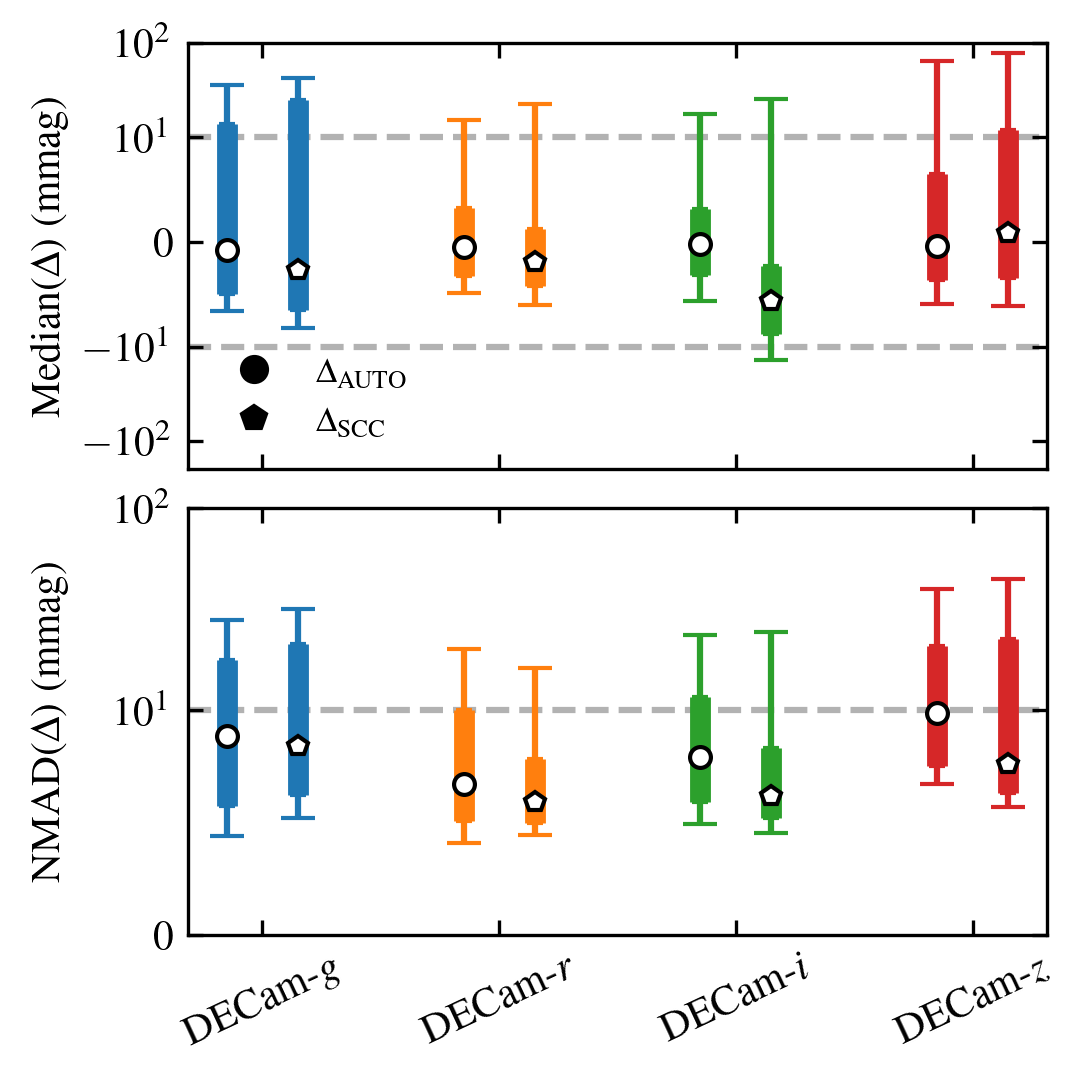}
    \caption{Same as Fig.~\ref{fig:dr1_edfn_photom}, but for EDF-S.}
    \label{fig:dr1_edfs_photom}
\end{figure}

\begin{figure}[hpt!]
    \includegraphics[width=\linewidth]{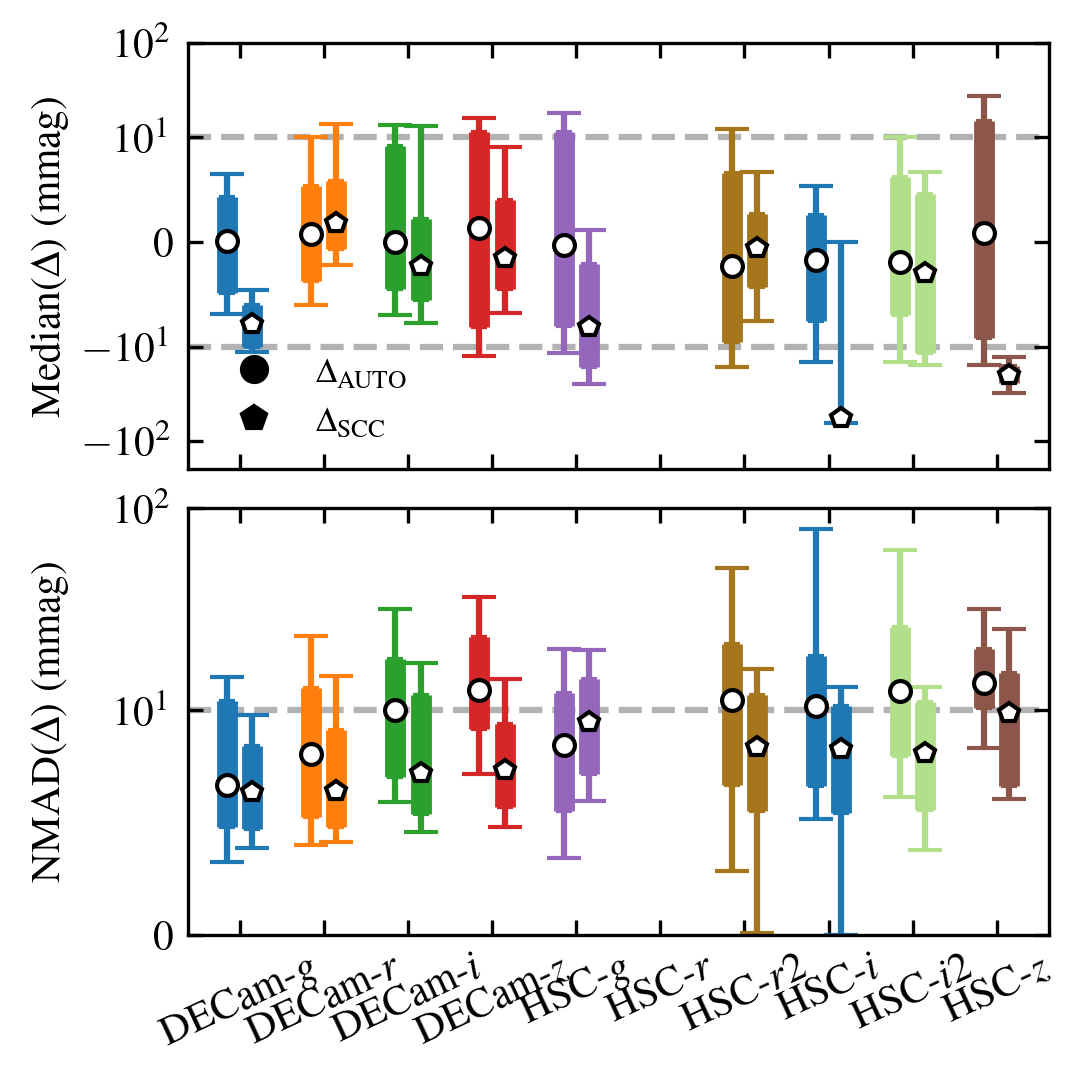}
    \caption{Same as Fig.~\ref{fig:dr1_edfn_photom}, but for EDF-F.}
    \label{fig:dr1_edff_photom}
\end{figure}

\begin{figure}[hpt!]
    \includegraphics[width=\linewidth]{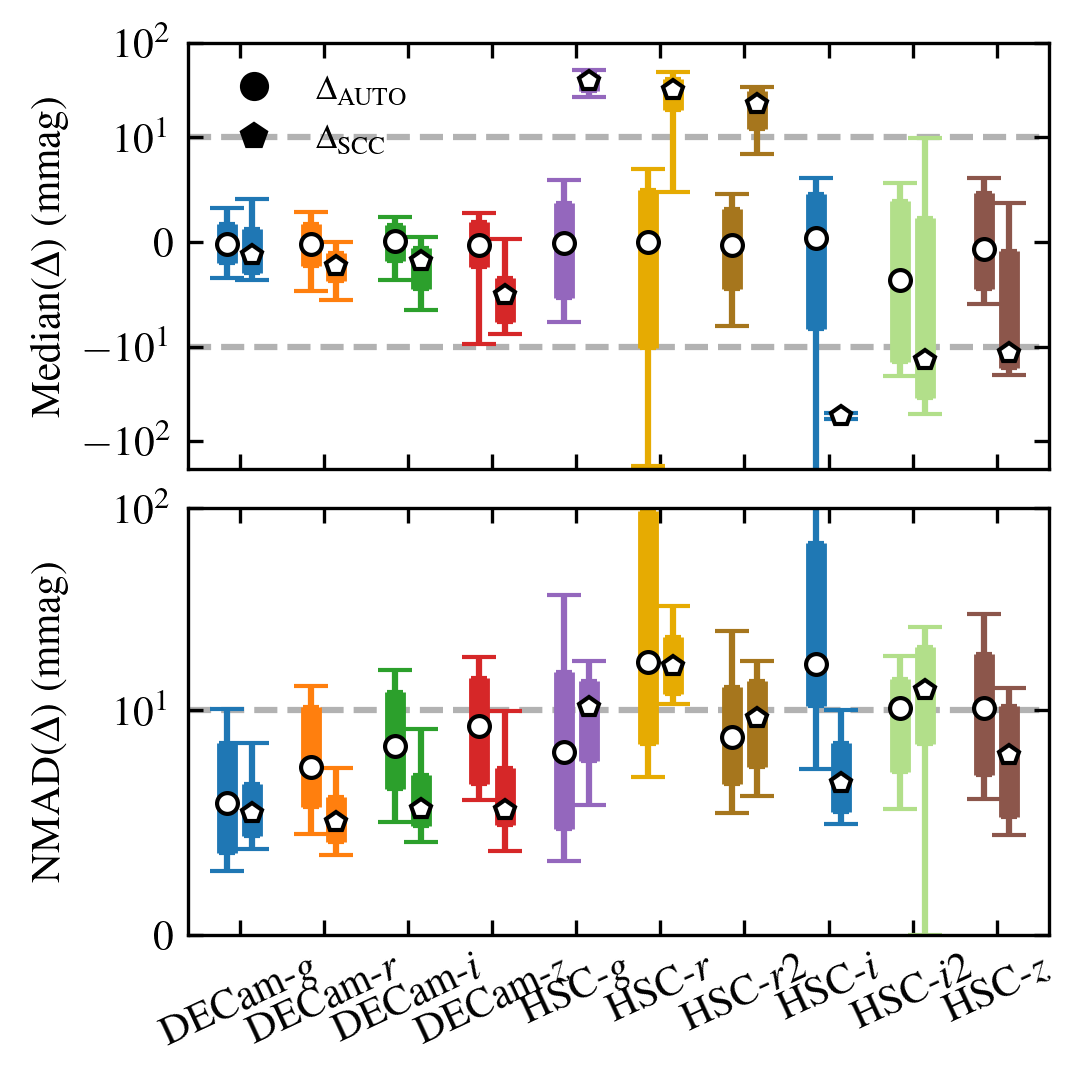}
    \caption{Same as Fig.~\ref{fig:dr1_edfn_photom}, but for COSMOS.}
    \label{fig:dr1_cosmos_photom}
\end{figure}

Similarly to the previous section, we performed the evaluation of the photometric stability of the deep StackedFrames by means of the $\Delta_{\rm AUTO}$ and $\Delta_{\rm SCC}$ proxies.
Given the reduced spatial extent of the fields compared to the EWS, here we focus on the global statistical results and not on spatial effects.
A detailed view of the spatial offset distribution (based on the SCC) can be found in Appendix~\ref{appendix:photom_stab_sky}.

Figures~\ref{fig:dr1_edfn_photom}~to~\ref{fig:dr1_cosmos_photom} show the photometric results from EDF-N, EDF-S, EDF-F, and the COSMOS auxiliary field, respectively.
Starting with EDF-N, we find a good internal consistency between PSF-based and Kron photometry, suggesting that the co-added PSF models -- resulting from combining a much higher number of SEFs than in the EWS -- are indeed a good description of the light distribution of stellar sources.
However, when comparing against the SCC we find significant systematic deviations well beyond the 10 mmag requirement.
This is indicative of a potential inconsistency between the absolute ZP definition at SEF and StackedFrame levels.
In terms of internal scatter (bottom panel of Fig.~\ref{fig:dr1_edfn_photom}), we also observe an increase with respect to the EWS.

Moving to EDF-S, the picture here is qualitatively different.
The majority of StackedFrames show values of $\Delta_{\rm AUTO}$ and $\Delta_{\rm SCC}$ within the requirements, although there are some tails extending beyond the 10 mmag limit.
The internal scatter in terms of NMAD($\Delta_{\rm SCC}$) suggests a good consistency between the zero-points, whereas the more elevated scatter in terms of NMAD($\Delta_{\rm AUTO}$) hints at challenges in terms of PSF model quality.
The outlier StackedFrames, visible in Fig.~\ref{fig:dr1_edfs_photom_stab_sky}, were identified as images with astrometric errors present in the input SEFs.

EDF-F is the field with the largest coverage in terms of number of bands.
The results shown in the top panel of Fig.~\ref{fig:dr1_edff_photom} show both good internal consistency in terms of calibration and PSF model quality in almost all bands.
In the case of HSC-$i$ and HSC-$z$, we observe strong deviations with respect to the SCC reference catalogue, suggesting problems in the ZP calibration.
Regarding the internal StackedFrame photometric scatter, both DECam and HSC observations yield similar results, reaching values below $\simeq$1\% when comparing to SCC.

Finally, the results from the COSMOS field are shown in Fig.~\ref{fig:dr1_cosmos_photom}.
In terms of differences between PSF and Kron magnitudes, we find that all bands perform well within the 10 mmag requirements.
When comparing the StackedFrame PSF photometry against the SEF PSF composite photometry, we find strong discrepancies in the HSC bands, hinting at a potential bias in the absolute ZP calibration of the StackedFrames.
These issues are currently being investigated and will be addressed in the next release.
The low internal scatter, however, reflects a good internal calibration and minimal issues arising from stacking the PSF models of a large number of input SEFs.
\change{We refer the reader to \citet{DR1-TP009} and \citet{DR1-TP011} for complementary photometric stability tests and photo-$z$ performance based on downstream products.}

\subsection{\label{s:stack_colourstab}StackedFrame colour-homogeneity}

To assess the photometric stability of the dataset, we first construct an
intrinsic, error-deconvolved model of the stellar locus in colour space and
subsequently evaluate localized deviations from this fiducial model across the
survey footprint. We begin by correcting the apparent magnitudes ($g, r, i, z$)
for interstellar extinction using \citet{Schlegel:1998} dust emission maps,
adopting the extinction coefficients derived by \citet{Cardelli:1989} and
corrections from \citet{Schlafly:2011}. The corrected magnitudes are used to construct
the de-reddened colour space ($g-r, r-i, i-z$) using a discrete transformation
matrix $\mathbf{W}$. To rigorously propagate the photometric uncertainties, the
corresponding error covariance matrices for each source are transformed via
$\mathbf{C}_{\mathrm{colour}} = \mathbf{W} \mathbf{C}_{\mathrm{mag}}
\mathbf{W}^{\rm T}$, where $\mathbf{C}_{\mathrm{mag}}$ is the covariance matrix, which in this case contains only the magnitude uncertainties (variance) in each band in its diagonal. We apply stringent quality control criteria, restricting the
sample to sources with robust measurements, low magnitude errors ($\sigma <
0.01$ mag), and bounded reduced chi-square statistics ($\chi^2 < 2$) as derived
in the StackedFrame catalogue, obtained using \sepp. This ensures the underlying
model is not unduly influenced by outliers or spurious detections. We note that
for the southern hemisphere, we adopt DECam $griz$ while for the north we adopt $g$
and $z$ from HSC, $r$ from MegaCam, and $i$ from Pan-STARRs. For simplicity, in
our figures and analysis we refer simply as
$griz$ both in north and south.

The intrinsic colour distribution of the surveys is modelled using the extreme
deconvolution (XD) algorithm. XD extends standard Gaussian mixture models (GMMs)
by explicitly accounting for the individual, heteroscedastic observational
uncertainties associated with each data point in the de-reddened colour space
described above. We fit an XDGMM utilising 26 components to a randomly selected,
representative sub-sample of the DR1 data. This sub-sample is selected from the entire DR1 North or South footprint and has the same quality control criteria as the sample described above. The number of components was
chosen using the Bayesian information criterion (BIC) such that we do not
over-fit the stellar locus. The XDGMM model yields a continuous, error-free
probability density function that serves as the global fiducial model of the
stellar locus. The robust nature of the XD algorithm ensures that the underlying
distribution is accurately disentangled from instrumental noise, capturing the
true astrophysical spread of the stellar colours, as well as possible outliers
in the distribution.

With the global fiducial XDGMM model established, we test the spatial stability
of the photometry by quantifying localized zero-point offsets. The survey
footprint is partitioned into a regular two-dimensional spatial grid in right
ascension and declination. For each spatial bin (indexed $i,j$) containing a
sufficient number of sources, $N_{*}(\alpha_i, \delta_j)$, we introduce a
colour offset vector, $\boldsymbol{\Delta} = \left(\Delta_{g-r}, \Delta_{r-i},
\Delta_{i-z}\right)$, as a free nuisance parameter. We determine the optimal offset
for each spatial bin by maximising the log-likelihood of the local data\change{, including their measurement uncertainties, }
evaluated against the global XDGMM model utilising the Nelder--Mead optimisation
algorithm. By mapping these derived colour offsets $\boldsymbol{\Delta}$ and
computing their local spatial derivatives, we can effectively map relative
photometric gradients, spatially varying systematics, and relative zero-point
variations across the observing footprint. To test how robust this method is, we
repeat the fit outlined above where we randomly pick a set of $N_{*}(\alpha_i,
\delta_j)$ stars with no spatial constraints. This effectively creates a dataset
bootstrapped from the original distribution used to build the XDGMM model -- for
each spatial bin -- thus by definition having the expected value of
$\boldsymbol{\Delta}_{\mathrm{expected}} = \boldsymbol{0}$. We compute the
standard deviation of the best-fit $\boldsymbol{\Delta}$ in this bootstrapping
experiment as the intrinsic error ($\sigma_{\rm intrinsic}$) in the method
outlined above.
In the south we find a $\sigma_{\rm intrinsic}$ of 5.0, 2.8, 2.5 mmag in
$g-r, r-i, i-z$ respectively.
While in the north $\sigma_{\rm intrinsic}$
was found to be 6.3, 3.2, 2.5 mmag in $g-r, r-i, i-z$. These numbers can be
interpreted as how well one can determine colour shifts in the underlying
dataset, using a typical number of stars given by $\langle N_{*}(\alpha_i, \delta_j)\rangle$.

 \begin{figure*}[hpt!]
     \includegraphics[width=\linewidth]{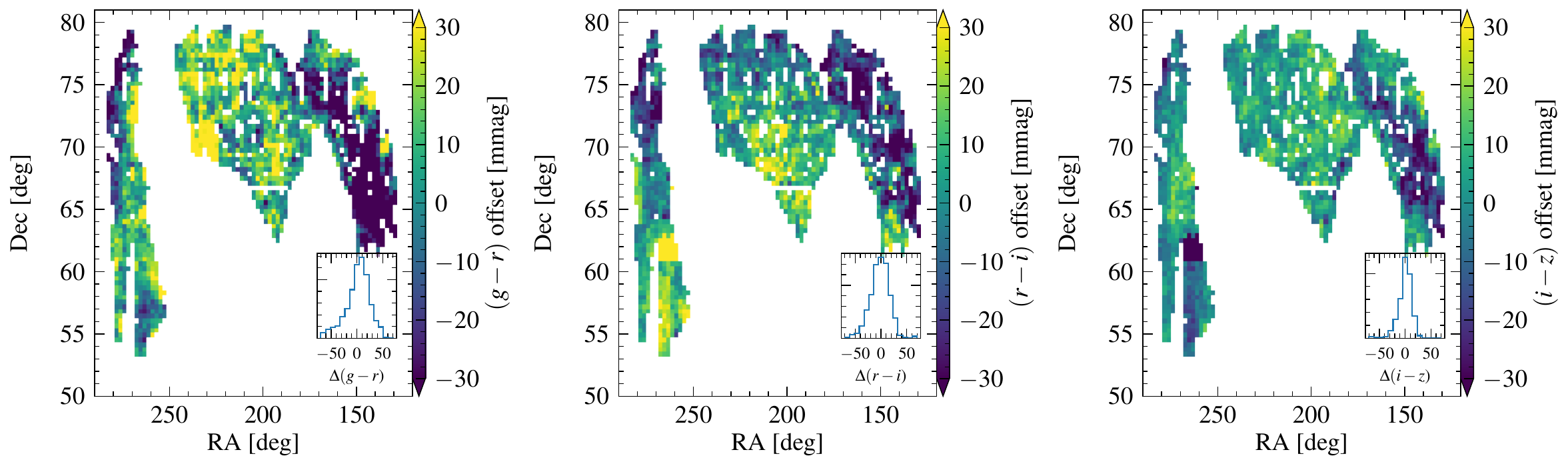}
     \includegraphics[width=\linewidth]{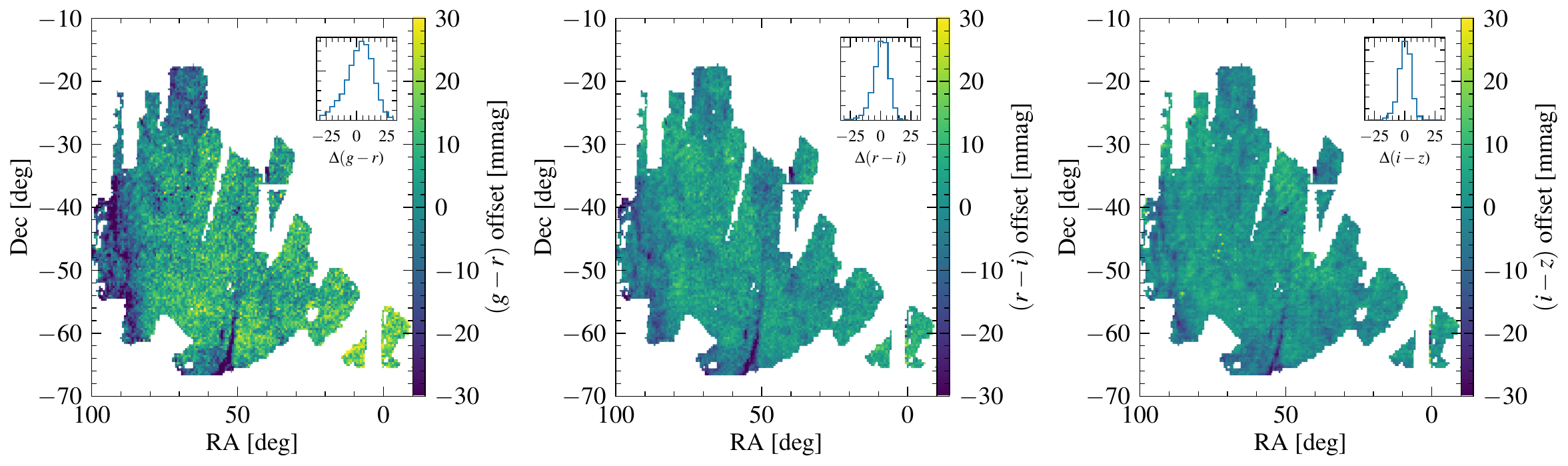}
     \caption{\emph{Top panels:} spatial distribution, in equatorial
     coordinates, of best-fit colour offsets in $g-r$ (left), $r-i$ (centre),
     and $i-z$ (right). The small inset in each panel shows the distribution of
     the values in the respective figure. The offsets are given in units of
     mmag. Note that the stellar locus analysis requires $\sim \, 200$ stars per
     spatial bin, thus the footprint displayed here differs from the EWS
     slightly (e.g. Fig. \ref{fig:dr1_ews_photom_stab_sky}), especially near the
     survey edges. \emph{Bottom panels:} the same as above, but for the southern
     hemisphere.}
     \label{fig:colorH_maps}
\end{figure*}

\begin{figure}[hpt!]
    \centering
     \includegraphics[width=0.79\linewidth]{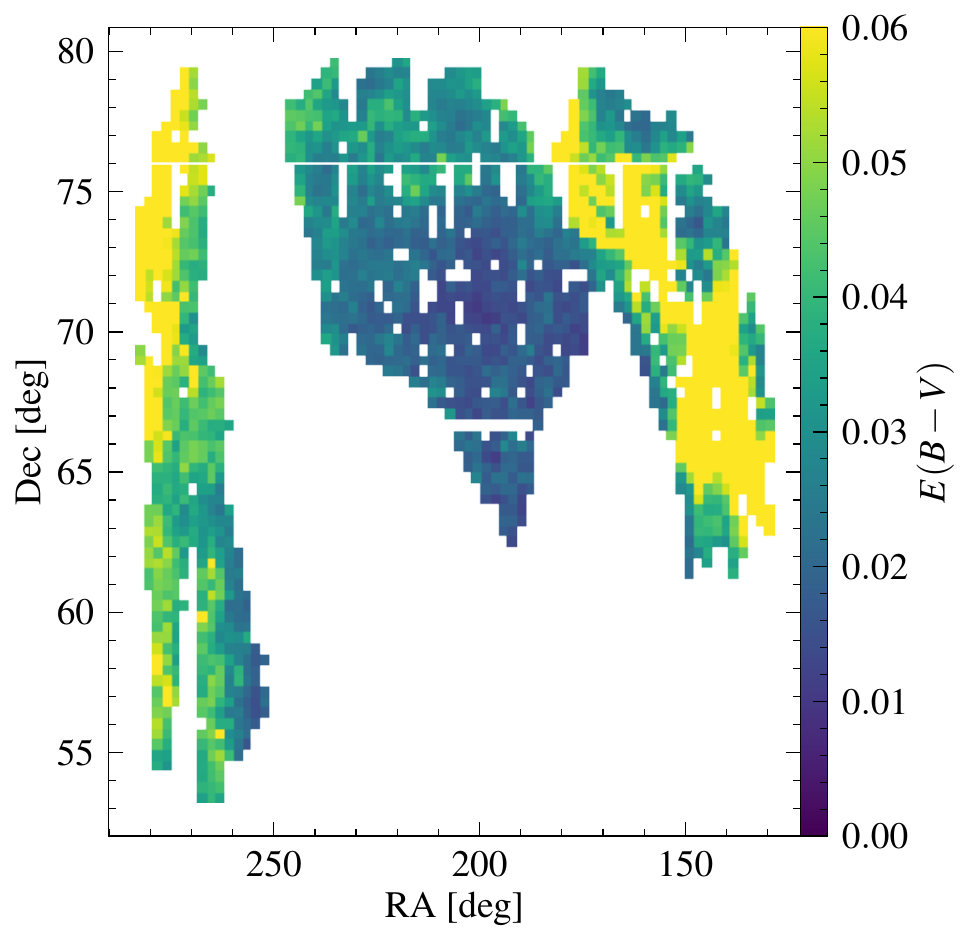}
     \includegraphics[width=0.80\linewidth]{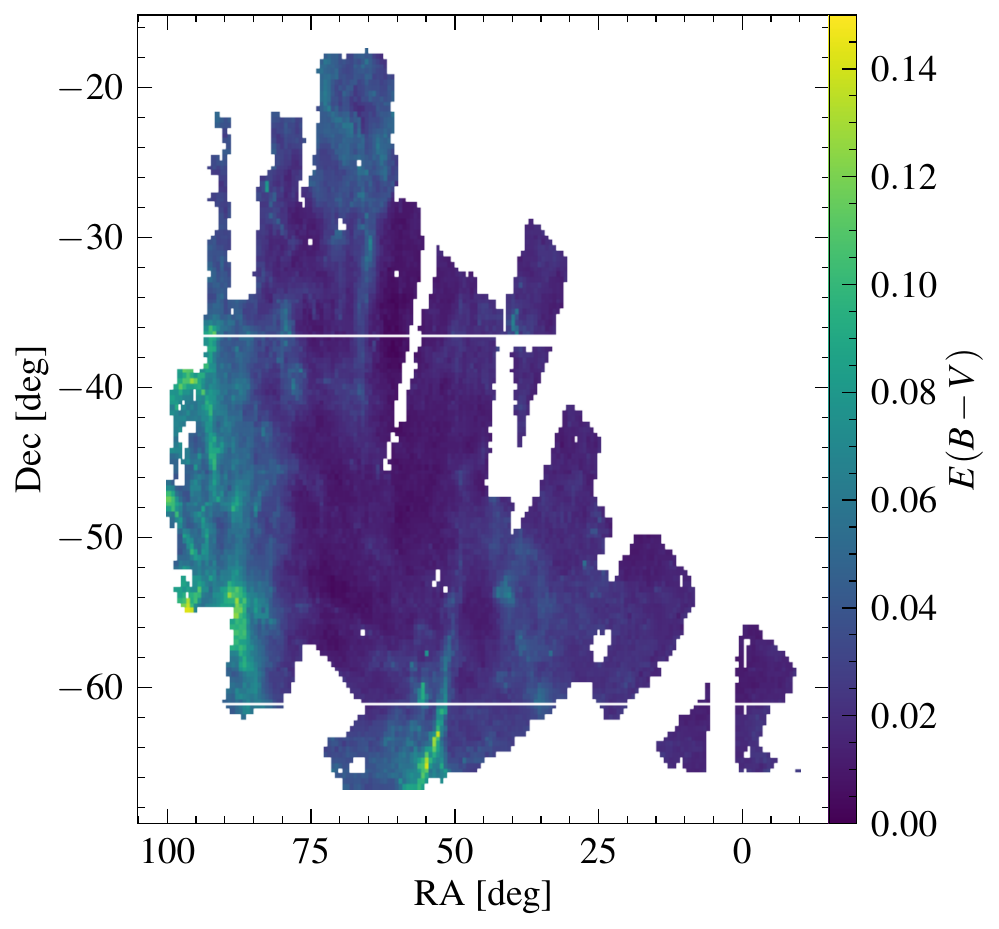}
     \caption{Extinction $E(B-V)$ maps \citep{Schlegel:1998, Schlafly:2011} for the EWS-North (top) and EWS-South (bottom).}
     \label{fig:ebv_maps}
\end{figure}

In Fig.~\ref{fig:colorH_maps} we show the spatial distribution of the three
components of the colour offset vector.  We observe that for the south
hemisphere, the distribution of offsets shows a smooth behaviour with deviations
$<$ 5--10 mmag. Some contiguous regions display larger/smaller offsets.
However, this can be attributed to regions of large Galactic extinction, as
evidenced by Fig~\ref{fig:ebv_maps}. The effect of extinction is still
visible, since correction for extinction applied here is not perfect due to dust
map interpolation, not accounting for the 3D structure of the dust, and
imperfect knowledge of the extinction law. 

In the same figure discussed above, we also see the results for the northern surveys.
In this case, the residuals in the colour offset are much larger, ranging from 10--25 mmag.
This behaviour is associated with the calibration and stacking effects discussed in Sect.~\ref{s:dr1_known_issues}.

\section{\label{s:discussion}Discussion}

DR1 represents the first large-scale integration of heterogeneous ground-based optical data within the \Euclid SGS framework.
Beyond product delivery, it served as an operational validation of the EXT calibration, production, and validation workflow.
In this section we discuss the lessons learned from DR1, the origin of the main validation failures, and the improvements planned for the OU-EXT PFs towards a fully mission-compliant ground-based dataset in future releases.

\subsection{\label{s:data_validation}DR1 data validation policy and statistical outcome}

\begin{table*}
\centering
\caption{\change{Summary of the DR1 StackedFrame validation results based on Eqs.~\eqref{eq:qc_cut1} and \eqref{eq:qc_cut2} by field and band.
Each entry gives $N_{\rm valid}/N_{\rm all}$ followed by the valid fraction
in percent. Validation was performed against either the external reference
catalogue (ERC; $\dagger$) or the SEF composite catalogue (SCC; $\ddagger$).}}
\label{tab:stacks_flagged}

% \small
\setlength{\tabcolsep}{3pt}

\begin{tabular}{@{}lcccccc@{}}
\toprule
Field & $g$ & $r$ & $r2$ & $i$ & $i2$ & $z$ \\
\midrule
EWS-North$^\dagger$
& 2488/2508 (99.2)
& 2527/2569 (98.4)
& --
& 2584/2646 (97.7)
& --
& 2364/2370 (99.7) \\

EWS-South$^\dagger$
& 6776/6878 (98.5)
& 6786/6878 (98.7)
& --
& 6783/6878 (98.6)
& --
& 6769/6879 (98.4) \\

EDF-N$^\ddagger$
& 518/520 (99.6)
& 0/127 (0.0)
& 491/507 (96.8)
& 0/77 (0.0)
& 496/520 (95.4)
& 490/520 (94.2) \\

EDF-S$^\ddagger$
& 461/597 (77.2)
& 553/597 (92.6)
& --
& 512/596 (85.9)
& --
& 509/597 (85.3) \\

EDF-F (HSC)$^\ddagger$
& 183/196 (93.4)
& --
& 161/167 (96.4)
& 3/94 (3.2)
& 121/125 (96.8)
& 106/114 (93.0) \\

EDF-F (DECam)$^\ddagger$
& 269/274 (98.2)
& 249/274 (90.9)
& --
& 250/274 (91.2)
& --
& 264/274 (96.4) \\

COSMOS (HSC)$^\ddagger$
& 77/84 (91.7)
& 78/85 (91.8)
& 85/85 (100.0)
& 2/85 (2.4)
& 78/82 (95.1)
& 80/85 (94.1) \\

COSMOS (DECam)$^\ddagger$
& 81/85 (95.3)
& 81/85 (95.3)
& --
& 76/85 (89.4)
& --
& 82/85 (96.5) \\
\bottomrule
\end{tabular}
\end{table*}

Although the primary goal of the \Euclid mission is cosmology, non-cosmological science remains a fundamental component of the project.
For this reason, the data-quality criteria adopted by OU-EXT to flag the DR1 \stacks were intentionally less restrictive than the strict requirements discussed in Sect.~\ref{s:qa}, allowing the scientific exploitation of data that may not yet meet the most demanding cosmological thresholds.

The flagging criteria are based on the photometric stability metrics defined in Sect.~\ref{s:stack_photomstab} and require
\begin{align}
 &|\Delta| < 50~\rm mmag\,, \label{eq:qc_cut1}\\
 &\left|\frac{\Delta - \rm median(\Delta)}{\rm NMAD(\Delta)}\right| < 3\,, \label{eq:qc_cut2}
\end{align}
where $\Delta$ denotes the global median magnitude offset computed per StackedFrame with respect to the SCC or ERC reference catalogues.
The quantities $\mathrm{median}(\Delta)$ and $\mathrm{NMAD}(\Delta)$ refer to the corresponding global per-band distributions in DR1.
In practice, the stacked zero-point must agree with the reference catalogue within 50~mmag and remain within three times the scatter of the global distribution.

The resulting statistics for each photometric band and field are summarised in Table~\ref{tab:stacks_flagged}.
The EWS shows a low fraction of rejected stacks in both hemispheres, at the level of $\simeq1$--1.5\%.
This demonstrates that the combination of Stage~1 recalibration and Stage~2 stacking achieves a largely homogeneous photometric solution across the wide footprint.

In contrast, the deep fields exhibit significantly higher overall failure rates, with rejected fractions reaching up to $\simeq$20\% in fields such as EDF-S.
The most severe failures occur for HSC-$i$ in EDF-N, EDF-F and COSMOS, and for HSC-$r$ in EDF-N.
By comparison, HSC-$r2$ and HSC-$i2$ perform significantly better in the same fields.
These higher failure rates reflect the increased complexity of the deep data, including the larger number of input SEFs, heterogeneous observing conditions, and greater sensitivity to small calibration inconsistencies.
The following subsection discusses the origin of these issues.

\subsection{\label{s:dr1_known_issues}Origin of DR1 validation failures}

A number of the rejected StackedFrames can be traced to well-identified production or calibration issues.

In the EWS-North footprint, a patch of Pan-STARRS stacks was found to suffer from incorrect photometric calibration (see Fig.~\ref{fig:dr1_ews_photom_stab_sky}).
These stacks were excluded from the public DR1 release.
\change{In addition, many of the stacks contain image artifacts that required the creation of an additional mask, described in \citet{DR1-TP028}.
These features were already present in the input Pan-STARRS SEF images but were not included in their pixel masks and therefore propagated into the StackedFrames, as \pf{1} did not apply any additional pixel-rejection step during co-addition, unlike the stacking procedure described by \citet{Gwyn+25}.}

In EWS-South a batch of DECam stacks around RA$\simeq5^\circ$ was removed due to a software issue in the SGS common library, which resulted in an incorrect selection of input \sefs products during stacking.
These problems were production-related and are not intrinsic to the calibration methodology.

In the deep fields, particularly EDF-N, EDF-F (HSC), and COSMOS (HSC), a non-negligible fraction of stacks fail the validation cuts.
In several cases, the discrepancies are associated with inconsistencies between the absolute zero-point definition at the \sef\ and \stack\ product levels.
The deep data combine a very large number of single-epoch exposures, increasing the sensitivity to small systematic offsets introduced during recalibration, failures in source-catalogue generation for individual exposures, or stacking.
In addition, astrometric imperfections in subsets of input SEFs have been identified as a contributing factor in some fields, such as EDF-S.
These effects are under investigation and will be addressed in future processing campaigns.
Overall, the DR1 experience highlights the importance of a tight integration between calibration, stacking, and validation workflows, across stage-1 and -2 teams, especially for deep, multi-layer datasets.

\subsection{\label{s:dr2_improvements}Pipeline limitations and improvements towards DR2}

Beyond the specific production issues identified above, DR1 has revealed several structural limitations of the current \change{Stage 2} pipeline that will be addressed in DR2.

\paragraph{Optimising StackedFrame production with \texttt{pySwarp}.}
The stacking pipeline generates substantial I/O load, particularly for deep fields where the number of input \sefs products can be very large.
To mitigate this limitation, we are developing a modified version of \swarp\ with a \texttt{Python} interface, referred to as \texttt{pySwarp}.
This implementation keeps resampled SEFs in memory rather than writing intermediate products to disk before final co-addition.
Prototype tests indicate a reduction of the read load by a factor of $\simeq2$ and of the write load by a factor of $\simeq10$, significantly improving scalability for large production campaigns such as DR2.
These gains will be particularly important when processing the expanded northern footprint and deep datasets.

\paragraph{Limited PSF model sampling.}
The number of PSF models generated per StackedFrame is typically smaller than the total number of MER detections.
For sources without a dedicated PSF model, the nearest model is adopted, which can affect PSF-based photometry in crowded or rapidly varying seeing conditions (e.g. as a result of multi-epoch SEF co-addition).
This limitation may be alleviated by expanding the input source catalogue used for PSF modelling, potentially leveraging the higher-resolution space-based detections.

\paragraph{Noise propagation.}
In DR1, only the background RMS component of the noise has been propagated from the input \sefs images into the final stacked products.
The absence of the Poisson contribution from source fluxes can bias model-fitting photometry and underestimate flux uncertainties.
For DR2, the full error budget -- including source Poisson noise -- will be propagated and made available to the MER pipelines, enabling more accurate photometric measurements and uncertainty estimates.

\paragraph{Passband effective-wavelength variations.}
Wide-field imagers such as HSC, Pan-STARRS, and DECam exhibit spatial and temporal variations in filter transmission.
Combined with field angle-dependent incidence effects, these variations induce colour-dependent biases that can impact SED reconstruction and photometric-redshift estimation \citep[see][]{EP-Paltani}.
The adopted requirement is to characterise effective-wavelength variations with a precision of $\simeq10$~\AA.
This correction was not included in DR1, but its modelling and propagation from the \sefs to the final \stacks products is planned for the next release.

In summary, DR1 demonstrates that the EXT production chain delivers a largely homogeneous wide-area dataset, while identifying specific areas -- particularly in deep fields and in noise and passband modelling -- where improvements are required to fully meet the most stringent cosmological requirements.
The developments described above are designed to ensure that DR2 achieves a higher level of uniformity, robustness, and scalability.

\section{\label{s:conclusions}Summary and conclusions}

This paper describes the ground-based optical imaging component of \Euclid DR1 produced by OU-EXT, covering the EWS footprint as well as the EDFs and EAFs.
The EXT data provide the optical complement to the \Euclid space observations, essential for constraining the SEDs of stars and galaxies and for robust photometric-redshift estimation in cosmological analyses.
To this end, we presented the end-to-end production chain, \change{from the initial} homogenisation and \gaia XP-based recalibration of heterogeneous external surveys to the stacking procedure, focusing on the quality of the final released products: $ugriz$ co-added MER-tiled images and their associated stacked PSF models, weight maps, and validation diagnostics.

\change{The DR1 production campaign} comprises about 2500 deg$^2$ of ground-based optical imaging distributed over the EWS and EDFs, combining data from CFHT/MegaCam, Subaru/HSC, Pan-STARRS1, and DECam.
A unified calibration, stacking and validation framework was applied across all instruments, enabling a homogeneous assessment of depth, photometric calibration, and PSF quality on a MER-tile basis.

In terms of photometric depth, the DR1 StackedFrames reach the end-of-mission requirements for the majority of EWS tiles in most bands.
In the northern EWS footprint, 78\%, 96\%, and 98\% of tiles satisfy the nominal requirements in $g$, $r$, and $i$, respectively, while $z$ remains the most challenging band in terms of uniform depth.

The photometric stability of the different EXT components is evaluated through comparisons with \gaia XP-based synthetic photometry, internal photometric metrics, and independent external catalogues.
In EWS-South, more than 95\% of stacks satisfy the 10~mmag relative calibration requirement (i.e. $\leq1\%$ relative flux calibration), demonstrating that the combination of Stage~1 recalibration and Stage~2 stacking yields a highly stable photometric solution over large contiguous areas.
EWS-North shows larger heterogeneity -- in part due to its multi-instrument nature --, yet the majority of tiles in each band meet the 10~mmag requirement.
Median tile-to-tile variations are generally within the 10~mmag requirement, although localized coherent structures remain in some northern bands.
Internal PSF-consistency tests further show that, for most bands, PSF and Kron-like photometry agree at the $\sim$1\% level.

\change{
The deep and auxiliary fields typically reach depths $1$--$2$~mag beyond those of the EWS, although with larger spatial variations reflecting the heterogeneous combination of observing campaigns and the larger number of contributing exposures.
Their photometric validation also indicates stronger dataset-dependent systematics, especially in the HSC-$r$ and HSC-$i$ bands.
}

\change{Despite the overall performance achieved in DR1, significant processing-related issues remain, arising from the OU-EXT recalibration and co-addition procedures and affecting a subset of the resulting products.}
In future releases, \change{the quality of both individual exposures and co-added images will be improved. The processing performance of the Stage~2 stacking pipeline will be enhanced by optimising the co-addition strategy to reduce I/O load, while data quality will be further improved through more robust PSF model reconstruction, propagation of a more complete noise model including Poisson contributions, and incorporation of passband effective-wavelength variations at the $\sim$10~\AA\ level.}
These developments will further enhance the uniformity and scientific exploitation of the EXT data in DR2 and beyond.

\begin{acknowledgements}
  %We thank XXX and YYY for helpful comments on the manuscript.
  \AckEC
  \AckDRone
  %\AckQone
  This work was enabled by observations made from the CFHT and Subaru telescopes. The scientific community is honored to have the opportunity to conduct astronomical research on Maunakea in Hawai‘i. We recognize and acknowledge the very significant cultural role and reverence of Maunakea to the Kanaka Maoli (Native Hawaiians) community.
  
\end{acknowledgements}

\bibliography{Euclid, ext, DR1}

\begin{appendix}

\clearpage
\onecolumn

\section{SEF flag map propagation}
\label{appendix:sef_flagmaps}
\change{This section describes the SEF pixel flags used during the DR1 \stack production. Table~\ref{tab:ext_sef_pixel_flags} lists the flags adopted for each contributing survey, together with their descriptions. A value of True (T) indicates that pixels carrying the corresponding flag are masked during co-addition. All SEF flags are propagated to the final co-added products.
}

\begin{table*}[htp!]
\centering
\caption{EXT SEF pixel flags for all contributing surveys in DR1.
$a)$ Flag bit value stored in the \texttt{FLAG} FITS extension of the SEFs.
$b)$ Flag description.
$c)$ If True (T), the science pixel is rejected during co-addition.
All SEF flags are propagated to the StackedFrame flag extension.}
\label{tab:ext_sef_pixel_flags}

% \scriptsize
\setlength{\tabcolsep}{3pt}
\renewcommand{\arraystretch}{0.95}
\begin{minipage}[t]{0.49\textwidth}
\vspace{0pt}
\centering
\begin{tabularx}{\linewidth}{@{}cXc@{}}
\toprule
Bit$^a$ & Description$^b$ & Mask$^c$ \\
\midrule

\multicolumn{3}{c}{\textbf{DES (DECam)}} \\
1   & Hot/dead pixel or column        & F \\
2   & Saturated pixel                 & F \\
4   & Interpolated pixel              & F \\
8   & Too little signal / poor read   & F \\
16  & Cosmic ray                      & F \\
32  & Bright star                     & F \\
64  & Bleed trail                     & F \\
128 & Bright star (pixel interpolated) & F \\
\addlinespace

\multicolumn{3}{c}{\textbf{CFIS (MegaCam)}} \\
1   & Hot/dead pixel or column        & T \\
2   & Saturated pixel                 & T \\
4   & Interpolated pixel              & T \\
8   & Too little signal / poor read   & T \\
16  & Cosmic ray                      & T \\
32  & Bright star                     & T \\
64  & Bleed trail                     & T \\
128 & Bad pixel                       & T \\
\addlinespace

\multicolumn{3}{c}{\textbf{Pan-STARRS}} \\
1     & A detector defect is present      & T \\
2     & Flat-field model unreliable       & T \\
4     & Dark model unreliable             & T \\
8     & Pixel does not contain valid data & T \\
16    & Poor charge transfer efficiency   & T \\
32    & Saturated pixel                   & T \\
64    & Lower value than expected         & T \\
128   & Pixel suspected of being bad      & F \\
256   & Cosmic ray present                & T \\
512   & Diffraction spike                 & F \\
1024  & Optical ghost                     & F \\
2048  & Streak                            & F \\
4096  & Bright star core                  & F \\
8192  & Bad pixel after convolution       & T \\
16384 & Poor pixel after convolution      & T \\
32768 & Internal temporary flag           & T \\
\addlinespace

\multicolumn{3}{c}{\textbf{WHIGS/WISHES (HSC-$g$, HSC-$z$)}} \\
1    & Detector defect                                  & T \\
2    & Saturated on CCD                                 & F \\
4    & Interpolated pixel                               & F \\
8    & Cosmic ray detected                              & T \\
16   & CCD edge                                         & T \\
32   & Detected as source footprint                     & F \\
64   & Detected as source footprint with negative flux  & T \\
128  & Nearly saturated (linearity corrected)           & F \\
256  & Field-edge pixel                                 & T \\
512  & Bright-star halo region                          & T \\
1024 & Cross-talk corrected                             & F \\
2048 & Deblending skipped                               & F \\
4096 & NaN pixel (masked later)                         & T \\
\bottomrule
\end{tabularx}
\end{minipage}
\hfill
\begin{minipage}[t]{0.49\textwidth}
\vspace{0pt}
\centering
\begin{tabularx}{\linewidth}{@{}cXc@{}}
\toprule
Bit$^a$ & Description$^b$ & Mask$^c$ \\
\midrule

\multicolumn{3}{c}{\textbf{H2O}} \\
\multicolumn{3}{c}{\textit{HSC-$g$, HSC-$r$, HSC-$r2$, HSC-$i$, HSC-$i2$, HSC-$z$}} \\
1    & Detector defect                                  & T \\
2    & Saturated on CCD                                 & T \\
4    & Interpolated pixel                               & F \\
8    & Cosmic ray detected                              & T \\
16   & CCD edge                                         & T \\
32   & Detected as source footprint                     & F \\
64   & Detected as source footprint with negative flux  & T \\
128  & Nearly saturated (linearity corrected)           & F \\
256  & Field-edge pixel                                 & T \\
512  & NaN pixel (masked later)                         & T \\
1024 & Cross-talk corrected                             & F \\
2048 & Deblending skipped                               & T \\
\addlinespace

\multicolumn{3}{c}{\textbf{SSP}} \\
\multicolumn{3}{c}{\textit{HSC-$g$, HSC-$r$, HSC-$r2$, HSC-$i$, HSC-$i2$, HSC-$z$}} \\
1    & Detector defect                                  & T \\
2    & Saturated on CCD                                 & T \\
4    & Interpolated pixel                               & F \\
8    & Cosmic ray detected                              & T \\
16   & CCD edge                                         & T \\
32   & Detected as source footprint                     & F \\
64   & Detected as source footprint with negative flux  & T \\
128  & Nearly saturated (linearity corrected)           & F \\
256  & Field-edge pixel                                 & T \\
512  & Bright-star halo region                          & T \\
1024 & Cross-talk corrected                             & F \\
2048 & Deblending skipped                               & T \\
4096 & NaN pixel (masked later)                         & T \\
\bottomrule
\end{tabularx}
\end{minipage}

\end{table*}

\clearpage
\twocolumn

\section{Reference catalogues}
\label{appendix:ref_catalogues}

The photometric calibration and validation of the \ext\ data products rely on a hierarchy of reference catalogues, each playing a complementary role in the Stage~1 recalibration and Stage~2 quality assessment.

The primary calibration reference adopted by \ext\ is \gaia~DR3 \citep{GaiaDR3+23}.
\gaia data are used to calibrate the astrometry of the SEFs and to derive their photometric zero-points through synthetic photometry computed from the \gaia XP low-resolution spectra \citep[see][for details]{Montegriffo+23}.
This approach provides a homogeneous, all-sky reference framework that enables the combination of heterogeneous ground-based datasets obtained with different instruments, passbands, and observing conditions (e.g. UNIONS, DES).
The \gaia XP-based synthetic photometry defines the global photometric system onto which all SEFs are recalibrated in Stage~1.
This catalogue is referred to as GRC.

Because \gaia-based reference stars are typically limited to $G \lesssim 17.5$, bright-star saturation can limit the number of available sources in the deep fields for validation purposes.
To extend the magnitude range of the validation sample, Stage~1 also produces SEF-composite catalogues by combining repeated measurements of sources observed in overlapping SEFs.
For each source, the composite photometry is computed as an inverse-variance weighted mean of the individual measurements after applying the Stage~1 zero-point solution.
These SEF-composite catalogues provide a deeper internal reference and are therefore used primarily for photometric validation in the Deep and Auxiliary fields, as discussed in Sect.~\ref{s:stack_photomstab}.
Throughout this paper, we refer to this catalogue as SCC.

For the external validation of the StackedFrame photometry, supplementary catalogues were constructed using survey-native PSF photometry.
In the northern hemisphere, Pan-STARRS1 \citep[PS1;][]{Chambers+16} $griz$ data are employed, while in the southern hemisphere the reference catalogue is based on the Dark Energy Survey Data Release~2 \citep[DESDR2;][]{Abbott+21}.
In both cases, the input photometry corresponds to forced PSF-model photometry performed on individual exposures \citep[see][]{Magnier+20, Abbott+21}.
These catalogues provide an independent validation layer that is not tied to the internal EXT calibration chain.
We refer to this catalogue as ERC.

To ensure consistency with the \gaia XP-based photometric system adopted in Stage~1, PS1 and DES photometry are transformed into the \gaia synthetic frame using a colour transformation based on a random forest regression implemented with \textsc{scikit-learn}.
The matched stellar samples are divided into training and test subsets, and the regressor hyper-parameters are optimised by cross-validation.
For each source $n$ in the training set $\mathcal{D}=\{(x_n,y_n)\}_n^N$, the input feature vector is defined as
\begin{equation}
    x_{n,\,\rm ref} = \left\{ g-r,\; r-i,\; i-z,\; (g-r)^2,\; (r-i)^2,\; (i-z)^2 \right\}_{\rm ref},
\end{equation}
where ``ref'' denotes either PS1- or DESDR2-native photometry.  

The regression target values are
\begin{equation}
\label{eq:rf_reg_target_val}
    y_n = \left\{ m_n - i_{\rm ref} \right\}_n^K,
\end{equation}
where $K$ is the number of EXT bands available in each hemisphere.
In the north, $K=8$ (HSC $g,r,r2,i,i2,z$, MegaCam $r$, and PS1 $i$), while in the south $K=4$ (DECam $griz$).
The predicted \gaia-frame magnitudes are then obtained as $\widehat{m_n} = \widehat{y_n} + i_{\rm ref}$.

\begin{figure}
    \centering
    \includegraphics[width=\linewidth]{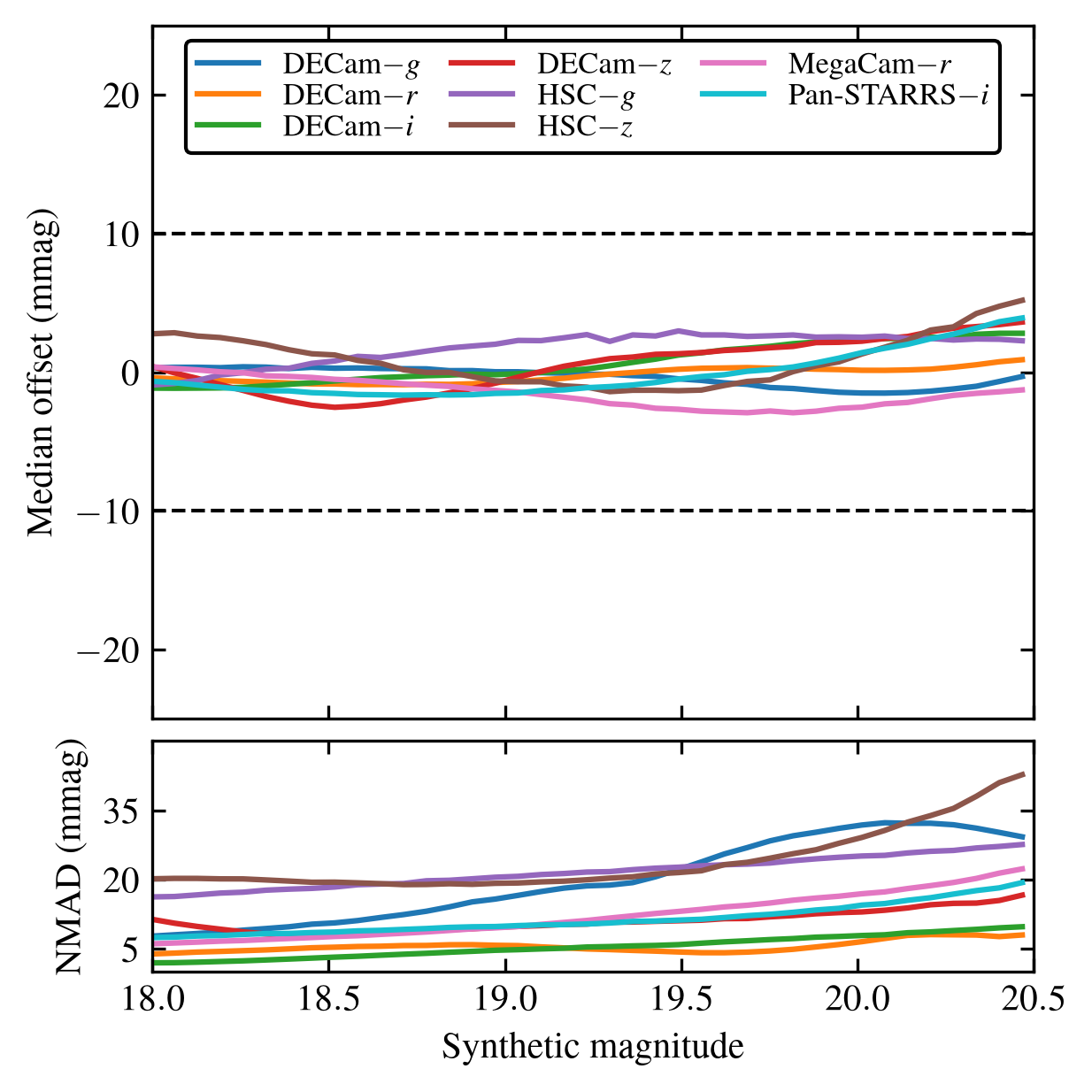}
    \caption{
    \change{
    Relative comparison of the ERC against UNIONS-native and DES DR2 photometry. The median magnitude residuals and corresponding NMAD are shown as a function of magnitude over the range used for StackedFrame validation ($18$--$20.5$~mag). For each band, the global median offset between the ERC and the corresponding native photometric system has been subtracted to highlight residual magnitude-dependent trends.}
    }
    \label{fig:erc_validation}
\end{figure}

The performance of the transformation is evaluated on the independent test set using the median residual and NMAD as a function of magnitude.
As shown in Figs.~\ref{fig:s1_ps_hsc1}--\ref{fig:desdr2_decam}, the global median offsets remain below $\simeq1$~mmag in all bands, with intrinsic scatter around $\sim10$~mmag over the magnitude range probed by the GRC.
\change{As a final validation step, we compared the ERC photometry with UNIONS-native photometry (S. Gwyn, internal communication) and DES DR2 over the magnitude range adopted for validation ($18$--$20.5$~mag). Figure~\ref{fig:erc_validation} shows the resulting median residuals and NMAD as a function of magnitude for the UNIONS and DES filters, after subtracting the global offset between the ERC and the corresponding native photometric system. These global offsets are 20.8, $-4.7$, $-17.3$, and $-1.7$~mmag for the UNIONS $griz$ bands, respectively, and 9.9, $-11.1$, $-3.1$, and 5.9~mmag for DES DR2. After removing these offsets, the median residuals remain within 10~mmag in all bands, with variations as a function of magnitude smaller than $\simeq5$~mmag, indicating that the transformation does not introduce significant magnitude-dependent systematics. The NMAD increases towards fainter magnitudes, exceeding 20~mmag in some HSC bands. However, because each StackedFrame typically contains $\gtrsim1000$ reference sources over this magnitude range, this source-to-source scatter has a much smaller impact on the uncertainty of the median photometric offset used for validation.}
These results demonstrate that the transformed PS1 and DES catalogues provide a stable and independent reference frame for validating the StackedFrame photometry.

In summary, the \gaia XP-based synthetic photometry defines the global calibration system, the SEF-composite catalogues extend the usable magnitude range in deep fields for internal photometric validation, and the transformed PS1/DES catalogues provide an external validation reference.
Together, these catalogues ensure both internal consistency and independent verification of the photometric performance reported in Sect.~\ref{s:stack_photomstab}.

\begin{figure}
    \centering
    \includegraphics[width=\linewidth]{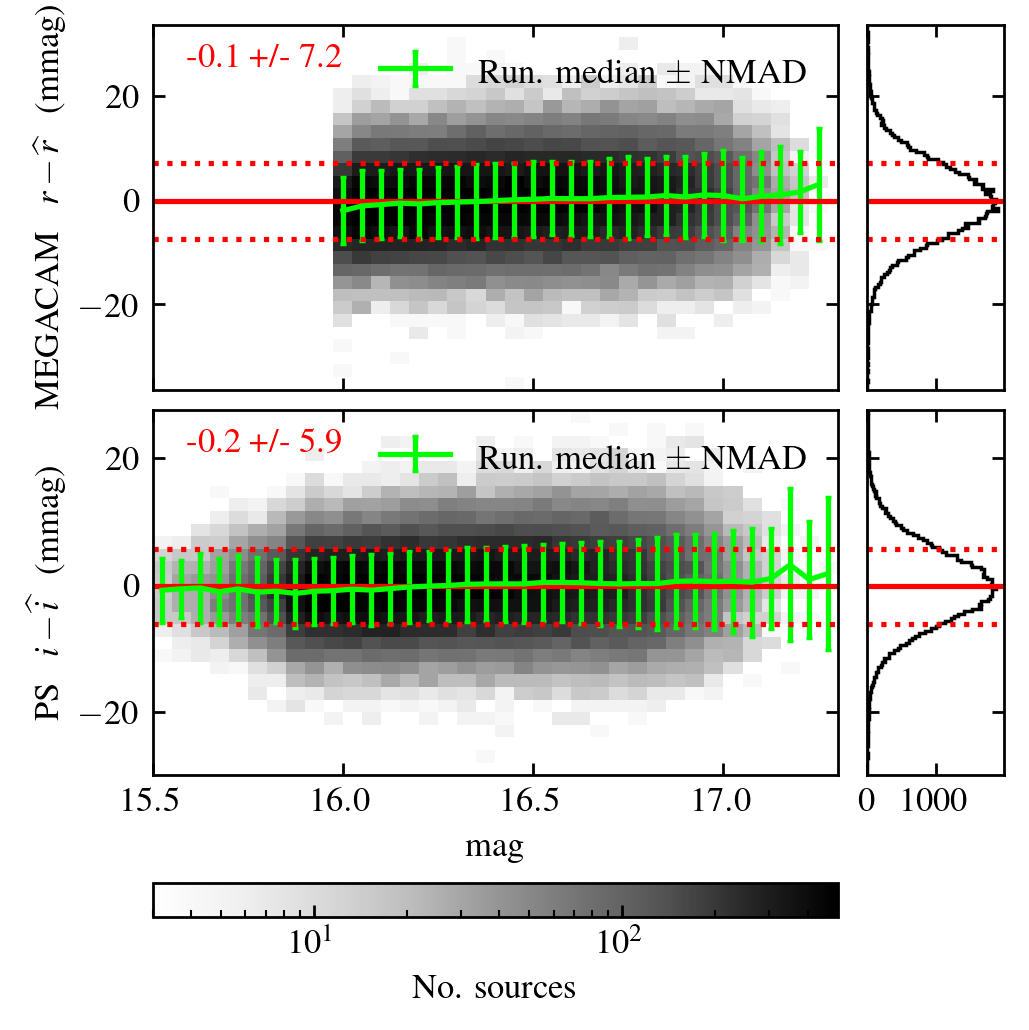}
    \caption{
    Residual photometric offsets between Stage 1 \gaia XP-based synthetic photometry and PS1-transformed photometry for MegaCam and PS EXT photometry. Green error bars denote the running median and NMAD values as function of magnitude, while red solid (dotted) lines denote the global median (NMAD) value, quoted on the top-left corner of each panel in mmag.
    % Same as \ref{fig:s1_ps_hsc1}, but for MegaCam and PS EXT photometry using as reference PS1.
    }
    \label{fig:s1_ps_megacam_ps}
\end{figure}

\begin{figure}
    \centering
    \includegraphics[width=\linewidth]{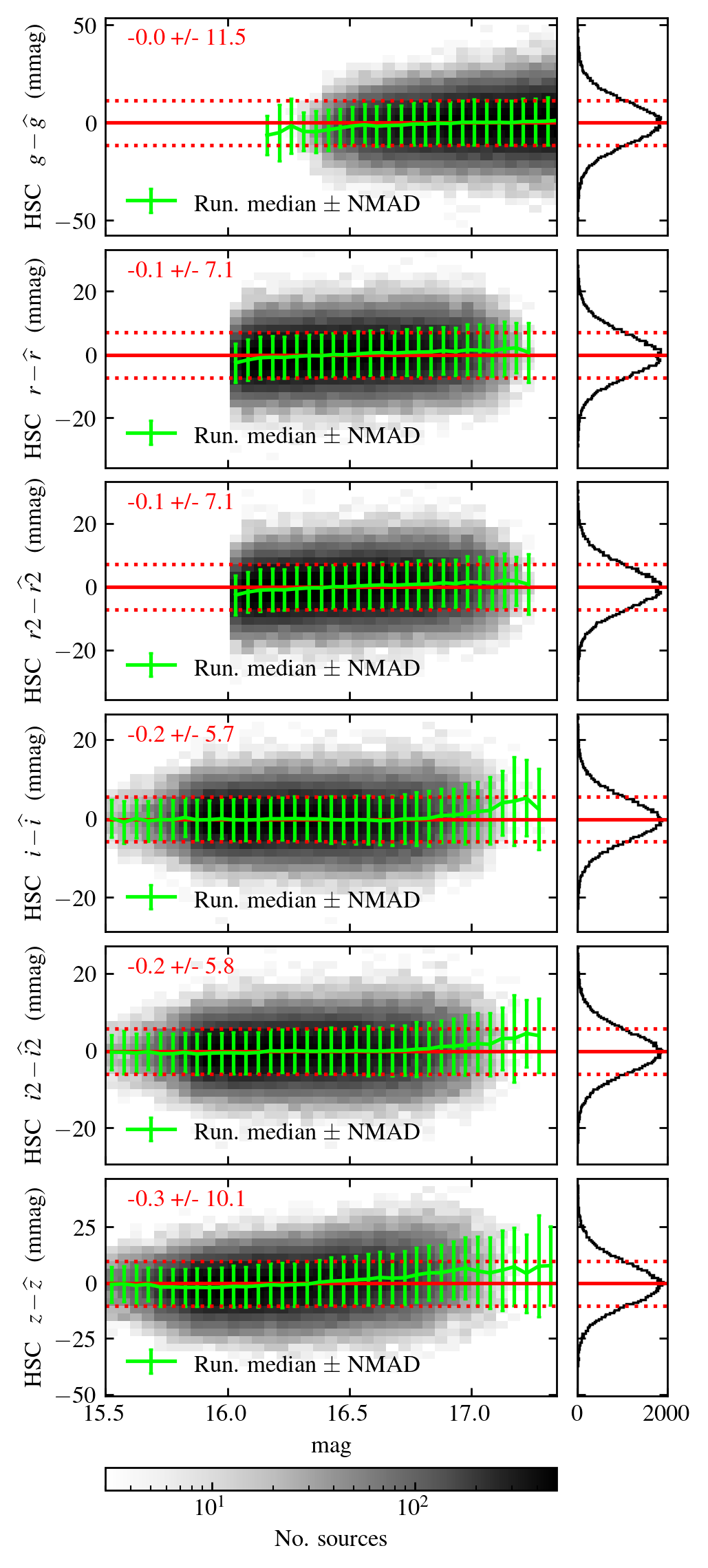}
    \caption{
    Same as \ref{fig:s1_ps_megacam_ps}, but for HSC EXT photometry using as reference PS1.
    }
    \label{fig:s1_ps_hsc1}
\end{figure}

\begin{figure}
    \centering
    \includegraphics[width=\linewidth]{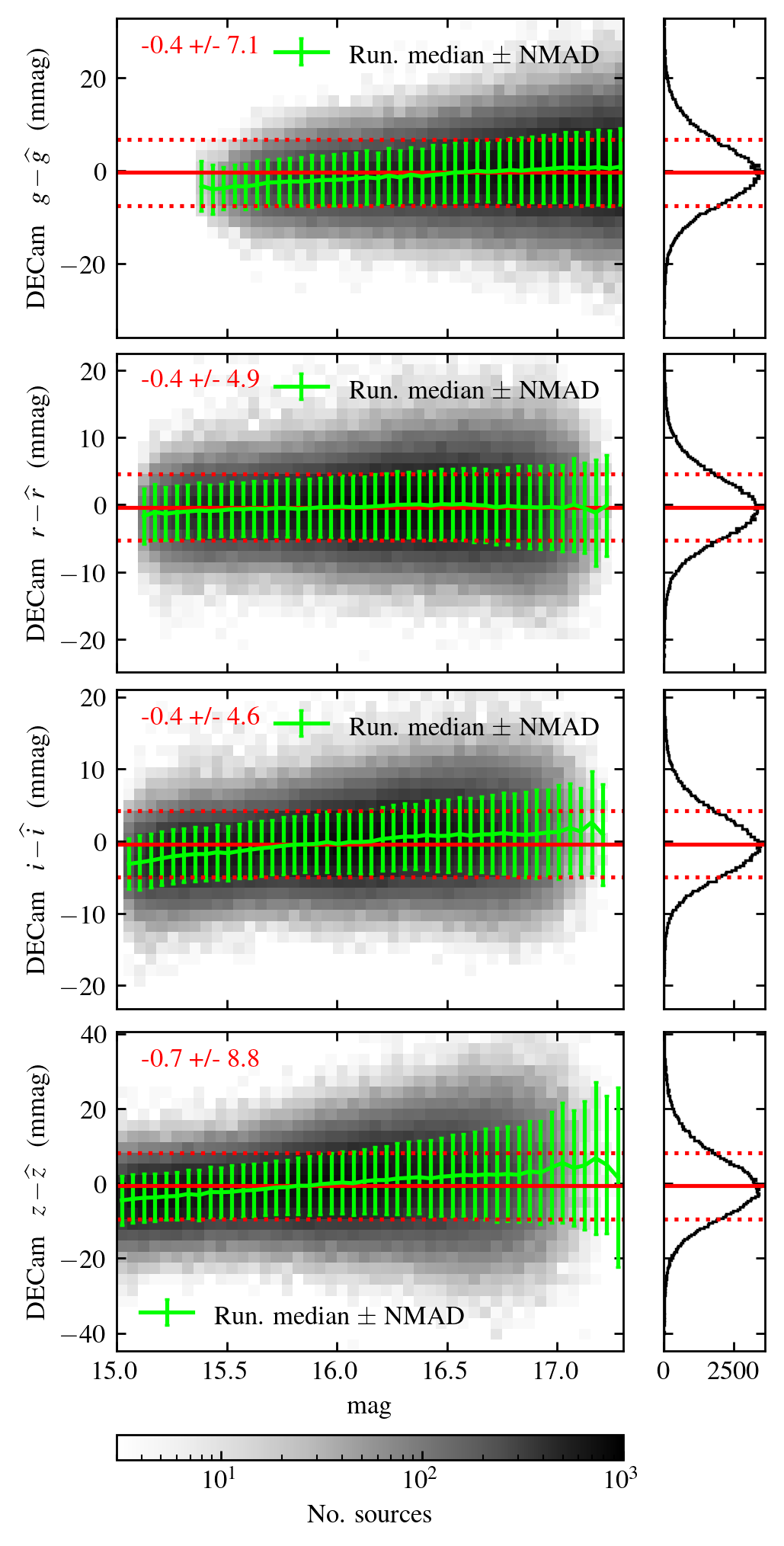}
    \caption{Same as \ref{fig:s1_ps_hsc1}, but for DECam EXT    photometry using as reference DESDR2.}
    \label{fig:desdr2_decam}
\end{figure}

% \onecolumn

\clearpage
\onecolumn

\section{Spatial distribution of photometric depth}
\label{appendix:depth_sky}

This appendix presents the spatial distribution of the achieved $10\sigma$ point-source depth (measured in $2''$ diameter apertures) across the DR1 footprint.
Each map shows the depth estimated per StackedFrame (i.e. per MER tile), following the methodology described in Sect.~\ref{s:stack_depth}.
The colour scale encodes the limiting magnitude in the corresponding band.

\begin{figure*}[htp!]
    \centering
    \includegraphics[width=0.9\linewidth]{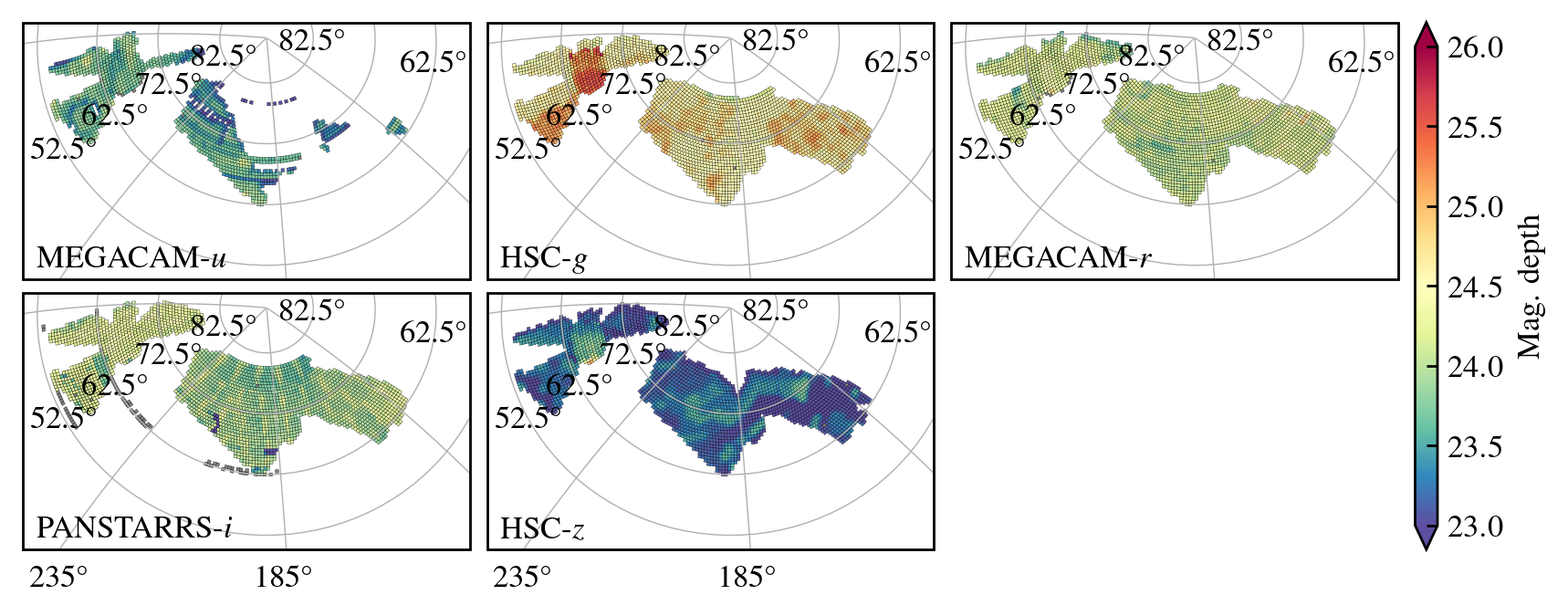}
    \includegraphics[width=0.9\linewidth]{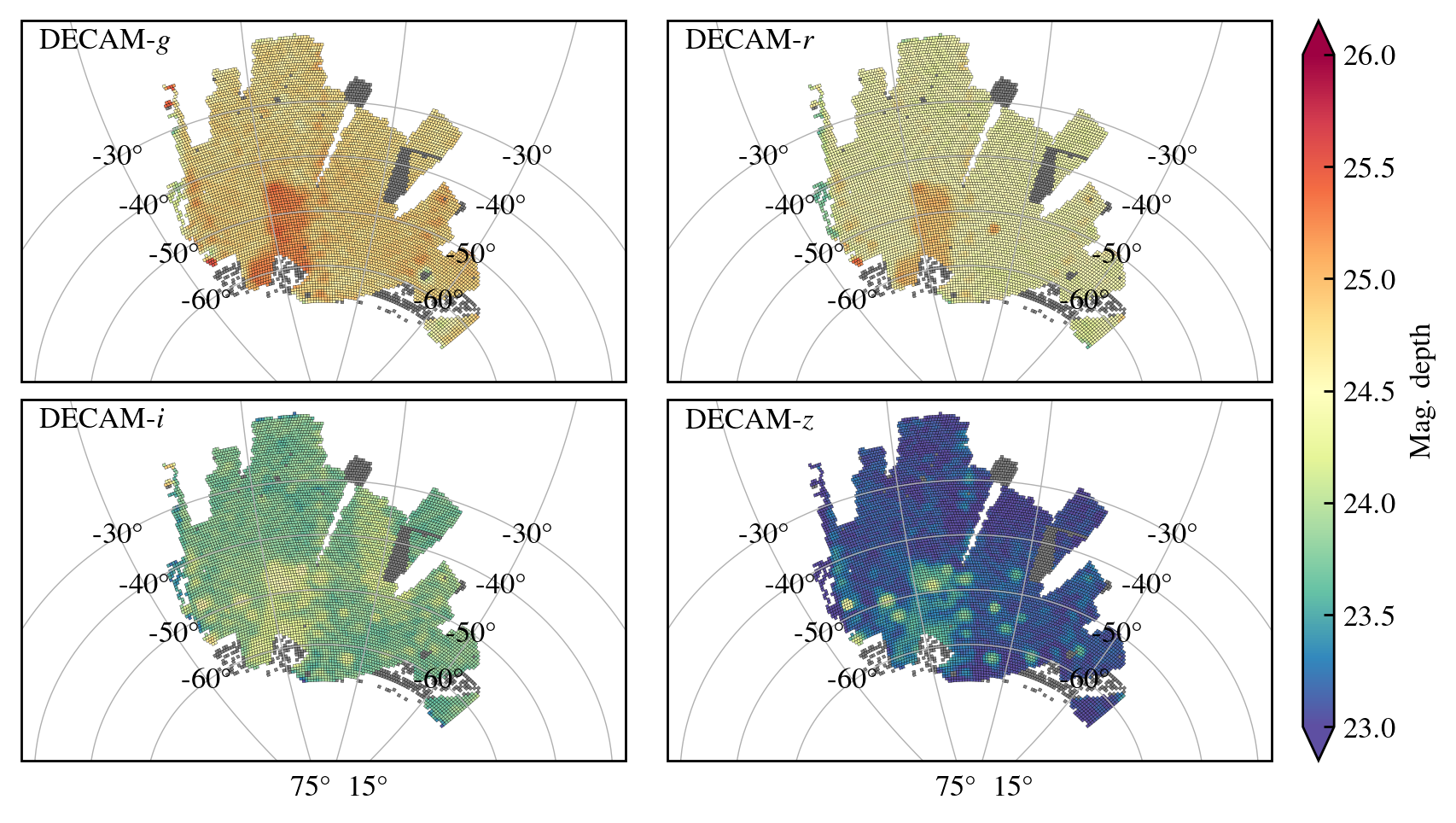}
    \caption{EWS StackedFrame spatial distribution of achieved photometric depths. The maps adopt a Lambert azimuthal equal-area projection in RA and Dec to facilitate the visualisation of the tiling scheme and the variations across tiles.}
    \label{fig:dr1_ews_depth_sky}
\end{figure*}

Figure~\ref{fig:dr1_ews_depth_sky} shows the distribution across the EWS footprint in the northern and southern hemispheres.
The EWS exhibits a largely homogeneous depth over contiguous regions, with variations primarily driven by differences in exposure time, observing conditions, and survey tiling strategy.
In the southern hemisphere (covered by DECam observations), the depth appears particularly uniform, with a small number of deeper regions associated with the DES Science Verification and supernovae fields.
In contrast, the northern footprint (UNIONS) combines multiple instruments and programmes.
Some of these programs (e.g. MegaCam-$u$) are still not fully completed and/or combine archival data with new observations (e.g. WHIGS, HSC-$g$), resulting in slightly larger tile-to-tile variations, that will be reduced in future releases.

\begin{figure}[htp!]
    \centering
    \includegraphics[width=\linewidth]{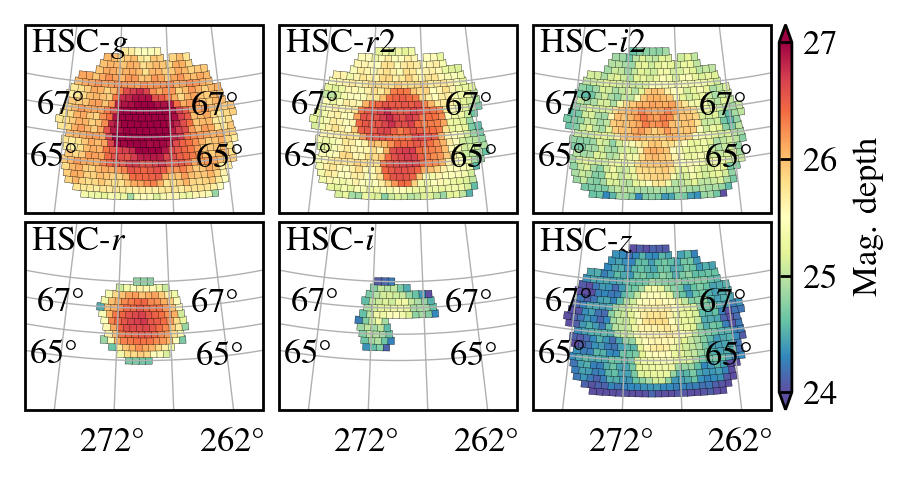}
    \caption{EDF-N StackedFrame spatial distribution of achieved photometric depths. The projection is the same as in Fig.~\ref{fig:dr1_ews_depth_sky}.}
    \label{fig:edfn_sky_depth}
\end{figure}

\begin{figure}[htp!]
    \centering
    \includegraphics[width=\linewidth]{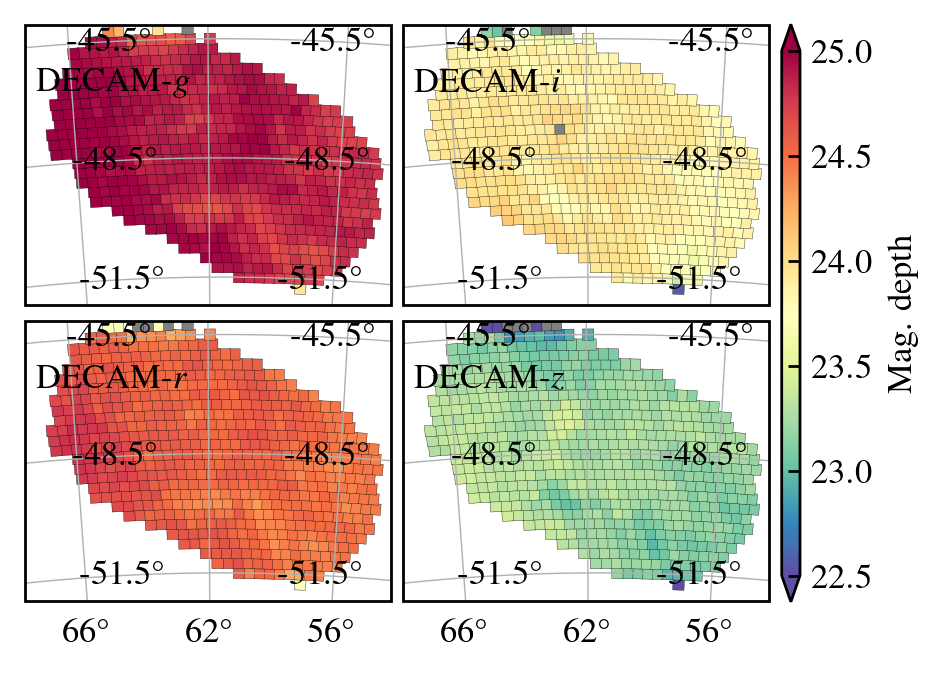}
    \caption{Same as Fig.~\ref{fig:edfn_sky_depth} for EDF-S.}
    \label{fig:edfs_sky_depth}
\end{figure}

\begin{figure}[htp!]
    \centering
    \includegraphics[width=\linewidth]{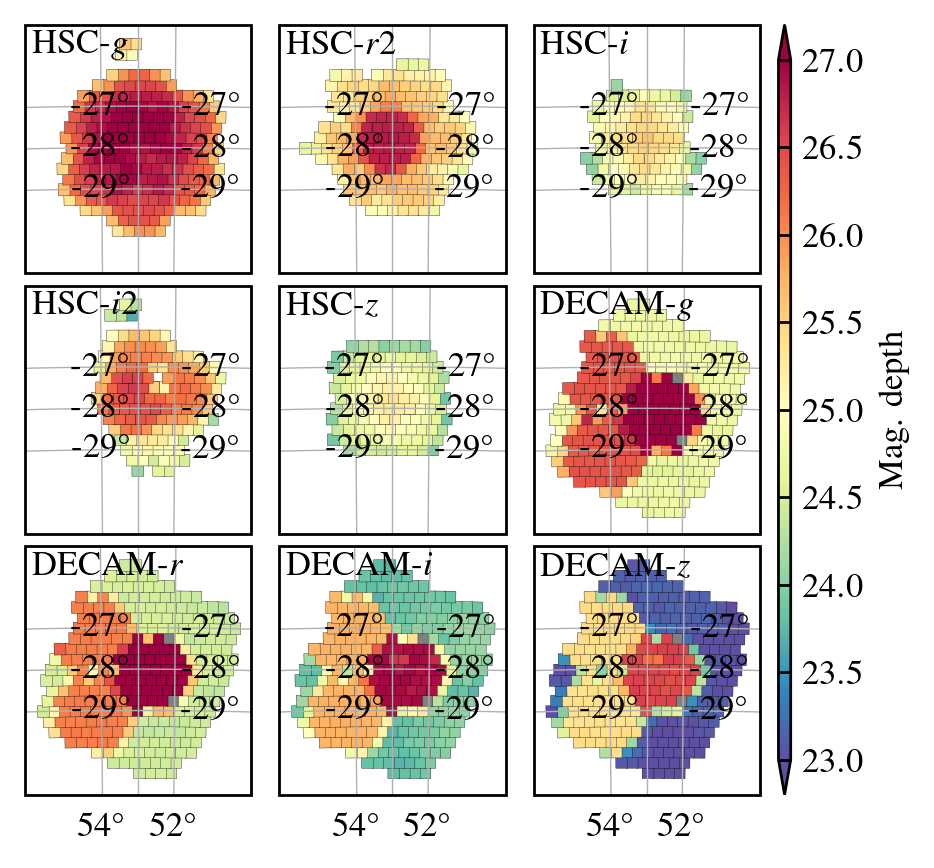}
    \caption{Same as Fig.~\ref{fig:edfn_sky_depth} for EDF-F.}
    \label{fig:edff_sky_depth}
\end{figure}

\begin{figure}[htp!]
    \centering
    \includegraphics[width=\linewidth]{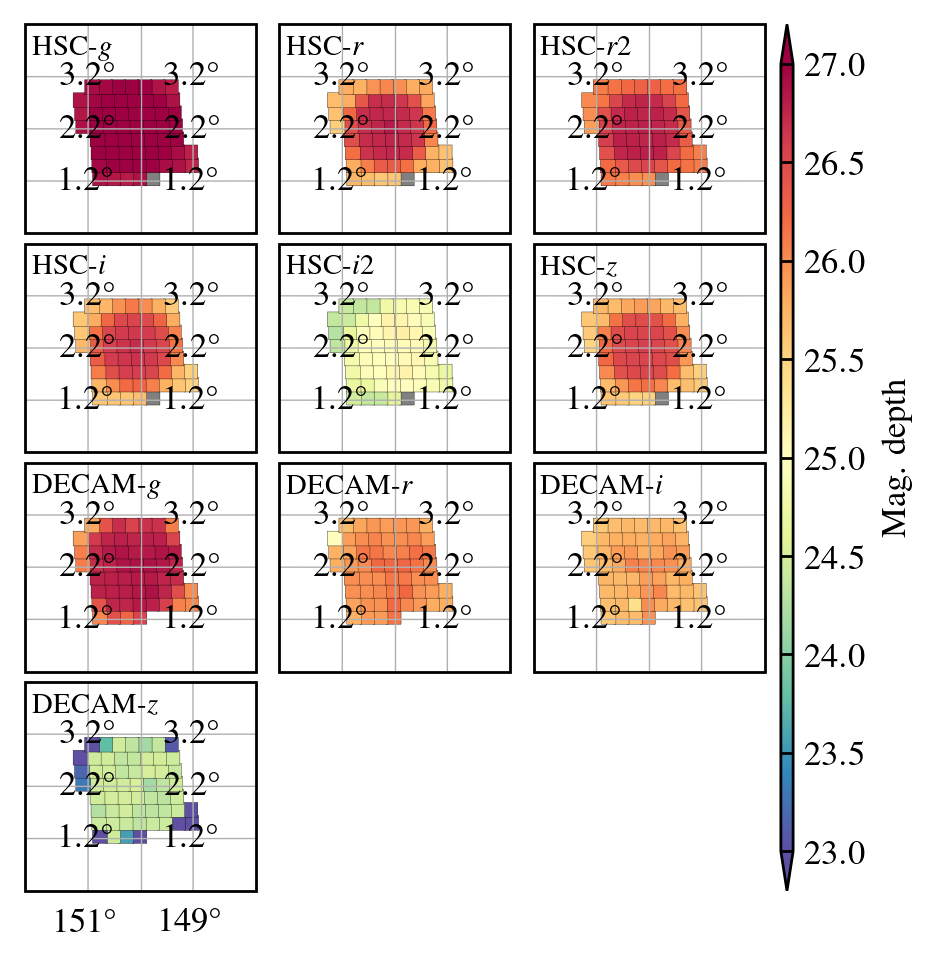}
    \caption{Same as Fig.~\ref{fig:edfn_sky_depth} for COSMOS.}
    \label{fig:cosmos_sky_depth}
\end{figure}

The deep and auxiliary fields are shown in Figs.~\ref{fig:edfn_sky_depth}--\ref{fig:cosmos_sky_depth}.
In these fields, the depth typically exceeds the EWS requirements by $1$--$2$ magnitudes, although the spatial distribution is more structured.
This reflects the complex layering of multiple observing campaigns, variable seeing conditions, and differences in the number of contributing SEFs.
%In some deep fields, depth gradients correlate with regions of higher SEF coverage, as expected from the co-addition process.

Overall, the depth maps confirm the statistical results presented in Sect.~\ref{s:stack_depth}: the majority of EWS tiles meet the nominal end-of-mission requirements, while deep fields achieve significantly greater depth at the expense of increased spatial heterogeneity.

\twocolumn

\clearpage
\onecolumn

\section{Spatial distribution of photometric stability}
\label{appendix:photom_stab_sky}

This appendix presents the spatial distribution of the photometric zero-point offsets measured per StackedFrame across the DR1 footprint.
The maps show the median magnitude difference between PSF photometry measured on the StackedFrame and the adopted reference catalogue (GRC in the EWS; SCC in the deep fields), as described in Sect.~\ref{s:stack_photomstab}.
The colour scale represents the median offset per tile in mmag.

\begin{figure*}[htp!]
    \centering
    \includegraphics[width=0.8\linewidth]{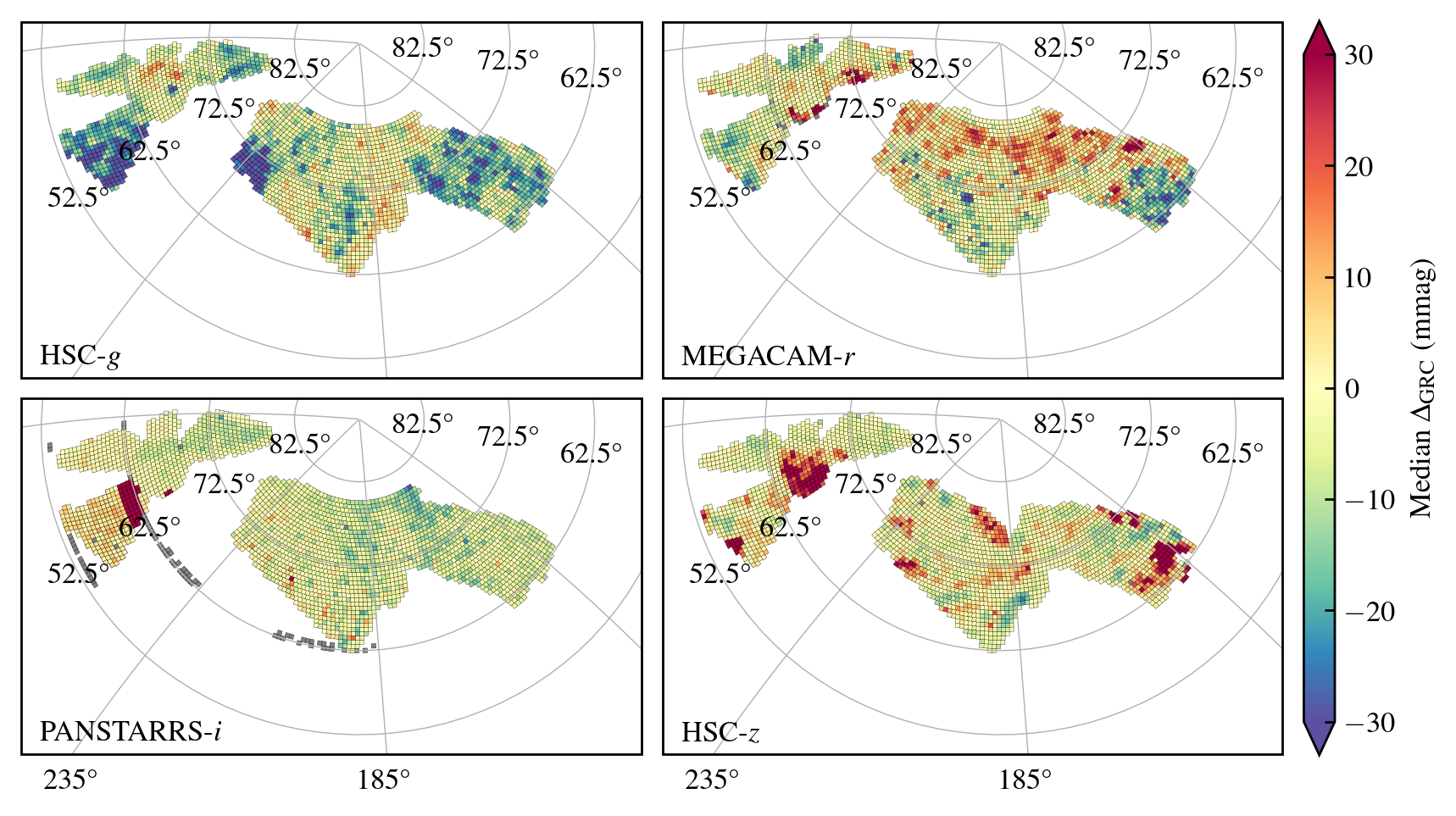}
    \includegraphics[width=0.8\linewidth]{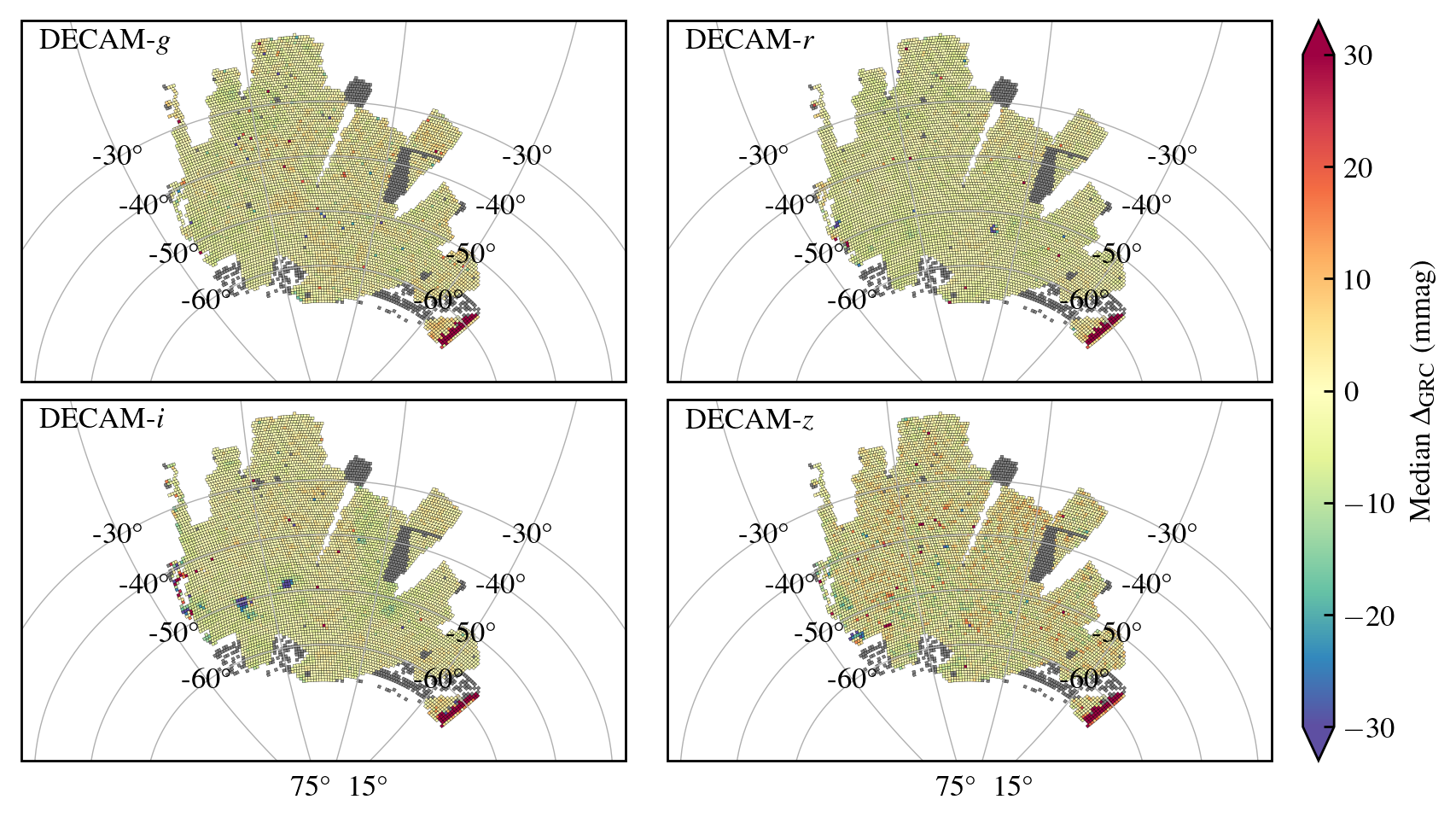}
    \caption{Spatial distribution of the photometric zero-point stability across the EWS, computed from the magnitude differences between StackedFrame PSF stellar photometry and the GRC. The projection is the same as in Fig.~\ref{fig:dr1_ews_depth_sky}.}
    \label{fig:dr1_ews_photom_stab_sky}
\end{figure*}

Figure~\ref{fig:dr1_ews_photom_stab_sky} illustrates the EWS footprint.
The majority of tiles exhibit offsets within $\pm10$~mmag, consistent with the mission-level requirement for relative photometric calibration.
Large-scale spatial gradients are generally weak, indicating that the Stage~1 recalibration and Stage~2 stacking do not introduce coherent calibration drifts over degree scales.
Localised structures are visible in some northern bands, especially in HSC-$g/z$ and MegaCam-$r$, corresponding to subsets of data affected by calibration inhomogeneities or production issues discussed in Sect.~\ref{s:dr1_known_issues}.
In the EWS-South region, the StackedFrames located along the RA$\approx5$ deg meridian were built with a wrong set of input SEFs and have been flagged as invalid in the release.

%These features are spatially confined and do not dominate the global calibration statistics.

\begin{figure}[htp!]
    \centering
    \includegraphics[width=\linewidth]{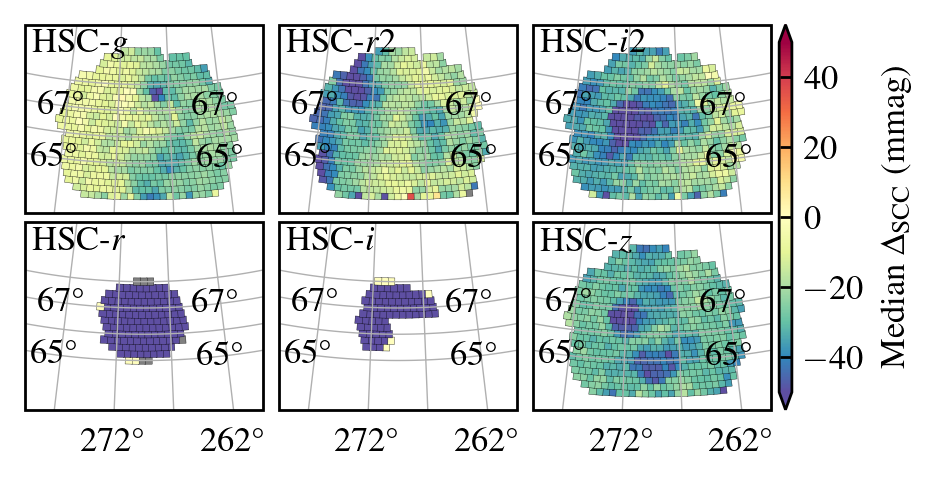}
    \caption{Spatial distribution of the photometric zero-point stability across the EDF-N, computed from the magnitude differences between StackedFrame PSF stellar photometry and the SCC. The projection is the same as in Fig.~\ref{fig:dr1_ews_depth_sky}.}
    \label{fig:dr1_edfn_photom_stab_sky}
\end{figure}

\begin{figure}[htp!]
    \centering
    \includegraphics[width=\linewidth]{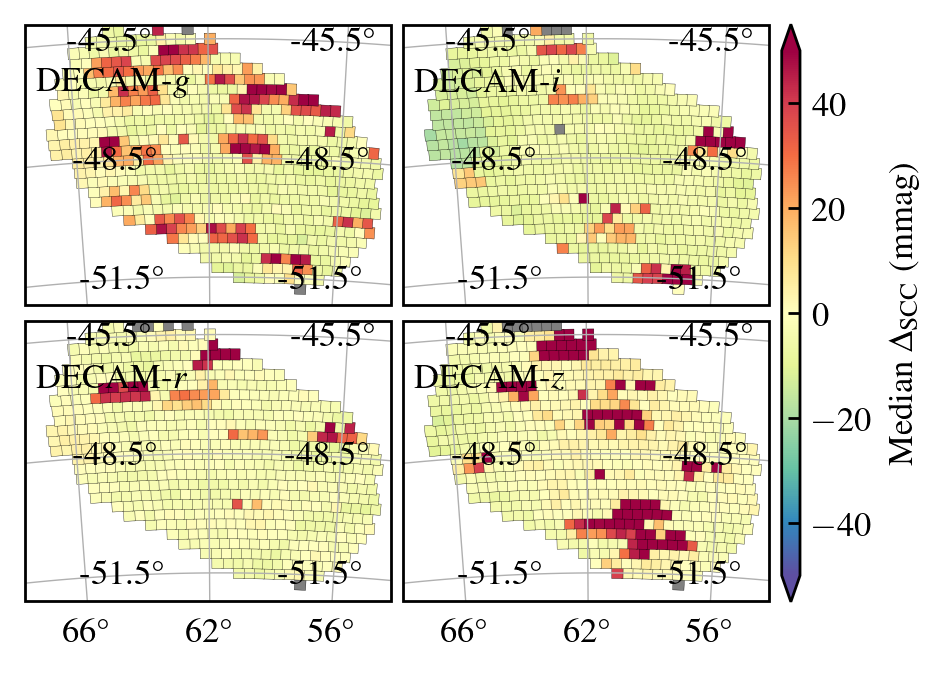}
    \caption{Same as Fig.~\ref{fig:dr1_edfn_photom_stab_sky}, but for EDF-S.}
    \label{fig:dr1_edfs_photom_stab_sky}
\end{figure}

\begin{figure}[htp!]
    \centering
    \includegraphics[width=\linewidth]{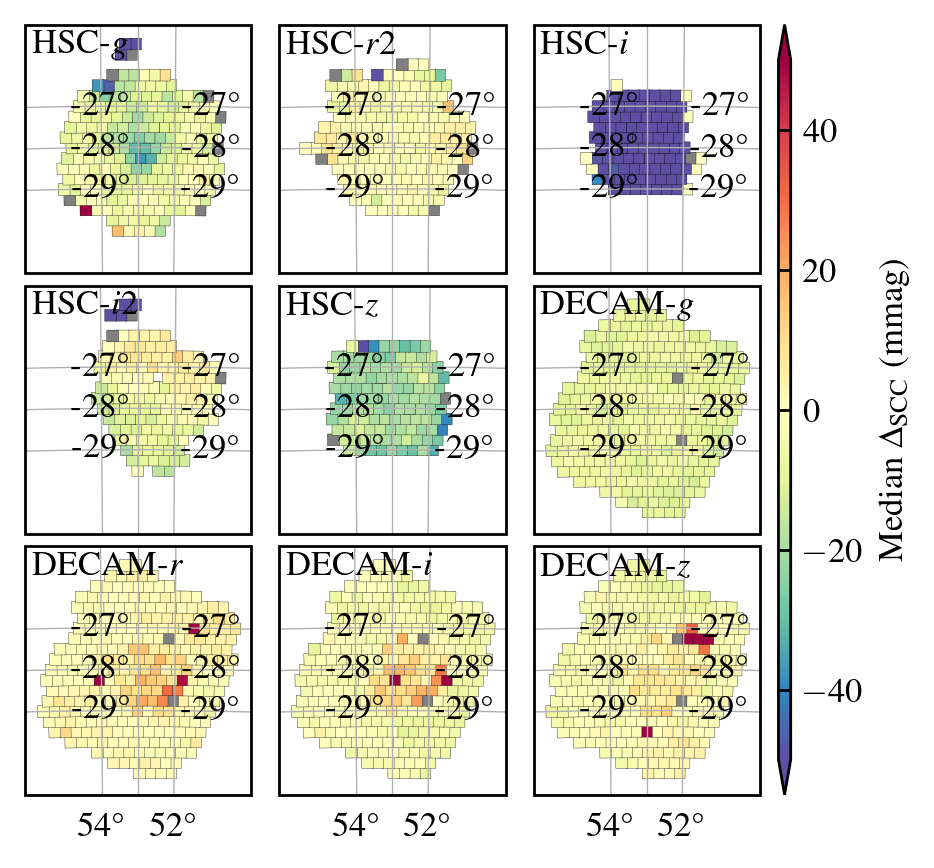}
    \caption{Same as Fig.~\ref{fig:dr1_edfn_photom_stab_sky}, but for EDF-F.}
    \label{fig:dr1_edff_photom_stab_sky}
\end{figure}

\begin{figure}[htp!]
    \centering
    \includegraphics[width=\linewidth]{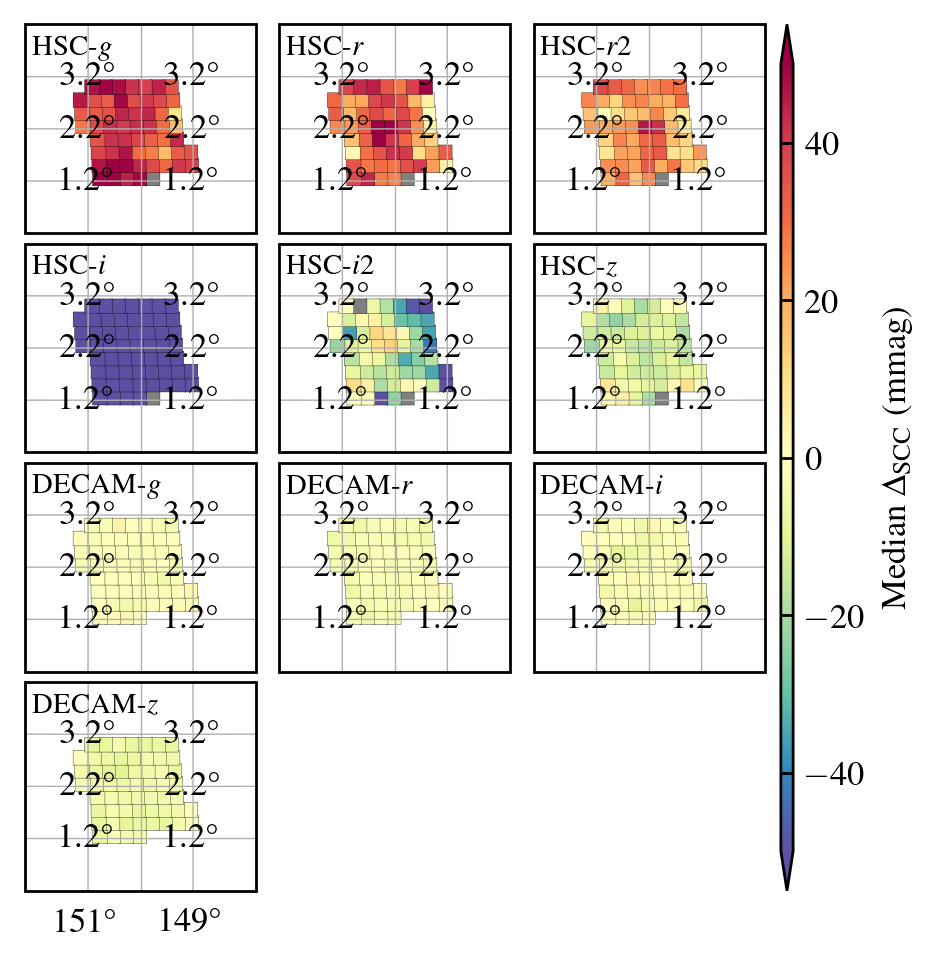}
    \caption{Same as Fig.~\ref{fig:dr1_edfn_photom_stab_sky}, but for COSMOS.}
    \label{fig:dr1_cosmos_photom_stab_sky}
\end{figure}

The deep-field maps (Figs.~\ref{fig:dr1_edfn_photom_stab_sky}--\ref{fig:dr1_cosmos_photom_stab_sky}) show a larger dynamic range of zero-point offsets.
This reflects the increased sensitivity of deep stacks to small systematic effects, as well as the combination of a large number of input SEFs.
In some fields covered by HSC, coherent regions of positive or negative offsets are visible, suggesting a potential problem with the StackedFrame zero-points, consistent with the elevated failure fractions reported in Table~\ref{tab:stacks_flagged}.
These patterns are currently under investigation and will be addressed in future processing iterations.

\twocolumn

\clearpage

\end{appendix}
\label{LastPage}

\end{document}